\documentclass[aps,prd,two column]{revtex4-2}

\usepackage{amsmath}
\usepackage{color}
\usepackage{graphicx}
\usepackage{epsfig}
\usepackage{bm}
\usepackage{hyperref}
\usepackage{natbib}
\usepackage{caption}
\usepackage{subcaption}

\def\be {\begin{equation}}
\def\ee {\end{equation}}
\def\nn {\nonumber}
\def\bea {\begin{eqnarray}}
\def\eea {\end{eqnarray}}

\def\e{E}

\newcommand{\om}{\omega} 

\newcommand{\vk}{\vec k}

\newcommand{\del}{\partial}

\begin{document}

\title{Electric field induction in quark-gluon plasma due to thermoelectric effects}


\author{Kamaljeet Singh}
\email{kspaink84@gmail.com}
\author{Jayanta Dey}
\email{jayantad@iitbhilai.ac.in}
\author{Raghunath Sahoo\footnote{Corresponding Author: Raghunath.Sahoo@cern.ch}}
\email{Corresponding Author: Raghunath.Sahoo@cern.ch}
\affiliation{ Department of Physics, Indian Institute of Technology Indore, Simrol, Indore 453552, India}

\begin{abstract}
Relativistic heavy-ion collisions produce quark-gluon plasma (QGP), which is locally thermalized. Due to electrically charged particles (quarks), QGP exhibits interesting thermoelectric phenomena during its evolution, resulting in an electromagnetic (EM) field in the medium. In this study, for the first time, we estimate the induced electric field in QGP due to the thermoelectric effect. This phenomenon can induce an EM field even in QGP produced by the head-on heavy-ion collision. In peripheral heavy-ion collisions, the presence of a spectator current generates a transient magnetic field at the early stage, which disrupts the isotropy of the induced electric field. For the numerical estimation, we use a quasiparticle-based model that incorporates the lattice quantum chromodynamics equation of state for QGP. The induced electric field is estimated with cooling rates derived from Gubser hydrodynamic flow. Thermoelectric coefficients such as Seebeck, magneto-Seebeck, and Nernst coefficients play a crucial role in determining the induced field.
Additionally, we account for the temperature evolution of QGP using different hydrodynamic cooling rates to calculate the transport coefficients. We also estimate the transport coefficients and the induced electric field in the presence of an external time-varying magnetic field, including the quantum effect of Landau quantization, and explore the effects of the intensity and decay parameter of the magnetic field on the induced electric field. Our findings reveal that the space-time profile of the induced electric field is zero at the center and increases as we go away from the center. During the early stages of evolution, the electric field can reach a maximum value of $eE \approx 1~m_\pi^2$, decreasing in strength over time. 
\end{abstract}

\maketitle
\section{Introduction}

The success of heavy-ion collision experiments is not limited to the discovery of phenomena of strong interaction at very high energy but also contributes to understanding fundamental physics applicable to many different areas, such as condensed matter physics. The quark-gluon plasma (QGP), which is a deconfined and locally thermalized state of quarks and gluons, produced in relativistic heavy-ion collisions (RHICs), undergoes many phases during its evolution~\cite{BREWER2021136189,Busza:2018rrf}. Some phenomena occurred during the evolution~\cite{Elfner:2022iae,PhysRevC.106.014905} significantly impact final state observables to be distinguished, and some might not yet be discovered~\cite{LIAO201699}. It is found that relativistic dissipative hydrodynamics is the most effective theory that can successfully explain final state observables such as spectra and flow~\cite{PhysRevLett.87.182301, PhysRevC.77.054901, article}. Recently, Ref.~\cite{Panda_2023} estimated the effect of the electric field in the flow harmonics considering different electric field configurations. 
The thermodynamic~\cite{PhysRevD.100.076009, PhysRevC.99.035210,Goswami:2023eol} and transport~\cite{Gavin:1985ph, PhysRevD.100.114004, PhysRevD.102.016016, PhysRevD.106.034008, PhysRevD.108.094007} properties of nearly baryon-free QGP matter created in the Large Hadron Collider (LHC) and that of QGP with finite baryon-chemical potential created in Relativistic Heavy-Ion Collider (RHIC) are explored very well. Moreover, going from head-on to peripheral collisions, an intense transient electromagnetic field is created that can be understood from laws of classical electrodynamics~\cite{Tuchin:2015oka, PhysRevD.90.034018,PhysRevLett.125.022301,KHARZEEV2008227,PhysRevX.14.011028}. The change in energy and centrality not only modifies the thermodynamic equation of state (EoS)~\cite{Kumar2024, Brandt2023, Borsanyi2012, Bali2014} of the created medium but many different phenomena, such as chiral magnetic effect~\cite{KHARZEEV2008227, PhysRevD.97.056024}, magnetic and inverse magnetic catalysis~\cite{PhysRevLett.73.3499, PhysRevD.55.6504}, and plausible transition through the quantum chromodynamics (QCD) critical point~\cite{PhysRevD.85.091901,Endrodi2011}. The electromagnetic (EM) field produced in peripheral RHICs due to the moving charged spectators magnetizes the QGP medium~\cite{Wang:2021eud, MCINNES2016173}. The magnitude of the field generated at the early stages of collisions could be as high as $10 - 15$~m$_\pi^2$ ($\approx 10^{18}$ Gauss), which submerged the QCD energy scale $\Lambda_{\rm QCD} \sim 1.5$~m$_\pi$~\cite{Simonov:2021eyt}. Here, m$_\pi$ is the mass of a pion, which is around $140$~MeV. For a physical comparison, the Earth's magnetic field at its surface is 0.25-0.65 Gauss. The created EM field decays in time, which can be explained by the medium's electrical conductivity~\cite{Dey:2021fbo} and the flux conservation as it expands. QGP being made of electrically charged particles (quarks), a nonzero EM field can be produced in the medium irrespective of the spectator current depending on the hydrodynamic flow and charge density profile. This is explored only recently by A. Dash and A. K. Panda in Ref.~\cite{DASH2024138342}. They have found that the produced electric and magnetic fields could be as high as $0.15$~m$_\pi^2$ for the SPS energy range and $1$~m$_\pi^2$ for RHIC energies, respectively. 

In the present work, we estimate the induced electric field in QGP using the thermoelectric effect. The creation of electric potential due to temperature gradient or vice versa is known as thermoelectric effect~\cite{goldsmid2010introduction}. In baryonic QGP, the heat flow is governed by the baryon current carried by the constituent quarks. As the quarks are electrically charged, a net nonzero electric potential develops, resulting in an electric field. This field is estimated using thermoelectric transport coefficients, such as Seebeck, magneto-Seebeck, and Nernst coefficients. In RHICs, these coefficients are studies for QGP~\cite{Das2021, PhysRevD.102.096011} and hadronic matter~\cite{PhysRevD.102.014030, PhysRevD.99.014015} in zero and finite external magnetic field cases. However, none of these investigations estimated the electromagnetic field created in QGP due to thermoelectric processes. In this work, we assessed the induced electric field and looked at the dynamic image of QGP, considering the influence of temperature evolution (or cooling). In a recent study~\cite{PhysRevD.109.014018}, we demonstrated the significance of temperature evolution in QGP observables like elliptic flow. It is essential to mention that unlike the EM field produced in peripheral heavy-ion collisions due to spectators, this field can be induced in QGP created in head-on collisions where a nonzero baryon chemical potential is expected. 
We also estimate the thermoelectric effect-induced electric field in a peripheral collision where a strong transient-external magnetic field in the early stage is expected to affect the QGP significantly. Where the quantum modification of energy states due to Landau quantization is incorporated. This is the first time an electric field generated solely by the thermoelectric effect has been estimated for QGP. The created spatiotemporal electric field can generate a spatiotemporal magnetic field, which is also discussed, along with possible limitations.      
 
The paper is organized in the following manner. First, we briefly derive the cooling rates for Bjorken and Gubser flow in Sec.\ref{R.hydro}. In Sec.\ref{head-on}, we derive the Seebeck coefficient for the head-on collision or $eB = 0$ case. Then, we estimate the Seebeck coefficient and induced electric field with QGP EoS. Then, we extend our study to peripheral collision in Sec.\ref{head-on} where the initial magnetic field is nonzero. We calculate the magneto-Seebeck and Nernst coefficients for the anisotropic system and calculate the induced electric field for the same in Sec.\ref{peripheral}. Finally, in Sec.\ref{sec-summary}, we have summarized our study with a possible outlook. Detailed calculations for thermoelectric coefficients are given in the appendix.  

\section{Relativistic hydrodynamics and cooling rates}\label{R.hydro}
The relativistic hydrodynamics is a successful macroscopic theory to understand the dynamics of relativistic fluid created in RHICs, namely QGP \cite{Rischke:1995mt,Okamoto2016}. Hydrodynamics and the correct EoS of a medium will provide us with a detailed dynamical evolution of the medium and its cooling rate or space-time-dependent temperature profile. Here, to estimate thermoelectric coefficients and induced electric field, an analytical expression of cooling rate is required, for which we use Bjorken and Gubser flow~\cite{PhysRevD.27.140,PhysRevD.82.085027}.   

The symmetric energy-momentum tensor ($T^{\mu \nu}$) of a fluid can be expressed in terms of hydrodynamic degrees of freedom as~\cite{PhysRevC.103.014905},
\begin{align} \label{EMT}
{T}^{\mu\nu}=\left(\epsilon +p\right){u}^{\mu}{u}^{\nu} - p{g}^{\mu\nu} + \pi^{\mu \nu} + W^\mu u^\nu + W^\nu u^\mu~.
\end{align}
Where $\epsilon$, $p$, and $u^\mu$ are the energy density, thermal pressure, and four-velocity of fluid, respectively. Dissipative quantities are viscous stress tensor $\pi^{\mu \nu}$ and energy flow $W^\mu$. The metric tensor of flat spacetime is $g^{\mu\nu}$ = $\rm{diag}(1,-1,-1,-1)$. A hydrodynamic system is governed by the conservation equations - energy-momentum conservation and number or charge conservation. Now, the projection of the energy-momentum conservation equation along the fluid velocity
\begin{align}
{u}_{\mu}{\partial}_{\nu}{T}^{\mu\nu}  =  0,
\label{continuity}
\end{align} 
leads to the continuity equation~\cite{PhysRevLett.99.172301}. One can obtain the required cooling rate by solving Eq.~(\ref{continuity}) with the relevant flow profile and EoS.
 

\subsection{Bjorken flow}
The Bjorken flow represents a boost invariant longitudinal expansion of the medium created in heavy-ion collisions~\cite{PhysRevD.27.140}. Along with longitudinal boost invariant, Bjorken considered translation and rotational invariance symmetry in the transverse plane, which leads to the velocity profile
\begin{equation}
    u^{\mu} = \gamma~\left(1,~0,~0,~\frac{z}{t}\right)~.
    \label{Eq Bjorken flow}
\end{equation}
One can solve the hydrodynamic equations with the flow profile to obtain the system's dynamics. Here, we will briefly describe the cooling rate for an ideal hydrodynamic system with and without a magnetic field.

\subsubsection{Ideal hydrodynamics at $B = 0$}
At the initial time of QGP formation, it behaves as a conformal fluid where all the dissipative quantities vanish, such that conservation of the entropy current is a good assumption. The energy-momentum tensor of such a perfect fluid in the absence of external magnetic fields can be expressed as
\bea
{T}^{\mu\nu}=\left(\epsilon +p\right){u}^{\mu}{u}^{\nu} - p{g}^{\mu\nu}~.
\label{ideal hydro}
\eea
Solving the hydrodynamic Eq.~(\ref{continuity}) by using Bjorken flow Eq.~(\ref{Eq Bjorken flow}) and ideal energy-momentum tensor Eq.~(\ref{ideal hydro}) we get~\cite{PhysRevD.27.140}
\begin{align} \label{ideal}
\partial_{\tau}\epsilon + \frac{\epsilon +  p}{\tau} &=0~,   
\end{align}
where $\tau$ is the proper time. Now, using ideal equation of state $p=\frac{\epsilon}{3} \propto T^{4}$, we can obtain cooling rate for ideal hydrodynamics as,
\begin{align}\label{idT} 
T = T_{0} {\left(\frac{\tau_{0}}{\tau}\right)}^{\frac{1}{3}}~.
\end{align}
Where, initial condition is $T(\tau_0) = T_0$, with initial temperature $T_0$ at the medium formation time $\tau_0$.   

\subsubsection{Ideal magnetohydrodynamics}
In peripheral RHICs, a huge magnetic field is created, which can magnetize the QGP. The energy-momentum tensor for an ideal magnetized fluid can be expressed as~\cite{ROY201545,PhysRevD.109.014018,Panda2021}
\begin{align} \label{eq:EMTensor}
{T}^{\mu\nu}=\left(\epsilon +p+{B}^{2}\right){u}^{\mu}{u}^{\nu}-
\left(p+\frac{{B}^{2}}{2}\right){g}^{\mu\nu}-{B}^{\mu}{B}^{\nu}~.
\end{align}
Depending on the medium's electrical conductivity, the magnetic field created during the collision will decay with time. Here, we consider an exponential decay profile~\cite{Satow:2014lia}
\bea
B(\tau) = B_0 \exp{[-\tau/ \tau_B]}~.
\label{magnetic field}
\eea 
where $B_0$ is the magnitude of initial magnetic field and $\tau_B$ is its decay parameter. The field is directed along the $y$ axis in the transverse plane, considering the longitudinal z-axis to align with the beam axis. 
Now, solving the hydrodynamic equation with ideal EoS, and after some rearrangement, we get 
\begin{align}
\frac{\partial T}{\partial\tau} + p(\tau) T = q(\tau) T^{-3}~,
\end{align}
where p$(\tau)$ = $\frac{1}{3\tau}$, q$(\tau)$ = ($\alpha \beta - \frac{\alpha}{\tau}$)exp(-2$\beta \tau$ ) with $\alpha = \frac{B_0^2}{12a}$, $\beta = \frac{1}{\tau_B}$, and $a = \left(16+\frac{21}{2}N_f\right)\frac{\pi^2}{90}$.
The above equation can be recognized as the Bernoulli differential equation and can be analytically solved to obtain
\begin{align} \label{Temp}
T &= \Big[T_{0}^4 {\left(\frac{\tau_{0}}{\tau}\right)}^{\frac{4}{3}} + \frac{4\alpha}{{(2\beta\tau)}^{\frac{4}{3}}} \Big\{\Gamma{(4/3, 2\beta\tau)} - \Gamma(4/3, 2\beta\tau_0) \Big\} \nn\\
&~~~~- \frac{2\alpha}{{(2\beta\tau)}^{\frac{7}{3}}}\Big\{\Gamma(7/3, 2\beta\tau) - \Gamma(7/3, 2\beta\tau_0) \Big\}\Big]^\frac{1}{4}~.
\end{align}
%
\subsection{Gubser flow}
\label{Subsection:Gubser}
Gubser generalizes the Bjorken flow to include radial expansion of the medium and longitudinal boost-invariant expansion~\cite{PhysRevD.82.085027}. Working with rapidity or the Milne coordinate system is convenient in heavy-ion collisions. Milne coordinates ($\tau, \eta, r, \phi$), can be related to the Cartesian coordinates $(t, x, y, z)$ as,
\begin{align}
    &\tau = \sqrt{t^2 - z^2},~~~~\eta = arctanh \frac{z}{t},\nn\\
    &r = \sqrt{x^2 + y^2},~~~~\phi = arctan \frac{y}{x}.
\end{align}
In this coordinate system, the four-velocity $u^\mu$ constructed from symmetry consideration with  boost, rotation, and reflection invariance ($\eta \rightarrow - \eta$) is given as
\begin{align}
    &u^{\tau} = \cosh{\kappa} = \gamma_{r},~~~ \frac{u^{r}}{u^{\tau}} = \tanh{\kappa} = v_r,~~~ u^\eta = u^\phi = 0.
\end{align}
Where Lorentz factor $\gamma_r$ = $\frac{1}{\sqrt{1-v_{r}^2}}$ and $v_r$ = $\sqrt{\Vec{v}_{x}^2 + \vec{v}_{y}^2}$, $v_r$ is known as transverse velocity. $\kappa$ is parametrized as $\kappa(\tau, r) = {arctanh} (\frac{2q^{2}\tau r}{1 + q^{2} \tau^{2} +  q^{2} r^{2}})$.
Here, we take the case of the viscous medium while preserving the conformal invariance of the theory, that is, 
$p = \epsilon / 3$ and shear viscosity $\eta $= $H_0 \epsilon^ {3/4}$. $H_0$ is a dimensionless quantity. 
Using the Gubser flow, we can solve the hydrodynamic equation for viscous medium to obtain the temperature profile as~\cite{PhysRevD.82.085027},
 \begin{align}\label{GubserT}
  T &= {1 \over \tau f_*^{1/4}} \Big[ {\frac{{T}_0}{(1+g^2)^{1/3}}} + \frac{{\rm H}_0 g}{\sqrt{1+g^2}} \Big\{1 - (1+g^2)^{1/6}\nn\\
  &~~~~~~~~~~~~~~~~~~~~~~{}_2F_1\left({1 \over 2}, {1 \over 6}; {3 \over 2}; -g^2\right) \Big\} \Big]~.    
 \end{align}  
Where ${T}_0$ is an integration constant and ${}_2F_1$ denotes a hypergeometric function with
$g = {1 - q^2 \tau^2 + q^2 r^2 \over 2q\tau}$. For numerical estimations, we used semirealistic numbers for a central gold-gold collision at $\sqrt{s_{\rm NN}} = 200\,{\rm GeV}$ are $\hat{T}_0 = 5.55$ and ${\rm H}_0 = 0.33$ if we choose $1/q = 4.3$~fm. 

\begin{figure}
	\centering
	\includegraphics[scale=0.44]{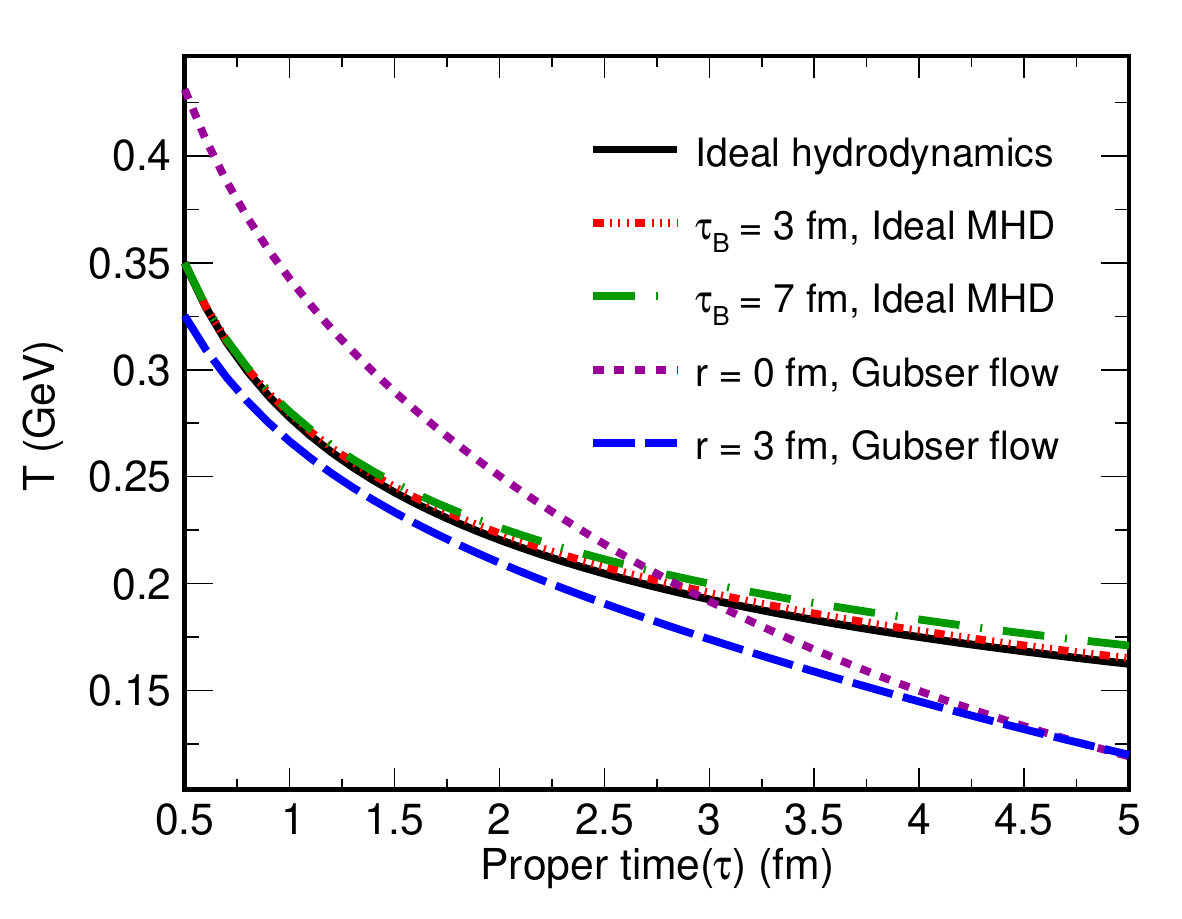}
\caption{Temperature ($T$) as a function of proper time ($\tau$) for ideal hydrodynamics, ideal magnetohydrodynamics (MHD) for $eB_0 = 5~m_\pi^2$ with two decay parameters at $\tau_{B} = 3,~7$~fm, and Gubser flow at $r = 0,~3$~fm.}
	\label{tempall}
\end{figure}

In Fig.~(\ref{tempall}), we have plotted the cooling rate (temperature as a function of proper time)
for all the three hydrodynamic cases discussed above. Note that the initial conditions are free to adjust within the Heisenberg uncertainty principle. The cooling rates are represented as follows. The ideal hydrodynamics at $B = 0$ with a black solid line, ideal magnetohydrodynamics (MHD) at $\tau_{B} = 3,~7$~fm with the red dashed-triple dot line and green dash-dot line, respectively, and for Gubser flow at $r = 0,~ 3$~fm by magenta dashed and blue dashed line, respectively. Interestingly, the temperature in Gubser flow falls faster due to the presence of transverse flow even at the finite viscosity of the medium. The cooling rates plotted here will be helpful in the analysis of results obtained in the subsequent sections.
Note that the magnetic decay parameter values considered here are high and unrealistic in HIC. In principle, it depends on many parameters, such as QGP electrical conductivity and impact parameters, and can be determined through MHD simulation. However, considering such a high parameter value allows us to distinguish the decay effect in the plots, which gives us a qualitative dependency on the decay parameter. For more insight into the lifetime of the electromagnetic field in QGP due to spectator current, see Refs.~\cite{PhysRevC.97.044906, Tuchin2013ie, Huang_2016, Satow:2014lia, Hongo:2013cqa}.

	
\section{Electric field induction in head-on collisions}
	\label{head-on}
 This section calculates the thermoelectric coefficient and induced electric field for the QGP medium created in a head-on collision. Due to the absence of spectator current, no magnetic field is expected to be created in the head-on collisions. Here, we will show that an electric field can be induced in the medium due to the thermoelectric effect, even though there is no initial EM field during the collision.
 
\subsection{Seebeck coefficient in evolving QGP}
QGP created in RHICs cools down rapidly as it goes through space-time evolution with rapid expansion. This rapid cooling affects the medium's transport coefficients, such as thermal conductivity and thermoelectric coefficients. In all the earlier studies of thermoelectric coefficients~\cite{Das2021, Abhishek:2020wjm, Kurian:2021zyb, PhysRevD.102.096011}, a static picture of QGP is considered where the local temperature gradients are present i.e. $\frac{\del {T}}{\del x^k} \neq$ 0, but the effects of cooling are not considered during its evolution i.e. $\frac{\del {T}}{\del t}$ = 0, which is a more realistic scenario. In Ref.~\cite{PhysRevD.109.014018}, for the first time, we have estimated the effect of temperature evolution on the thermal conductivity of QGP. Here, we will calculate the thermoelectric coefficient for evolving QGP using a similar approach of Ref.~\cite{PhysRevD.109.014018}. We will also show the difference between the expressions obtained considering a static QGP.  

To study the fluid properties of QGP medium, we consider a system of relativistic fluid consisting of particles with mass $m_i$, momentum $\vec{k_i}$, energy $\om_i = \sqrt{\vec k_i^2+m_i^2}$, and chemical potential $\mu_i = {\rm b}_i \mu_B$ for i$^{\rm th}$ species, with b$_i$ as baryon quantum number and $\mu_B$ as baryon chemical potential. 
The single particle distribution function at equilibrium for i$^{\rm th}$ species is
\begin{align}\label{Dis-f}
f^0_i= \frac{1}{e^{\frac{\om_i-{\rm b}_i\mu_B}{T}}\pm 1}~,
\end{align}
where $\pm$ stands for fermion and boson, respectively. The total single-particle distribution function ($f_i$) for a system slightly out of equilibrium ($\delta f_i$) can be written as $f_i=f^0_i+\delta f_i$. When a system is out of equilibrium, net nonzero currents arise in the system. In kinetic theory, electric current density for such a system can be expressed as~\cite{PhysRevD.108.094007}
\begin{align}\label{Cur-den}
	\vec{j} =  \sum_{i} q_i g_i \int \frac{d^3|\vk_i|}{(2\pi)^3} \frac{\vec{k_i}}{\om_i} \delta f_i~. 
\end{align}
Here, $q_i$ is the electric charge, and $g_i$ is the degeneracy of the i$^{\rm th}$ species particles. Similarly, the microscopic definition of heat flow~\cite{PhysRevD.108.094007, PhysRevD.109.014018},
\begin{align}\label{mic1}
{\vec {I}}_i &= \sum_{i} g_i\int \frac{d^3|\vk_i|}{(2\pi)^3} \frac{\vec {k}_i}{\om_i} (\om_i -{\rm b}_i h)\delta f_i~.
\end{align}
Where $h$ is the total enthalpy per particle.
 To find the expression of $\delta f_i$, we solve the Boltzmann transport equation (BTE) with the help of relaxation time approximation (RTA)~\cite{PhysRevD.102.016016}
\begin{align}\label{BTE-RTA0}
	\frac{\del f_i}{\del t} + \frac{\vk_i}{\om_i} \cdot \frac{\del f_i}{\del \vec{x_i}} + q_i \vec{E}\cdot \frac{\del f_i}{\del \vec{k_i}}
	= -\frac{\delta f_i}{\tau^i_R}~,
\end{align}
where $\tau^i_R$ is the relaxation time of the particle. $\Vec{E}$ represents the external electric field. Here, the deviation of the distribution function from equilibrium is driven by the electric field and temperature gradients. With leading order contribution, we can consider an ansatz of $\delta f_i$ as~\cite{PhysRevD.104.094037}
\begin{align}\label{delta-f0}
	\delta f_i = (\vec{k_i} \cdot \vec{\Omega}) \frac{\del f^0_i}{\del \om_i}~.
\end{align}  
As QGP is a rapidly cooling system, to account for all leading order contributions from gradient forces, a general form of unknown vector ${\vec \Omega}$ can be assumed as \begin{align}\label{Omega00}
\vec{\Omega} = &~\alpha_1 \vec{E} +\alpha_2 \vec{\nabla}T + \alpha_3 \vec{\nabla}\dot{T}~.
\end{align}  
The unknown coefficients $\alpha_j$ ($j=1,2,3$) determine the strength of the respective gradient force fields driving the system away from equilibrium. The coefficient $\alpha_3$ corresponds to the term that arises due to the evolving picture of the medium. For the case of a static picture, where the effects of cooling are not considered, only $\alpha_1$ and $\alpha_2$ contribute, whereas $\alpha_3$ vanishes in Eq.~(\ref{Omega00}) as in Ref.~\cite{Das2021, Kurian:2021zyb}. We can solve Eq.~(\ref{BTE-RTA0})  using Eqs.~(\ref{delta-f0}) and (\ref{Omega00}), and get the expression of $\alpha_j$'s. After getting the expressions for $\alpha_j$'s, the simplified form of $\delta f_i$ is (see Appendix~\ref{appendix2}),
\begin{align}
 \delta f_i &= -q_i \tau_R^i
 (\vec{v_i}.\vec{E})
 + \frac{(\om_i - {\rm b}_ih) \tau_R^i}{T}\Big[\Big(\vec{v_i}\cdot\vec{\nabla}{T}\Big)\nn\\
&-\tau_R^i\Big(\vec{v_i}\cdot\vec{\nabla}{\dot T}\Big)\Big]\frac{\partial f^0_i}{\partial \om_i}.
\end{align}
Using the expressions of $\delta f_i$ in Eq.~(\ref{Cur-den}), we can express electric current as 
\begin{align}\label{Jl}
j^l = &\sum_{i} \frac{q_i g_i}{3} \int \frac{d^3|\vk_i|}{(2\pi)^3} v_i^2 \tau_R^i\Big[-q_iE^l+ \frac{(\om_i - {\rm b}_ih)}{T}\nn\\
&\Big\{
\frac{\del {T}}{\del x^l} - \tau_R^i \frac{\del {\dot T}}{\del x^l}\Big\}\Big]\frac{\partial f^0_i}{\partial \om_i}.
\end{align}

%
The components of the electric current in the x,y, and z directions are given as,
\begin{widetext}
\begin{align}\label{J0c}
j^x &= \sum_{i} \frac{q_i g_i}{3} \int \frac{d^3|\vk_i|}{(2\pi)^3} v_i^2 \tau_R^i\Big[-q_iE^x+ \frac{(\om_i - {\rm b}_ih)}{T}\Big\{
\frac{\del {T}}{\del x} - \tau_R^i \frac{\del {\dot T}}{\del x}\Big\}\Big]\frac{\partial f^0_i}{\partial \om_i}.\nn\\
j^y &= \sum_{i} \frac{q_i g_i}{3} \int \frac{d^3|\vk_i|}{(2\pi)^3} v_i^2 \tau_R^i\Big[-q_iE^y+ \frac{(\om_i - {\rm b}_ih)}{T}\Big\{
\frac{\del {T}}{\del y} - \tau_R^i \frac{\del {\dot T}}{\del y}\Big\}\Big]\frac{\partial f^0_i}{\partial \om_i}.\nn\\
j^z &= \sum_{i} \frac{q_i g_i}{3} \int \frac{d^3|\vk_i|}{(2\pi)^3} v_i^2 \tau_R^i\Big[-q_iE^z+ \frac{(\om_i - {\rm b}_ih)}{T}\Big\{
\frac{\del {T}}{\del z} - \tau_R^i \frac{\del {\dot T}}{\del z}\Big\}\Big]\frac{\partial f^0_i}{\partial \om_i}.  
\end{align}
\end{widetext}
We introduce the following integrals to write Eqs.~(\ref{J0c}) in compact form as
 \begin{align}\label{Li}
&L_{1i} = \frac{g_i }{3T} \int \frac{d^3|\vk_i|}{(2\pi)^3}
\frac{\vk^2_i}{\om_i^2}  \tau_R^if^0_i(1-f^0_i)~, \nn\\
&L_{2i} = \frac{g_i}{3T}   \int \frac{d^3|\vk_i|}{(2\pi)^3}\frac{\vec{k}^2_i}{\om_i^2}(\om_i - {\rm b}_i h) \tau_R^i f^0_i(1 - f^0_i)~,\nn\\    
&L_{3i} = \frac{g_i}{3T} \int \frac{d^3|\vk_i|}{(2\pi)^3}\frac{\vec{k}^2_i}{\om_i^2}(\om_i - {\rm b}_i h) \tau_R^{i2} f^0_i(1 - f^0_i).
\end{align}
Hence,  Eqs.~(\ref{Jl}) can be written  as,
\begin{align} \label{j0}
    j^l &= \sum_i q_i^2 L_{1i} E_l  - \frac{1}{T} \sum_i q_i \Big(L_{2i} - L_{3i} \frac{d}{dt}\Big) \frac{dT}{dx^l}~.
\end{align}
In above Eq., by setting $j_x$ = $j_y$ = $j_z$=0, we can get $E_x$, $E_y$ and $E_z$ in terms of temperature gradients $\frac{dT}{dx}$, $\frac{dT}{dy}$ and $\frac{dT}{dz}$. Here, we rearrange  the Eqs.~(\ref{j0}) in the matrix form as
\begin{align}
    \mathbf{\boldsymbol{\sigma}~E = L~X},
\end{align}
where,
\begin{align}
&\boldsymbol{\sigma} = 
\begin{pmatrix}
\sigma_e & 0 & 0 \\
0 & \sigma_e & 0\\
0& 0 & \sigma_e \\
\end{pmatrix},~\hspace{10mm}
\mathbf{E} = 
 \begin{pmatrix}
E_x\\E_y\\E_z
\end{pmatrix},~\nn\\
 &\mathbf{L} = \frac{1}{T} 
\begin{pmatrix}
L_{23} & 0 & 0 \\
0 & L_{23}  & 0\\
0 & 0 & L_{23} \\
\end{pmatrix},~
\mathbf{X} =
\begin{pmatrix}
\frac{dT}{dx}\vspace{2mm}\\
\frac{dT}{dy}\vspace{2mm}\\
\frac{dT}{dz}\\
\end{pmatrix}.\nonumber    
\end{align}
Hence, after finding the inverse of $\boldsymbol{\sigma}$ matrix, we can obtain the components of the electric field as,
\begin{align}\label{matrix}
    \mathbf{E} &= \mathbf{(\boldsymbol{\sigma}^{-1} L)X} \nn\\
\Rightarrow \begin{pmatrix}
E_x\\E_y\\E_z
\end{pmatrix}
 &= 
\begin{pmatrix}
S & 0 & 0 \\
0 & S & 0\\
0 & 0 & S \\
\end{pmatrix} 
\begin{pmatrix}
\frac{dT}{dx}\vspace{2mm}\\
\frac{dT}{dy}\vspace{2mm}\\
\frac{dT}{dz}\\
\end{pmatrix}~.   
\end{align}
Where $\sigma_{el}$ = $\sum_i q_i^2 L_{1i}$ is electrical conductivity. Further, we defined the integrals $L_{23}$ = $\sum_i q_i \Big(L_{2i} - L_{3i} \frac{d}{dt}\Big)$.
Note that it is a generalized study of thermoelectric coefficients, i.e., we did not consider any preferred direction of the temperature gradient or a particular cooling rate. Therefore, we get the Seebeck coefficient($S$) in three directions. 
Here, we identify the Seebeck coefficient $S$ as,
\begin{align}\label{s0}
   S &=  \frac{(\frac{L_{23}}{T^2})}{(\frac{\sigma_{el}}{T})}~. 
\end{align}
Due to the boost invariance, $T$ is invariant along the longitudinal $z$ axis, therefore $\frac{\partial T}{\partial z} = 0$. Hence, only $x$ and $y$ components of the electric field $\vec E$ survive in the transverse plane. 
 
\subsubsection{Analytical solution for massless noninteracting system}\label{ana}
Here, we address the analytic solution of Seebeck coefficient Eq.~(\ref{s0}) for massless ($\omega_i = k_i$) and noninteracting system. 
In this case, the Seebeck coefficient can be solved analytically using the Riemann zeta function $\zeta(s)$ and
Dirichlet eta $\eta(s)$ function:
\begin{align}
    \zeta(s) = \frac{1}{\Gamma(s)}\int_{0}^{\infty} \frac{x^{s-1}}{e^{x}-1}dx~,~~&~~
    \eta(s) = \frac{1}{\Gamma(s)}\int_{0}^{\infty} \frac{x^{s-1}}{e^{x}+1}dx~,
\end{align}
where, $\Gamma(s)$ is the gamma function. Now, using the relation $\eta(s) = 1-2^{(1-s)}\zeta(s)$, Eqs.~(\ref{Li}) can be write in terms of the $\zeta(s)$ as,
\begin{align}
    L_{1i} &= \sum_i \frac{g_i \tau_R^i T}{6 \pi^2} \Big\{ 2 \zeta(2)+ \Big(\frac{b_i \mu_{B}}{T}\Big)^2\Big\},\nn\\
    L_{2i} &= \sum_i \frac{g_i \tau_R^i T^2}{6 \pi^2} \Big(\frac{b_i^3 \mu_{B}^3}{T^2} + 6 b_i \mu_{B} \zeta(2) - 2 b_i h \zeta(2)  - \frac{b_i^3 \mu_{B}^2 h}{T^2}\Big),\nn\\
    L_{3i} &= \sum_i \frac{g_i \tau_R^{i2} T^2}{6 \pi^2} \Big(\frac{b_i^3 \mu_{B}^3}{T^2} + 6 b_i \mu_{B} \zeta(2) - 2 b_i h \zeta(2)  - \frac{b_i^3 \mu_{B}^2 h}{T^2}\Big).
\end{align}
Hence, using the values of $L_{1i,2i,3i}$ we get
\begin{align}
    \sigma_{el} &= \sum_i \frac{g_i q_i^{2}\tau_R^i T}{6 \pi^2} \Big\{ 2 \zeta(2)+ \Big(\frac{b_i \mu_{B}}{T}\Big)^2\Big\},\nn\\
    L_{23} &= \sum_i q_i~\frac{g_i \tau_R^i T^2}{6 \pi^2} \Big(\frac{b_i^3 \mu_{B}^3}{T^2} + 6 b_i \mu_{B} \zeta(2) - 2 b_i h \zeta(2)  \nn\\
    & -\frac{b_i^3 \mu_{B}^2 h}{T^2}\Big)\left(1 - \tau_R^i \frac{d}{dt} \right).
\end{align}
Note that, $L_{23}$ is proportional to electric charge $q_i$. Degeneracy is the same for all three flavors, $g_i = 6$ (for a spin and color degeneracy). $\tau_R^i = \tau_R$ (say) is the same for all the flavors for massless and noninteracting systems of particles. Therefore, the charge of the three flavors (up, down, and strange) adds up to zero, i.e., $L_{23} = 0$. As a result, the Seebeck  coefficient for the massless noninteracting system is
\begin{align}
   S =  \frac{(\frac{L_{23}}{T^2})}{(\frac{\sigma_{el}}{T})} = 0~. 
\end{align}

\subsubsection{Numerical solution for QGP EoS}
\begin{figure*}
	\centering
	\includegraphics[scale=0.40]{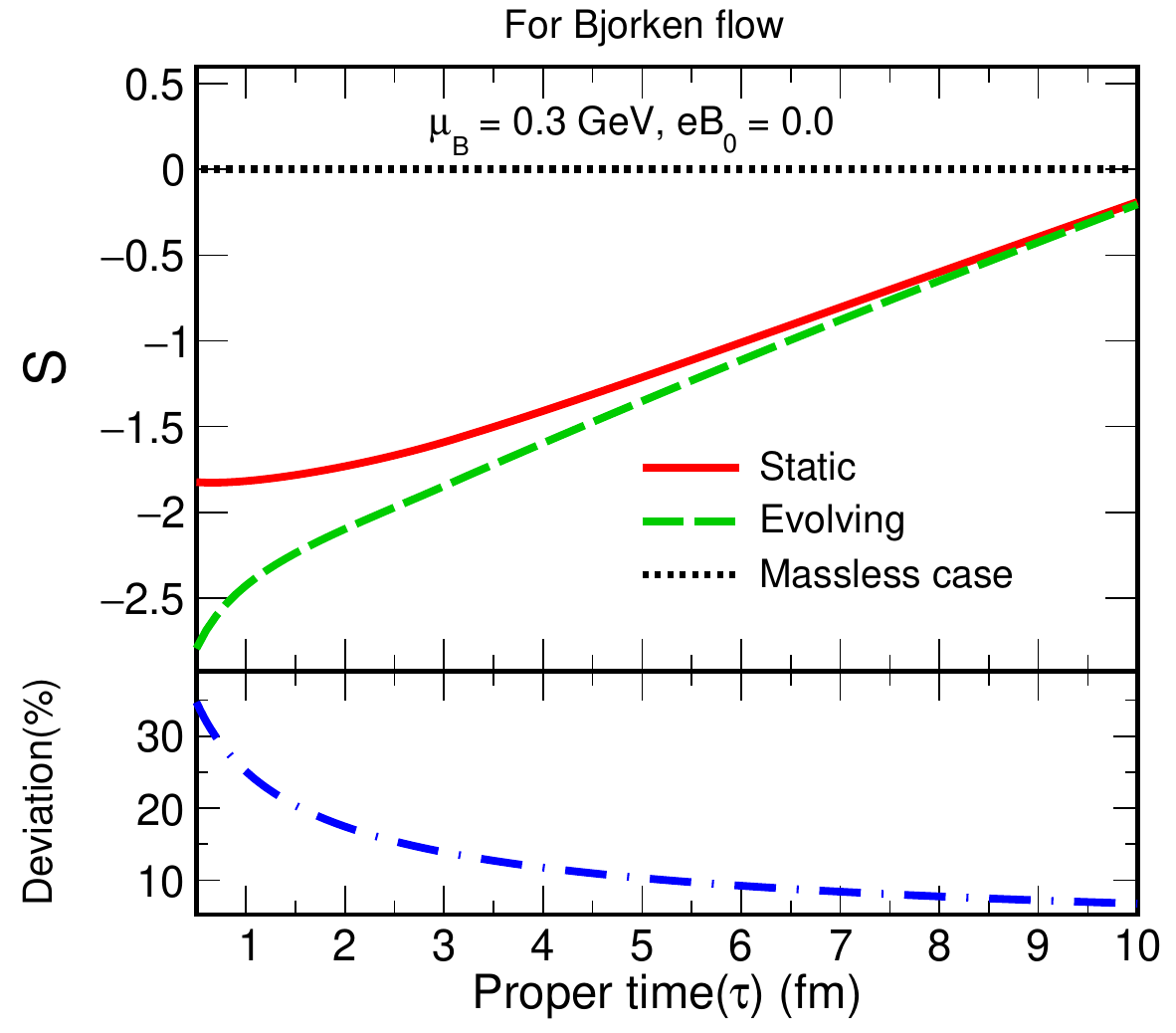}
 \includegraphics[scale=0.40]{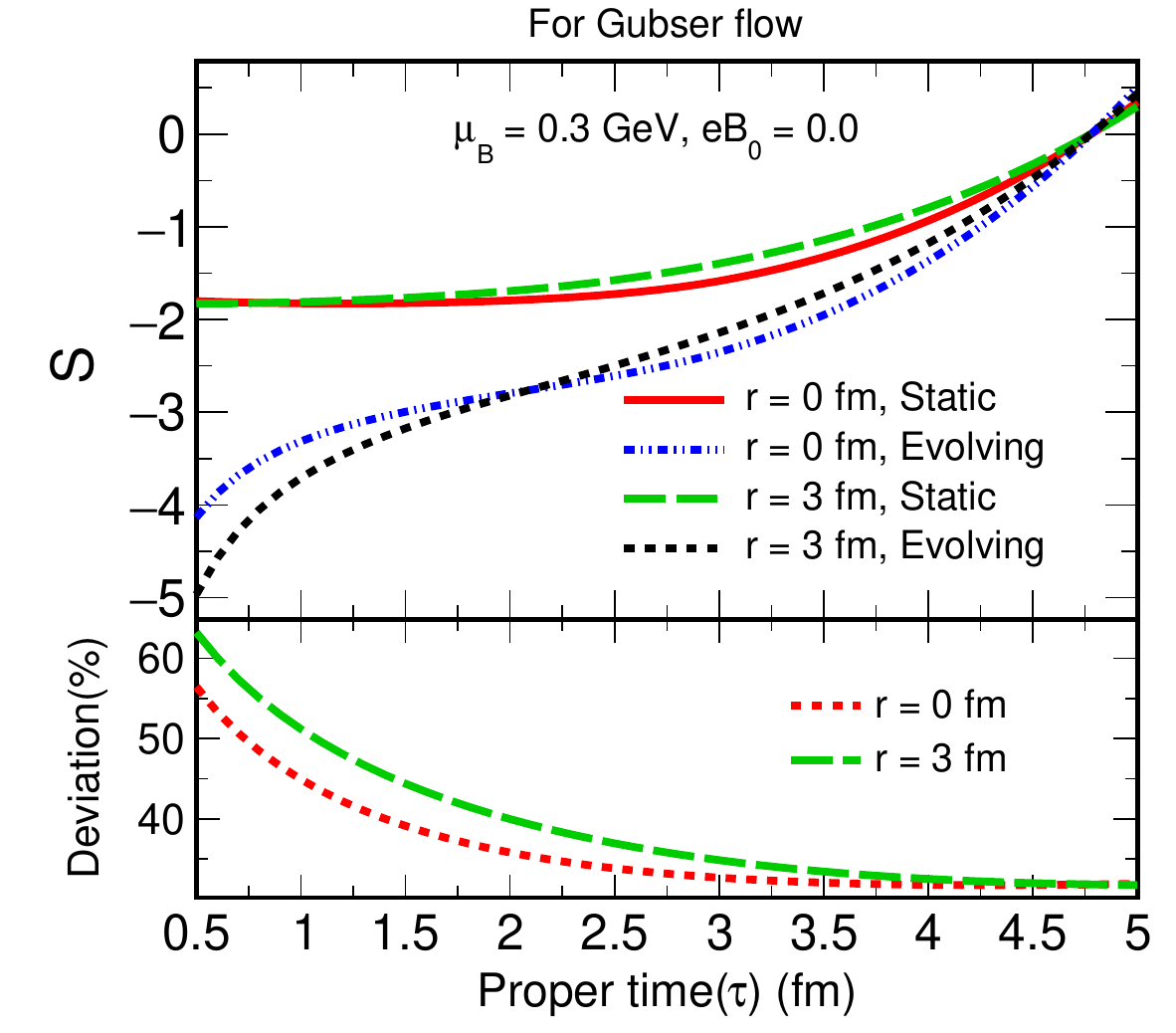}
\caption{Seebeck coefficient ($S$) as a function of proper time ($\tau$) at $\mu_B = 0.3$~GeV. Left: Upper panel represents $S$ for Bjorken flow,  bottom panel represents its percentage deviation of static from evolving picture \Big($ = \frac{(S^{e} - S^{s})}{S^{e}}\times 100 \Big)\% $. Right: Similar to the left plot but for Gubser flow at two different values of $r = 0,~3$~fm.}
	\label{Fig-up}
\end{figure*}
We use a quasiparticle model formulated by Gorenstein and Yang~\cite{Gorenstein:1995vm} for the numerical estimation. It is a phenomenological model where the lattice QCD equation of state for QGP is achieved by considering the thermal masses of the partons. The thermal mass $m(T)$ arises from the interactions among the partons. The thermodynamic consistency is achieved by introducing a bag constant arising from vacuum energy~\cite{Srivastava:2010xa}.  
The dispersion relation of the particle having energy $\om_i$ and momentum $k_i$ is $\om_i^{2}(k_i, T) = k_i^{2} + m_i^{2}(T)$. Where $m_i$ is the total effective mass of ith quark flavor and can be parametrized as 
\begin{align}
    m_i^2 = m_{i0}^2 + \sqrt{2}m_{i0}m_{iT} + m_{iT}^2~.
\end{align}
$m_{i0}$ and $m_{iT}$ are bare mass and thermal mass of the ith flavor with,
\begin{align}
    m_{iT}^2(T,\mu_B) = \frac{N_c^2 - 1}{8N_c}\Big( T^2 + \frac{\mu_B^2}{9 \pi^2}\Big)g^2(T).
\end{align}
The effective mass of gluon $(m_g)$ in this model can be represented as
\bea
m_{g}^2(T,\mu_B) = \frac{N_c}{6} g^2(T) T^2 \left(1+\frac{N_f + \frac{\mu_B^2}{\pi^2 T^2}}{6}\right).
\eea
$N_c$ represents the number of color degrees of freedom and $g^2(T) = 4\pi \alpha_s(T)$, $\alpha_s(T)$ is running coupling constant. We have used a One-loop leading-order running coupling constant \cite{Berrehrah:2013mua, PhysRevD.84.094004}, which is similar to the one used in lattice calculations for QCD.
For more details of the model, one can follow references~\cite{Gorenstein:1995vm, Srivastava:2010xa, Bannur:2006hp, Peshier:1995ty, Peshier:1999ww, Sambataro:2024mkr}. Note that, for the estimation of thermoelectric coefficients and the induced electric field, gluons contribute only to the enthalpy of the system.  
Now, for the relaxation time of quarks, we use a momentum-independent expression obtained for QCD matter~\cite{Hosoya:1983xm}
\begin{align}
    \tau_R = \frac{1}{5.1T\alpha_s^{2}\log(1/\alpha_s)[1+0.12(2N_f+1)]}.
\end{align}
For numerical estimation, the value of the strong coupling constant is taken to be fixed at $\alpha_s = 0.5$. 

In Fig.~(\ref{Fig-up}), we plot the Seebeck coefficient ($S$) as a function of proper time ($\tau$) at $\mu_B$ = 0.3 GeV in head-on collisions (or $eB_0 = 0$) for the static and evolving picture of QGP. Plots in the left panels are obtained in Bjorken flow with the ideal hydrodynamic cooling rate, which in the right panels are for Gubser flow. In both the flow cases, we observed that the magnitude of coefficient $S$ decreases with an increase in the values of $\tau$. In both plots, we observe that coefficient $S$ approaches zero at later stages of evolution, whereas, for the limiting case of massless partons (black dotted line), the value of $S$ is zero throughout the evolution. Note that the negative sign in the value of $S$ implies that the direction of the induced electric field is opposite to the temperature gradient. With time, the temperature of the medium falls off, and $S$ reduces in both the hydrodynamic pictures. $S$ for Gubser flow reduces comparatively faster because of its cooling rate. From Eq.~(\ref{s0}), $S$ is the ratio of two different quantities; for the static picture where integral $L_{23}$ reduces to $L_{2i}$, the coefficient $S$ becomes independent of relaxation time $\tau_R$. This happens because we consider the same relaxation time for all the quarks regardless of their flavor. Whereas, for the evolving picture, $S$ shows its dependence on $\tau_R$ through integral $\L_{23}$. 
In the bottom panels of Fig.~(\ref{Fig-up}), the percentage deviation of static from the evolving picture is plotted against $\tau$ to quantify the cooling effect. In both the hydrodynamic cases, $S$ is enhanced due to temperature evolution. The effect is high in the early times and decreases with time. Moreover, this is higher in the Gubser flow case, which shows the impact is sensitive to the cooling rates.
\subsection{Induced electric field in isotropic system}
\label{Subsection:electricB0}
\begin{figure*}
	\centering
	\includegraphics[scale=0.55]{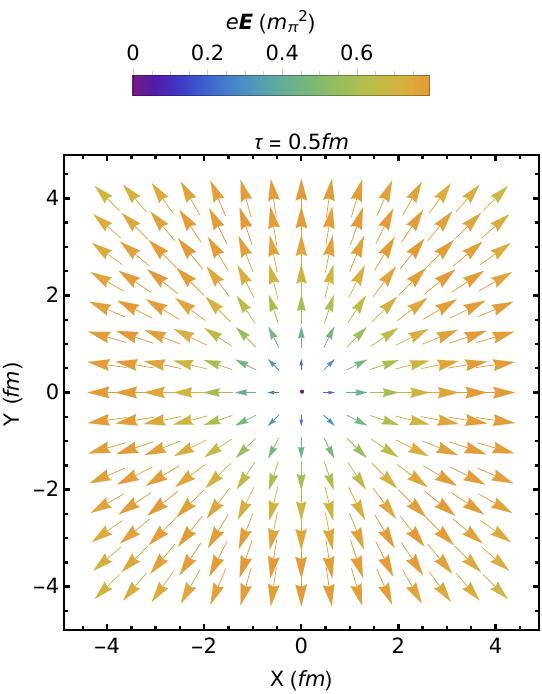}
        \includegraphics[scale=0.55]{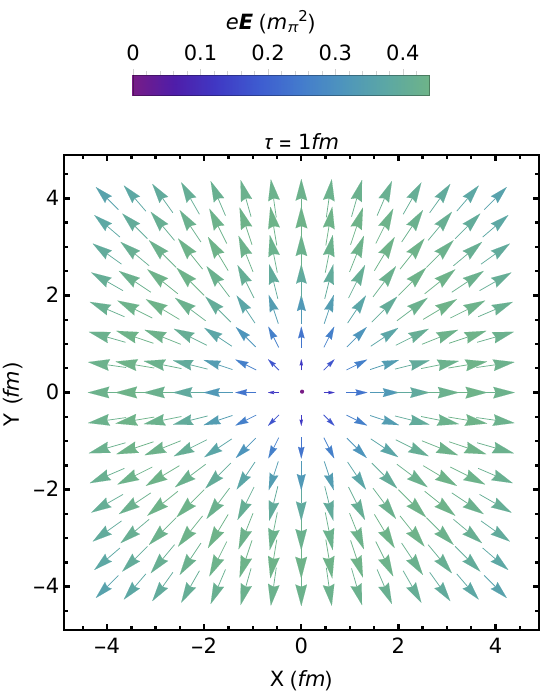}
        \includegraphics[scale=0.55]{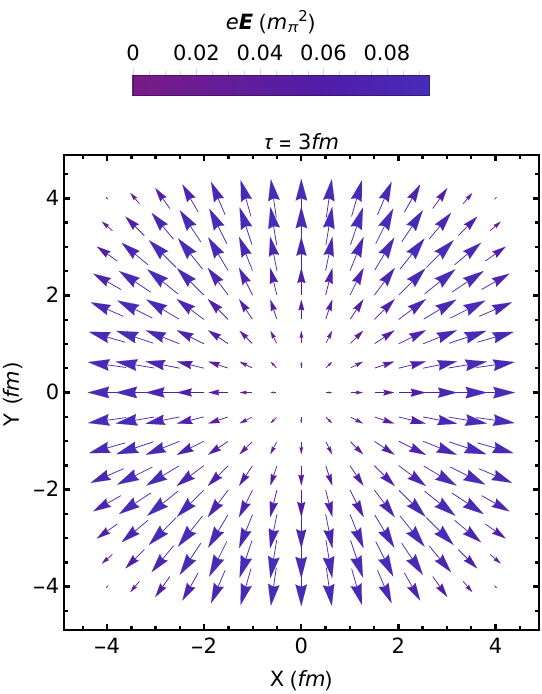}
    \caption{Time evolution of the induced electric field in the QGP in the head-on collisions with $eB_0 = 0$. Left: at $\tau = 0.5$ fm, middle: at $\tau = 1$ fm, and right: at $\tau = 3$ fm.}
\label{E0}
\end{figure*}
To estimate the induced electric field ($\vec E$) in the head-on collisions, we considered baryon chemical potential $\mu_B = 0.3$~GeV, attained at RHIC energies. Here, we used the Gubser cooling rate for the temperature gradient. From Gubser flow, $T = T(\tau, ~r)$, with $r = \sqrt{x^2 + y^2}$. Due to Gubser symmetry considerations as mentioned in subsection~(\ref{Subsection:Gubser}), $T$ is symmetric in the transverse ($x - y$) plane and remains invariant along the longitudinal direction or $z$ axis. Here,  we used the Seebeck coefficient calculated from the Bjorken flow for the case of evolving QGP.
In head-on collisions, there is no spectator current; the magnetic field is not produced during the collision. In this paper, by the terms ``isotropy" and ``anisotropy," we indicate momentum isotropy or anisotropy due to magnetic field only. We will discuss the induced electric field created in an isotropic QGP (without an external magnetic field) due to the thermoelectric effect. 

In Fig.~(\ref{E0}), we have shown the vector plot of the induced electric field in the transverse plane for head-on collision. In an isotropic medium, the thermoelectric transport coefficient matrix is isotropic as obtained in Eq.~(\ref{matrix}), only diagonal elements, Seebeck coefficient ($S$) are nonzero. Due to symmetry consideration, $T$ and its gradient are symmetric in the transverse ($x - y$) plane. Moreover, because of boost invariant symmetry along $z$ axis, $\frac{\partial T}{\partial z} = 0$. Therefore, only $x$ and $y$ components of the electric field survive, and $\vec E$ is symmetric in the transverse plane. Time evolution of $\vec E$ is shown by three panels in the Fig.~(\ref{E0}), from left to right $\tau = 0.5,~1,~3$~fm. The direction and magnitude of $\vec E$ are represented in the figures by the arrow tip and its length, respectively. Magnitude is also colored for clarity; the color bar on the top of the figures is given for magnitude in the unit of $m_\pi^2$. $\vec E$ is zero at the center (as $ \vec{\nabla} T$ vanishes) and directed radially outward with increasing magnitude. 
At $\tau = 0.5$~fm, the induced field is as high as $eE \approx 0.8~m_\pi^2$, and its strength decreases as the medium evolves in time. The direction of $\vec E$ can be understood as follows. From Eq.~(\ref{matrix}), $E_x = S \frac{dT}{dx}$, $E_y = S \frac{dT}{dy}$. From the right panel of Fig.~(\ref{Fig-up}), $S$ is negative at the early time. $\vec{\nabla} T$ is also negative, as $T$ decreases as we go away from the center ($x = y =0, ~{\rm or},~r = 0$). This explains the direction of $\vec E$. Value of $S$, $\vec{\nabla} T$, and hence $\vec E$ is sensitive to the cooling rate. As an analytical expression of the cooling rate is required for the calculation of thermoelectric coefficients and induced field, we used Gubser flow. However, in (3+1)D hydrodynamic simulations, one can incorporate the current study to estimate induced fields in QGP produced in heavy-ion collisions.

\section{Electric field induction in Peripheral collisions}\label{peripheral}
In peripheral heavy-ion collisions, a strong transient magnetic field is produced. In this section, we calculate an evolving relativistic fluid's thermoelectric coefficients and induced electric field in an external time-varying magnetic field. For detailed calculations, see Appendix~\ref{appendix2}. 

\subsection{Magneto-Seebeck and Nernst coefficients in evolving QGP}
The magnetic field breaks the rotational symmetry of the system, due to which the thermoelectric coefficient matrix becomes anisotropic with two independent elements - magneto-Seebeck and Nernst coefficients. Moreover, according to the Landau quantization, the energy level of charged particles gets quantized in the presence of the magnetic field.
However, in the RTA formalism, Landau quantization is not included. Therefore, deriving the transport coefficients in the presence of a magnetic field in the upcoming subsection, we introduce phase-space quantization with the modified dispersion relation for the Landau quantization in the subsequent subsection. 

\subsubsection{Without Landau quantization}
In the presence of a magnetic field, to find the expression of the deviated part of distribution function $\delta f_i$, which is driven by the temperature gradient, we use the RTA method with a Lorentz force term in BTE. In the presence of an external electromagnetic field,  BTE under RTA can be expressed as~\cite{PhysRevD.102.016016}
\begin{align}\label{BTE-RTA}
	\frac{\del f_i}{\del t} + \frac{\vk_i}{\om_i} \cdot \frac{\del f_i}{\del \vec{x_i}} + q_i \left(\vec{E} + \frac{\vk_i}{\om_i} \times \vec{B}\right) \cdot \frac{\del f_i}{\del \vec{k_i}}
	= -\frac{\delta f_i}{\tau^i_R}~,
\end{align}
where $\tau^i_R$ is the relaxation time of the particle. We consider a time-varying electromagnetic field of the form ~\cite{Satow:2014lia, Hongo:2013cqa}
\begin{align}
	B = B_0 \exp{\left(-\frac{\tau}{\tau_B}\right)},~~&~~
	E = E_0 \exp{\left(-\frac{\tau}{\tau_E}\right)},\label{Mag-Field}
\end{align}
where $B_0,~ E_0$ are the magnitudes of the initial fields having decay parameters of $\tau_B$ and $\tau_E$, respectively, and $\tau$ is the proper time. 
The deviation of the distribution function from equilibrium is driven by the electromagnetic field, so leading order contribution in $\delta f_i$ can be~\cite{PhysRevD.104.094037}
\begin{align}\label{delta-f}
	\delta f_i = (\vec{k_i} \cdot \vec{\Omega}) \frac{\del f^0_i}{\del \om}~.
\end{align}  
The general form of unknown vector ${\vec \Omega}$ can be assumed as (up to first-order in time derivative)
\begin{align}\label{Omega}
\vec{\Omega} = &~\alpha_1 \vec{E} + \alpha_2 \dot{\vec E} +\alpha_3 \vec{\nabla}T + \alpha_4 \vec{\nabla}\dot{T} + \alpha_5 (\vec{\nabla}T \times \vec{B}) \nn\\
 &+ \alpha_6 (\vec{\nabla}T\times \dot{\vec{B}})
 + \alpha_7 (\vec{\nabla}\dot{T} \times \vec{B})  + \alpha_8 (\vec{E} \times {\vec{B}}) \nn\\
 &+ \alpha_9 (\vec{E} \times {\dot{\vec{B}}})
 + \alpha_{10}(\dot{\vec{E}} \times {\vec{B}}),
\end{align}  
where the unknown coefficients $\alpha_j$ ($j=1,2,..7$) are helpful to determine the strength of the respective field in driving the system away from the equilibrium. The contribution of coefficients $\alpha_4$ and $\alpha_7$ is due to the considerations of the cooling effects, i.e., for the case of the evolving picture of the medium. Whereas for the static picture, both these coefficients will vanish. Similarly, for the constant electromagnetic field the coefficients $\alpha_2$, $\alpha_6$, $\alpha_9$, and $\alpha_{10}$ vanishes. The terms $\vec{B}$, $\dot{\vec B}$ do not contribute to the current for the case where the chiral chemical potential is zero ~\cite{Satow:2014lia}. 

In the current work, we introduce the cooling rate using hydro-dynamical theories and study the thermoelectric response of evolving QGP in the presence of a slowly varying magnetic field, where we can consider that the inverse of cyclotron frequency is approximately equal to the magnetic field decay parameter, i.e.,  $\tau_B \approx \frac{\om_i}{q_i B}$. 
Therefore,  we can solve Eq.~(\ref{BTE-RTA})  using Eqs.~(\ref{delta-f}) and (\ref{Omega}), and get the expression of $\alpha_j$'s. After getting the expressions for $\alpha_j$'s, the simplified form of $\delta f_i$ is(see Appendix~\ref{appendix2}),
\begin{align}
 \delta f_i &= \frac{-q_i \tau_R^i}{(1+\chi_i)(1+ \chi_i + \chi_i^2)}\Big[\Big\{(1+\chi_i)+\frac{\chi_i(1+ \chi_i - \chi_i^2)}{(1+ \chi_i^2)}\Big\}\nn\\
 &(\vec{v_i}.\vec{E}) \pm \Big\{\chi_i(1+\chi_i)
 + \frac{\chi_i^2(2+ \chi_i)}{(1+ \chi_i^2)}\Big\}\Big(\vec{v_i}.(\vec{E}\times \hat{b})\Big)\Big]\nn\\ 
 &+ \frac{(\om_i - {\rm b}_ih) \tau_R^i}{T(1+\chi_i)(1+ \chi_i + \chi_i^2)}\Big[(1+ \chi_i)\Big(\vec{v_i}\cdot\vec{\nabla}{T}\Big)\nn\\
 &-\tau_R^i\frac{(1+ \chi_i - \chi_i^2)}{(1+ \chi_i^2)}\Big(\vec{v_i}\cdot\vec{\nabla}{\dot T}\Big)\pm\chi_i(1+\chi_i)
 \Big(\vec{v_i}\cdot(\vec{\nabla}{T} \times \hat{b})\Big)\nn\\
 &\mp\tau_R^i \frac{\chi_i(2+\chi_i)}{(1+ \chi_i^2)}
 \Big(\vec{v_i}\cdot(\vec{\nabla}{\dot T} \times \hat{b})\Big)\Big]\frac{\partial f^0_i}{\partial \om_i}.
\end{align}
For simplicity we have considered $\tau_E = \tau_B$ and $\chi_i = \frac{\tau_R^i}{\tau_B} = \frac{\tau_R^i}{\tau_E}$. Here $\pm$ indicates a positive sign for positively charged particles (antiparticles) and a negative sign for negatively charged particles (antiparticles).
Using the above expressions of $\delta f_i$ in Eq.~(\ref{Cur-den}) and (\ref{mic1}), we can express electric current as (see Appendix~\ref{appendix2})
\begin{widetext}
\begin{align}\label{jl}
j^l &= \sum_{i} \frac{q_i g_i}{3} \int \frac{d^3|\vk_i|}{(2\pi)^3} v_i^2 \frac{\tau_R^i}{(1+ \chi_i + \chi_i^2)(1+\chi_i)}\Big[-q_i\Big\{ \Big(
(1+\chi_i)+\frac{\chi_i(1+ \chi_i - \chi_i^2)}{(1+ \chi_i^2)}\Big)\delta^{lk}E^k 
\pm \Big(\chi_i(1+\chi_i)+
\frac{\chi_i^2(2+ \chi_i)}{(1+ \chi_i^2)} \Big)\nn\\
&\epsilon^{ljk}h^jE^k\Big\}+ \frac{(\om_i - {\rm b}_ih)}{T}\Big\{(1+\chi_i)\delta^{lk}
\frac{\del {T}}{\del x^k}
- \tau_R^i \frac{(1+ \chi_i - \chi_i^2)}{(1+ \chi_i^2)}\delta^{lk}\frac{\del {\dot T}}{\del x^k}\pm\chi_i(1+\chi_i)\epsilon^{ljk}h^j
\frac{\del {T}}{\del x^k}\mp
\tau_R^i \frac{\chi_i(2+\chi_i)}{(1+ \chi_i^2)}\nn\\ &\epsilon^{ljk}h^j\frac{\del {\dot T}}{\del x^k}\Big\}\Big]\frac{\partial f^0_i}{\partial \om_i}.
\end{align}   
%
Hence, components of the electric current in the three spatial directions are given as,
\begin{align}\label{Jx}
j^x =& \sum_{i} \frac{q_i g_i}{3} \int \frac{d^3|\vk_i|}{(2\pi)^3} v_i^2 \frac{\tau_R^iq_i}{(1+ \chi_i + \chi_i^2)(1+\chi_i)}\Big\{\Big(
\frac{(1+ \chi_i^2)+\chi_i(2+ \chi_i)}{(1+ \chi_i^2)}\Big)E^x \pm \chi \Big(\frac{(1+ \chi_i^2)(1+\chi_i) + \chi_i(2+ \chi_i)}{(1+ \chi_i^2)} \Big)(E^z - E^y)\Big\} \nn\\
&\Big(-\frac{\partial f^0_i}{\partial \om_i}\Big)
+  \sum_{i} \frac{q_i g_i}{3} \int \frac{d^3|\vk_i|}{(2\pi)^3} v_i^2 
\frac{\tau_R^i(\om_i - {\rm b}_ih)}{T(1+ \chi_i + \chi_i^2)(1+\chi_i)}
\Big\{(1+\chi_i)
\frac{\del T}{\del x} 
- \tau_R^i \frac{(1+ \chi_i - \chi_i^2)}{(1+ \chi_i^2)}\frac{\del\dot T}{\del x }\pm\chi_i(1+\chi_i)
\Big(\frac{\del T}{\del z} - \frac{\del T}{\del y}\Big)\nn\\
&\mp \tau_R^i \frac{\chi_i(2+\chi_i)}{(1+ \chi_i^2)}\Big(\frac{\del \dot T}{\del z} - \frac{\del \dot T}{\del y}\Big)\Big\}\frac{\partial f^0_i}{\partial \om_i}~.
\end{align}
The components $j^y$ and $j^z$ can be obtained from the above equation by changing x, y, and z in cyclic order.
By defining the following integrals
 \begin{align}\label{l01}
&L_{1i} = \frac{g_i }{3T} \int \frac{d^3|\vk_i|}{(2\pi)^3}
\frac{\vk^2_i}{\om_i^2} f^0_i(1-f^0_i) \tau_R^i
\frac{(1+ \chi_i^2)+\chi_i(2+ \chi_i)}{(1+\chi_i) \left(1+\chi_i^2\right) \left(1+ \chi_i + \chi_i^2\right)}~, \nn\\
&L_{2i} = \frac{g_i}{3T}  \int \frac{d^3|\vk_i|}{(2\pi)^3}
\frac{\vk^2_i}{\om_i^2} f^0_i(1-f^0_i) \tau_R^i
\chi_i\frac{(1+\chi_i)(1+ \chi_i^2)+\chi_i(2+ \chi_i)}{(1+\chi_i) \left(1+\chi_i^2\right) \left(1+ \chi_i + \chi_i^2\right)}~,\nn\\
&L_{3i} = \frac{g_i}{3T}   \int \frac{d^3|\vk_i|}{(2\pi)^3}\frac{\vec{k}^2_i}{\om_i^2}(\om_i - {\rm b}_i h)f^0_i(1 - f^0_i) \tau_R^i 
\frac{1}{(1+\chi_i + \chi_i^2)}~,\nn\\    
&L_{4i} = \frac{g_i}{3T} \int \frac{d^3|\vk_i|}{(2\pi)^3}\frac{\vec{k}^2_i}{\om_i^2}(\om_i - {\rm b}_i h)f^0_i(1 - f^0_i) \tau_R^{i2}
\frac{(1+ \chi_i - \chi_i^2)}{(1+\chi_i) \left(1+\chi_i^2\right) \left(1+ \chi_i + \chi_i^2\right)}~, \nn\\
&L_{5i} = \frac{g_i}{3T} \int \frac{d^3|\vk_i|}{(2\pi)^3}\frac{\vec{k}^2_i}{\om_i^2}(\om_i - {\rm b}_i h)f^0_i(1 - f^0_i) \tau_R^i
\frac{\chi_i}{(1+\chi_i + \chi_i^2)}~,\nn\\
&L_{6i} = \frac{g_i}{3T} \int \frac{d^3|\vk_i|}{(2\pi)^3}\frac{\vec{k}^2_i}{\om_i^2}(\om_i - {\rm b}_i h)f^0_i(1 - f^0_i) \tau_R^{i2} 
\frac{\chi_i(2+ \chi_i)}{(1+\chi_i) \left(1+\chi_i^2\right) \left(1+ \chi_i + \chi_i^2\right)}~, 
\end{align}
we can write Eq.~(\ref{Jx}) as
\begin{align}
    j^x &= \sum_i q_i^2 L_{1i} E_x  - \sum_i q_i^2 L_{2i} E_y + \sum_i q_i^2 L_{2i} E_z  - \frac{1}{T} \sum_i q_i \Big(L_{3i} - L_{4i} \frac{d}{dt}\Big) \frac{dT}{dx} + \frac{1}{T} \sum_i q_i 
    \Big(L_{5i} - L_{6i} \frac{d}{dt}\Big) \frac{dT}{dy} \nn\\
   & - \frac{1}{T} \sum_i q_i \Big(L_{5i} - L_{6i} \frac{d}{dt}\Big) \frac{dT}{dz}~.
\end{align}
Now, by setting $j_x$ = $j_y$ = $j_z$ = 0, i.e., when net current due to the external field is zero, we get 
\begin{align}
    j^x = 0 &= \sigma_e E_x  - \sigma_H E_y + \sigma_H E_z  - \frac{L_{34}}{T} \frac{dT}{dx} + \frac{L_{56}}{T} \frac{dT}{dy} - \frac{L_{56}}{T} \frac{dT}{dz}~,\nn\\
    j^y = 0 &= \sigma_H E_x + \sigma_e E_y - \sigma_H E_z - \frac{L_{56}}{T}
     \frac{dT}{dx} - \frac{L_{34}}{T} \frac{dT}{dy} + \frac{L_{56}}{T} \frac{dT}{dz} ~,\nn\\ 
    j^z = 0 &= -\sigma_H E_x + \sigma_H E_y + \sigma_e E_z + \frac{L_{56}}{T}
     \frac{dT}{dx} - \frac{L_{56}}{T} \frac{dT}{dy} - \frac{L_{34}}{T} \frac{dT}{dz}~.
     \label{EqElec}
\end{align}
\end{widetext}
Where, $\sigma_{el}$ = $\sum_i q_i^2 L_{1i}$ and $\sigma_{H}$ = $\sum_i q_i^2 L_{2i}$ are the Ohmic and the Hall components of electrical conductivity, respectively. Further, we defined the integrals $L_{34}$ = $\sum_i q_i \Big(L_{3i} - L_{4i} \frac{d}{dt}\Big)$ and $L_{56}$ = $\sum_i q_i \Big(L_{5i} - L_{6i} \frac{d}{dt}\Big)$.
To solve the Eqs.~(\ref{EqElec}), we rearrange them in the matrix form of
\begin{align}
    \mathbf{\boldsymbol{\sigma}~E = L~X},
\end{align}
where,
\begin{align}
&\boldsymbol{\sigma} = 
\begin{pmatrix}
\sigma_e & -\sigma_H & \sigma_H \\
\sigma_H & \sigma_e & -\sigma_H\\
-\sigma_H & \sigma_H & \sigma_e \\
\end{pmatrix},~\hspace{6mm}
\mathbf{E} = 
 \begin{pmatrix}
E_x\\E_y\\E_z
\end{pmatrix},~\nn\\
 &\mathbf{L} = \frac{1}{T} 
\begin{pmatrix}
-L_{34} & L_{56} & -L_{56} \\
-L_{56} & -L_{34}  & L_{56}\\
L_{56} & -L_{56} & -L_{34} \\
\end{pmatrix},~
\mathbf{X} =
\begin{pmatrix}
\frac{dT}{dx}\vspace{2mm}\\ 
\frac{dT}{dy}\vspace{2mm}\\ 
\frac{dT}{dz}\\ 
\end{pmatrix}.\nonumber    
\end{align}
Hence, finding the inverse of $\boldsymbol{\sigma}$ matrix, we can obtain the components of the electric field as,
\begin{align}\label{matrix1}
    \mathbf{E} &= \mathbf{(\boldsymbol{\sigma}^{-1} L)X} \nn\\
\Rightarrow \begin{pmatrix}
E_x\\E_y\\E_z
\end{pmatrix}
 &= 
\begin{pmatrix}
S_B & \overline{NB} & NB \\
NB & S_B & \overline{NB}\\
\overline{NB} & NB & S_B \\
\end{pmatrix} 
\begin{pmatrix}
\frac{dT}{dx}\vspace{2mm}\\
\frac{dT}{dy}\vspace{2mm}\\
\frac{dT}{dz}\\
\end{pmatrix}~,    
\end{align}
where the transport coefficients are
\begin{align}
 S_B &= \frac{(\sigma_{e}^2 + \sigma_{H}^2)L_{34} + (2\sigma_{e} \sigma_{H})L_{56}}{T(\sigma_{e}^3 + 3\sigma_{e} \sigma_{H}^{2})},  \nn\\
 \overline{NB} &= \frac{\sigma_{H} (\sigma_{e} + \sigma_{H})L_{34} - \sigma_e(\sigma_e + \sigma_H)L_{56}}{T(\sigma_{e}^3 + 3\sigma_{e} \sigma_{H}^{2})}, \nn\\
 NB &= \frac{-\sigma_{H} (\sigma_{e} - \sigma_{H})L_{34} + \sigma_{e}(\sigma_{e} - \sigma_{H})L_{56}}{T(\sigma_{e}^3 + 3\sigma_{e} \sigma_{H}^{2})}. 
\end{align}
Note that this is the generalized case of the magneto-thermoelectric effect, i.e., we did not consider any preferred direction of the magnetic field. Therefore, we obtained three independent normalized transport coefficients - one magneto-Seebeck coefficient $S_B$ and two Nernst coefficients $NB$ and $\overline{NB}$.  

Now, in heavy-ion collisions, the net magnetic field could possibly have a preferred direction~\cite{Hongo:2013cqa}. Therefore, for a special case when the magnetic field is directed along the y-axis (say), i.e., the magnetic field unit vector $h^j \equiv (0, 1, 0)$,  Eq.~(\ref{jl}) gives,
\begin{align}
    j^x &= \sigma_e E_x  + \sigma_H E_z  - \frac{L_{34}}{T} \frac{dT}{dx}  - \frac{L_{56}}{T} \frac{dT}{dz}~,\nn\\
    j^y &=  \sigma_e E_y  - \frac{L_{34}}{T} \frac{dT}{dy}  ~,\nn\\ 
    j^z &= -\sigma_H E_x + \sigma_e E_z + \frac{L_{56}}{T}
     \frac{dT}{dx}  - \frac{L_{34}}{T} \frac{dT}{dz}~.
     \label{redElec}
\end{align}
Hence, by setting $j_x$ = $j_y$ = $j_z$ = 0, we can get the components of the induced electric field as,
\begin{align}\label{Ex}
 E_x &= \frac{\Big\{\sigma_{e} L_{34} + \sigma_{H}L_{56} \Big\}\frac{dT}{dx}}{T\Big\{(\sigma_{e})^2 + (\sigma_{H})^2\Big\}} 
 +  \frac{\Big\{\sigma_{e}L_{56} - \sigma_{H} L_{34} \Big\}\frac{dT}{dz}}{T\Big\{(\sigma_{e})^2 + (\sigma_{H})^2\Big\}}~, \nn\\
 E_y &= \frac{L_{34}}{T\sigma_{e}}\frac{dT}{dy}~, \nn\\
 E_z &= \frac{\Big\{\sigma_{H} L_{34} - \sigma_{e}L_{56} \Big\}\frac{dT}{dx}}{T\Big\{(\sigma_{e})^2 + (\sigma_{H})^2\Big\}}
 +  \frac{\Big\{\sigma_{e} L_{34} + \sigma_{H}L_{56} \Big\}\frac{dT}{dz}}{T\Big\{(\sigma_{e})^2 + (\sigma_{H})^2\Big\}}~. 
\end{align}
The above equations can be written in matrix form as,
\begin{align}
\label{matrix2}
\begin{pmatrix}
E_x\\E_y\\E_z
\end{pmatrix}
 &= 
\begin{pmatrix}
S_B & 0 & NB \\
0 & \overline{S}_B & 0\\
-NB & 0 & S_B \\
\end{pmatrix} 
\begin{pmatrix}
\frac{dT}{dx}\vspace{2mm}\\
\frac{dT}{dy}\vspace{2mm}\\
\frac{dT}{dz}\\
\end{pmatrix}.   
\end{align}
Here, we identify the dimensionless transport coefficients - two magneto-Seebeck coefficients ($S_B$, $\overline{S}_{B}$) and one Nernst coefficient ($NB$) as,
\begin{align}
    S_B &= \frac{(\frac{\sigma_{el}}{T})(\frac{L_{34}}{T^2}) + (\frac{\sigma_H}{T})(\frac{L_{56}}{T^2})}{(\frac{\sigma_{el}}{T})^2 + (\frac{\sigma_{H}}{T})^2}, ~~~
    \overline{S}_{B} =  \frac{(\frac{L_{34}}{T^2})}{(\frac{\sigma_{el}}{T})},\nn\\
    NB &= \frac{(\frac{\sigma_{el}}{T})(\frac{L_{56}}{T^2}) - (\frac{\sigma_H}{T})(\frac{L_{34}}{T^2})}{(\frac{\sigma_{el}}{T})^2 + (\frac{\sigma_{H}}{T})^2}~.\label{sb}
\end{align}
Note that, at vanishing Hall conductivity ($\sigma_H = 0$) the magneto-Seebeck coefficient ${S_{B}}$ becomes the same as $\overline{S}_{B}$. This can be realized as follows. The coefficient $\overline{S}_{B}$ is responsible for the electric field in the y-axis or along the magnetic field, where the Lorentz force has a null contribution. Therefore, the expression of $\overline{S}_{B}$ is similar to that of $eB_0 = 0$ case as in Eq.~(\ref{s0}), except the fact that the integral in $L_{34}$ depends on magnetic decay parameter $\tau_B$ at finite magnetic field. Furthermore, for a static picture, where cooling effects are absent $\left(\frac{dT}{dt} = 0 \right)$, the integrals $L_{34}$ and $L_{56}$ reduce to $L_{34}$ = $\sum_i q_i L_{3i}$ and $L_{56}$ = $\sum_i q_i L_{5i}$, then the expression of Seebeck and Nernst coefficients becomes similar to that obtained in earlier studies~\cite{PhysRevD.102.096011, Das2021}.

\subsubsection{With Landau quantization}
Now, we study the effect of Landau quantization on the thermoelectric coefficients of the QGP medium and see how it deviates from classical results. Basic modifications will occur in the dispersion relation for medium constituents and the phase space integration. As we considered the magnetic field along the y direction therefore, momentum quantization will occur in its perpendicular plane, i.e., the x-z plane. Hence,

		\bea
		\om = (\vk^2+m^2)^{1/2} &\rightarrow&  \om_{l} = (k_y^2+m^2+2l|{q_i}|B)^{1/2} ,\nn\\
		2\int \frac{d^3\vk}{(2\pi)^3} &~\rightarrow~&  
		\sum_{l=0}^\infty \alpha_l \frac{|{q_i}|B}{2\pi} 
		\int\limits^{+\infty}_{-\infty} \frac{dk_y}{2\pi} ,
		\label{CM_QM}
		\eea
		The factor 2 of spin degeneracy in left hand side of last line will be converted to $\alpha_l$, which will be 2 for all Landau levels $l$, except lowest Landau level (LLL) $l=0$, where $\alpha_l=1$. In general, one can write $\alpha_l = 2 - \delta_{l,0}$. Under the conditions of an extremely high magnetic field for which all medium constituents occupy the lowest Landau energy level $l=0$.  It means that the perpendicular motion of medium constituents completely vanishes as $k_x\approx k_z\approx 0$ at $l=0$. However, below that strong magnetic field limt, $l>0$ energy levels might have some non-negligible contributions, and the lowest Landau level (LLL) approximation is not sufficient enough~\cite{PhysRevC.106.044914, Dey:2021fbo}. This is the case in our study, where the magnetic field is not constant but decays with time. Here, we also consider, $k_x^2\approx k_z^2\approx (\frac{k_x^2+k_z^2}{2})=\frac{2l|{q_i}|B}{2}$,
		the integrals in Eq.~(\ref{l01}) can be expressed as,
  \begin{widetext}
  \begin{align}
&L_{1i} = \frac{g_i }{T} \sum_{l=0}^\infty \alpha_l \frac{|{q_i}|B}{2\pi} 
		\int\limits^{+\infty}_{-\infty} \frac{dk_y}{2\pi} \frac{{l|{q_i}|B}}{\om^2_{l}} f^0_i(1-f^0_i) \tau_R^i~\frac{(1+ \chi_i^2)+\chi_i(2+ \chi_i)}{(1+\chi_i) \left(1+\chi_i^2\right) \left(1+ \chi_i + \chi_i^2\right)}~, \nn\\
&L_{2i} = \frac{g_i}{T}  \sum_{l=0}^\infty \alpha_l \frac{|{q_i}|B}{2\pi} 
		\int\limits^{+\infty}_{-\infty} \frac{dk_y}{2\pi} \frac{{l|{q_i}|B}}{\om^2_{l}} f^0_i(1-f^0_i) \tau_R^i~\chi_i\frac{(1+\chi_i)(1+ \chi_i^2)+\chi_i(2+ \chi_i)}{(1+\chi_i) \left(1+\chi_i^2\right) \left(1+ \chi_i + \chi_i^2\right)}~,\nn\\
&L_{3i} = \frac{g_i}{T}   \sum_{l=0}^\infty \alpha_l \frac{|{q_i}|B}{2\pi} 
		\int\limits^{+\infty}_{-\infty} \frac{dk_y}{2\pi} \frac{{l|{q_i}|B}}{\om^2_{l}}(\om_i - {\rm b}_i h)\tau_R^i~f^0_i(1 - f^0_i) \frac{1}{(1+\chi_i + \chi_i^2)}~,~\nn\\    
&L_{4i} = \frac{g_i}{T} \sum_{l=0}^\infty \alpha_l \frac{|{q_i}|B}{2\pi} 
		\int\limits^{+\infty}_{-\infty} \frac{dk_y}{2\pi} \frac{{l|{q_i}|B}}{\om^2_{l}}(\om_i - {\rm b}_i h)\tau_R^{i2}~f^0_i(1 - f^0_i) \frac{(1+ \chi_i - \chi_i^2)}{(1+\chi_i) \left(1+\chi_i^2\right) \left(1+ \chi_i + \chi_i^2\right)}~,~ \nn\\
&L_{5i} = \frac{g_i}{T} \sum_{l=0}^\infty \alpha_l \frac{|{q_i}|B}{2\pi} 
		\int\limits^{+\infty}_{-\infty} \frac{dk_y}{2\pi} \frac{{l|{q_i}|B}}{\om^2_{l}}(\om_i - {\rm b}_i h)\tau_R^i~f^0_i(1 - f^0_i) \frac{\chi_i}{(1+\chi_i + \chi_i^2)}~,~\nn\\
&L_{6i} = \frac{g_i}{T} \sum_{l=0}^\infty \alpha_l \frac{|{q_i}|B}{2\pi} 
		\int\limits^{+\infty}_{-\infty} \frac{dk_y}{2\pi} \frac{{l|{q_i}|B}}{\om^2_{l}}(\om_i - {\rm b}_i h)\tau_R^{i2}~f^0_i(1 - f^0_i) \frac{\chi_i(2+ \chi_i)}{(1+\chi_i) \left(1+\chi_i^2\right) \left(1+ \chi_i + \chi_i^2\right)}.~ 
\end{align}
\end{widetext}
Hence, the thermoelectric coefficients obtained in Eq.~(\ref{sb}) will also be modified under the conditions of Landau quantization. 

\subsubsection{Analytical solution for massless noninteracting system}
To find the analytical expressions for the massless noninteracting system, we follow the same procedure mentioned in Sec.~(\ref{ana}). Hence, for the case of a massless noninteracting system, the integral in Eq.~(\ref{l01}) turns into the following forms:
\begin{widetext}
\begin{align}
    L_{1i} &= \sum_i \frac{g_i \tau_R^i}{6\pi^2} \Big( 2\zeta(2)T^2 + b_i^2 \mu_{B}^2\Big)\frac{(1+ \chi_i^2)+\chi_i(2+ \chi_i)}{(1+\chi_i) \left(1+\chi_i^2\right) \left(1+ \chi_i + \chi_i^2\right)}~,\nn\\
    L_{2i} &= \sum_i \frac{2g_i \tau_R^i}{3 \pi^2}       ~ \eta(1) b_i \mu_{B} T \frac{(1+\chi_i)(1+ \chi_i^2)+\chi_i(2+ \chi_i)}{(1+\chi_i) \left(1+\chi_i^2\right) \left(1+ \chi_i + \chi_i^2\right)}~,\nn\\
    L_{3i} &= \sum_i \frac{g_i \tau_R^i}{6 \pi^2} \Big(b_i^3 \mu_{B}^3 + 6b_i \mu_{B} \zeta(2) T^2 - 2b_i h \zeta(2) T^2  - b_i^3 \mu_{B}^2 h\Big)\frac{1}{(1+\chi_i + \chi_i^2)}~,\nn\\
    L_{4i} &= \sum_i \frac{g_i \tau_R^i}{6 \pi^2} \Big(b_i^3 \mu_{B}^3 + 6b_i \mu_{B} \zeta(2) T^2 - 2b_i h \zeta(2) T^2  - b_i^3 \mu_{B}^2 h\Big)\frac{(1+ \chi_i - \chi_i^2)}{(1+\chi_i) \left(1+\chi_i^2\right) \left(1+ \chi_i + \chi_i^2\right)}~,\nn\\
    L_{5i} &= \sum_i \frac{g_i \tau_R^{i}}{2 \pi^2} \Big(3T^{3}\zeta(3) + 2 b_i^{2} \mu_{B}^{2} T \eta(1) - \frac{4}{3} h b_i^{2} \mu_{B} T \eta(1)  \Big)\frac{\chi_i}{(1+\chi_i + \chi_i^2)}~,\nn\\
    L_{6i} &= \sum_i \frac{g_i \tau_R^{i2}}{2 \pi^2} \Big(3T^{3}\zeta(3) + 2 b_i^{2} \mu_{B}^{2} T \eta(1) - \frac{4}{3} h b_i^{2} \mu_{B} T \eta(1)  \Big)\frac{\chi_i(2+ \chi_i)}{(1+\chi_i) \left(1+\chi_i^2\right) \left(1+ \chi_i + \chi_i^2\right)}.
\end{align}
\end{widetext}
Here, the integrals $L_{34} $ and $ L_{56}$ in Eq.~(\ref{sb}) are proportional to $q_i$, hence for the massless noninteracting system $L_{34} $ and $ L_{56}$ both individually summed upto zero. Therefore, dimensionless transport coefficients - two magneto-Seebeck coefficients ($S_B$, $\overline{S}_{B}$) and one Nernst coefficient ($NB$) in Eq.~(\ref{sb}) vanishes for the massless noninteracting system, hence
\begin{align}
    S_B &= \overline{S}_{B} = NB = 0.
\end{align} 
%

\subsubsection{Numerical solution for QGP EoS}
\begin{figure*}
     \centering
     \begin{subfigure}[b]{0.47\textwidth}
         \centering
         \includegraphics[width=\textwidth]{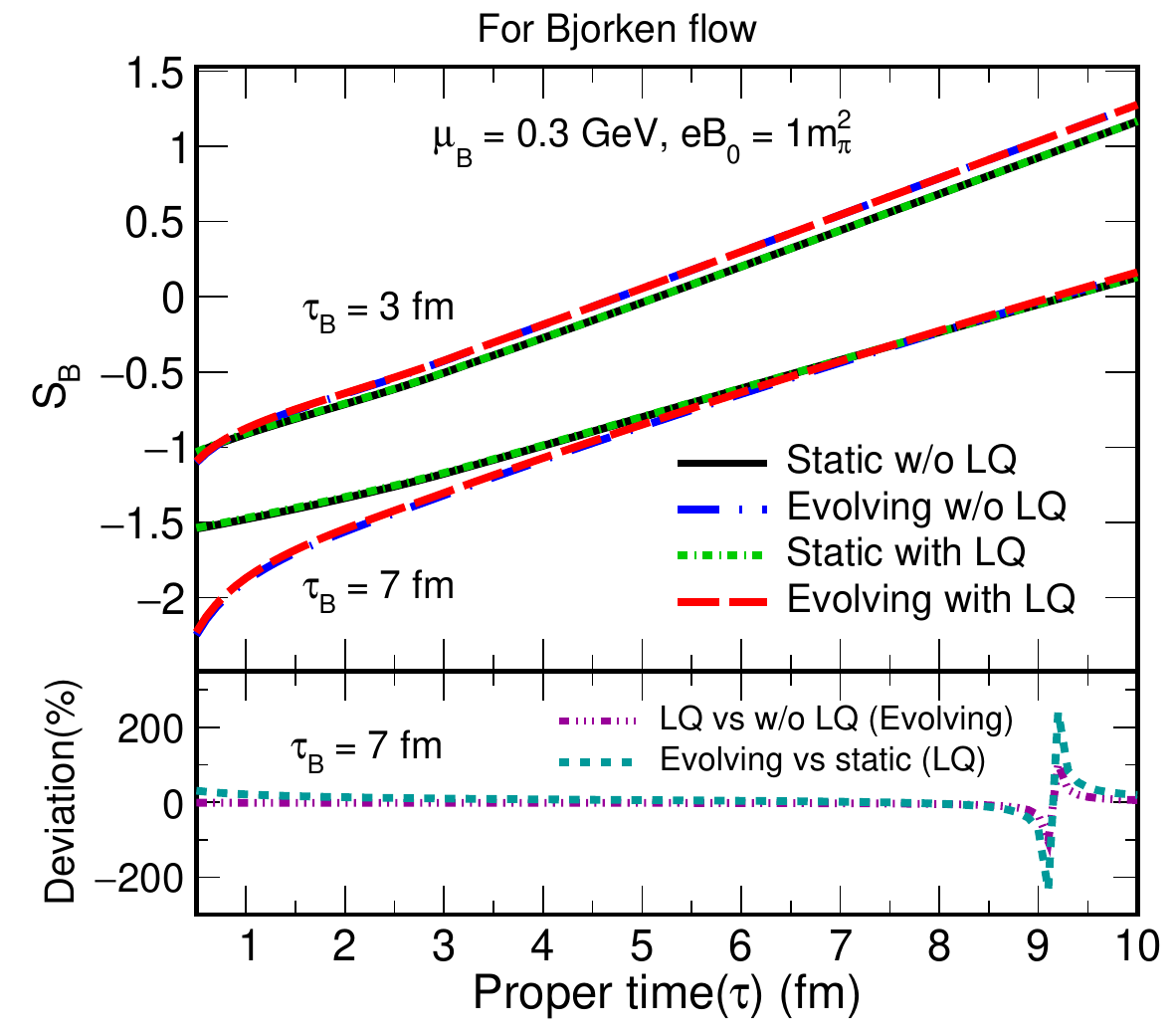}
          \caption{}
     \end{subfigure}
     \hfill
     \begin{subfigure}[b]{0.47\textwidth}
         \centering
         \includegraphics[width=\textwidth]{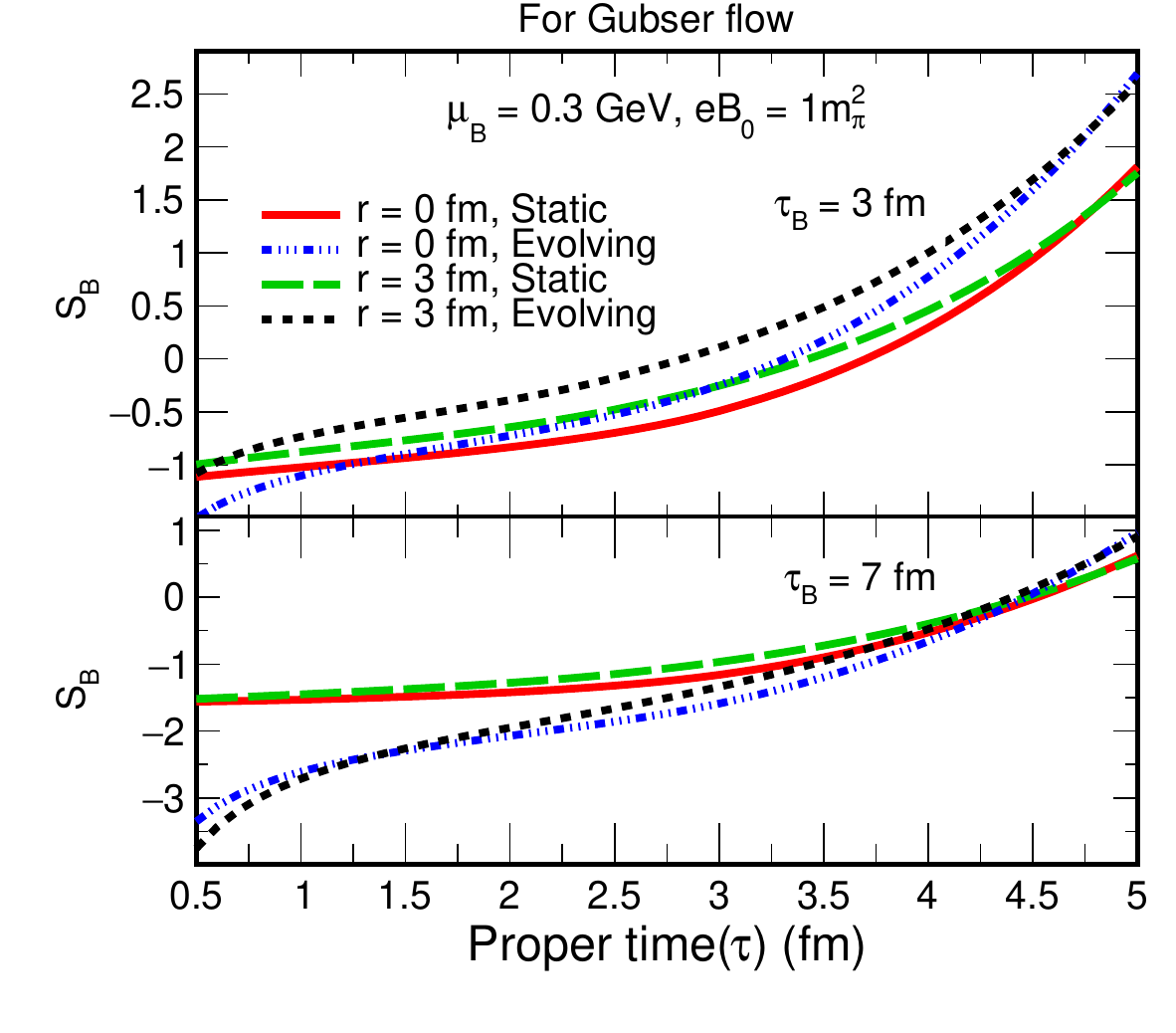}
          \caption{}
     \end{subfigure}
     \hfill
     \begin{subfigure}[b]{0.47\textwidth}
         \centering
         \includegraphics[width=\textwidth]{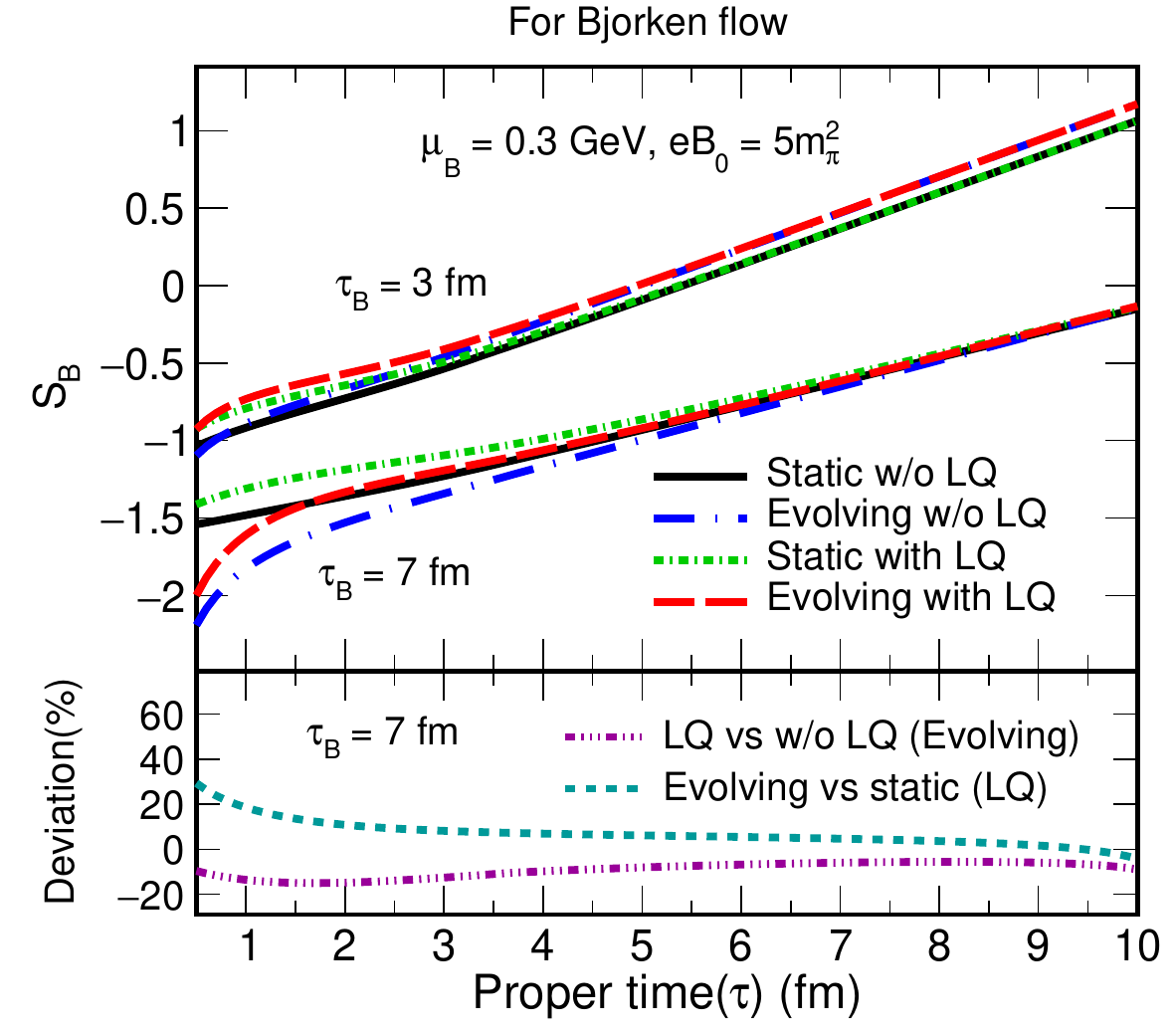}
          \caption{}
     \end{subfigure}
      \hfill
     \begin{subfigure}[b]{0.47\textwidth}
         \centering
         \includegraphics[width=\textwidth]{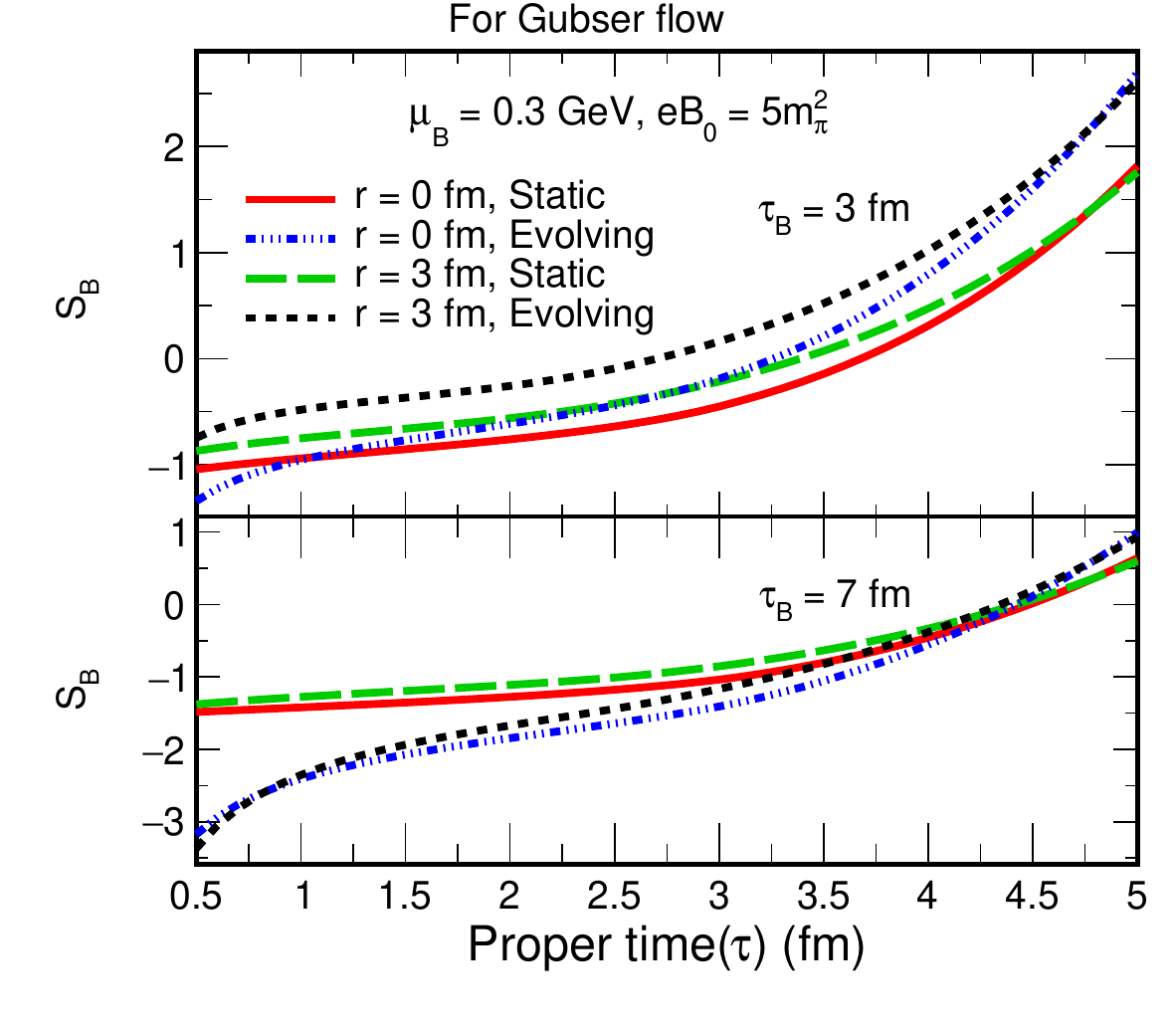}
          \caption{}
     \end{subfigure}
        \caption{Magneto-Seebeck coefficient ($S_B$) as a function of proper time ($\tau$) at $\mu_{B} = 0.3$ GeV.
            (a) $S_B$ for static, evolving, with and without (w/o) Landau quantization (LQ), four cases are plotted here in the upper panel for $\tau_B = 3,~7$~fm for Bjorken flow at $1~m_\pi^2$. The lower panel represents their percentage deviation at $\tau_B = 7~$fm. 
            (b) $S_B$ in Gubser flow for static and evolving picture at $r=0$ and $3$~fm with LQ is plotted at $\tau_B = 3~$fm (upper panel) and $\tau_B = 7~$fm (lower panel) at $eB_0 = 1~m_\pi^2$. 
            (c) and (d) are the same as (a) and (b), respectively, but at $eB_0 = 5~m_\pi^2$.}
	\label{Fig-temp2}
\end{figure*}
Figure ~(\ref{Fig-temp2}) represents the magneto-Seebeck coefficient ($S_B$) as a function of proper time ($\tau$) at $\mu_B = 0.3$ GeV in the presence of time-varying magnetic field with initial value $eB_0 = 1~m_{\pi}^{2}$ (a) and (b) and $eB_0$ = 5~$m_{\pi}^{2}$ (c) and (d) with its decay parameter $\tau_{B}$ = 3 fm, 7 fm. In all the plots, we observe similar behavior of $S_B$. For the case of Bjorken flow, at the early times, the coefficient $S_{B}$ is negative, which means that the direction of the produced electric field is opposite to the temperature gradient of the medium. Later, for a particular value of $\tau$, the $S_B$ becomes zero and then increases with positive value as $\tau$ increases. Here, we take the static and evolving pictures for both cases along with Landau quantization (LQ) and without Landau quantization (w/o LQ). For the Bjorken flow case, the ideal MHD cooling rate is used as mentioned in Eq.~(\ref{Temp}). The black solid line represents the static picture without Landau quantization, and the green dash-dot-dash line represents the static picture with Landau quantization. On the other hand, the blue dash-dot line represents the evolving picture w/o LQ, and the red dashed line represents the evolving picture with LQ. $S_B$ is highly sensitive to magnetic decay parameter $\tau_{B}$. 
In the lower panels of (a) and (c), the percentage deviation without LQ from LQ is plotted with a magenta dash-dotted line. As expected, the effect of LQ improves with the strength of the magnetic field ($eB_0$). 
The cyan dashed line represents the percentage deviation from static to evolving picture. In both plots, this deviation is nearly 30\%-35\%  in the early evolution and decreases as the medium evolves. 
As the value of $S_{B}$ becomes positive, this deviation shows a peak by shifting from a negative value to a positive value. Once this percentage deviation achieves its maximum value, later on, it starts to decrease with proper time. 
In Fig.~(\ref{Fig-temp2}) (b) and (d), the upper and bottom panels correspond to decay parameter $\tau_{B}$ = 3 fm and $\tau_{B}$ = 7 fm, respectively. Here, we use the cooling rate obtained from Gubser flow as mentioned in Eq.~(\ref{GubserT}). 
Here, we also study the effect of radial expansion of medium on coefficient $S_{B}$. The red solid line corresponds to $r=0$ fm, and the green dashed line corresponds to $r=3$ fm in the static picture. On the other hand, The blue dash-triple dot line corresponds to $r=0$ fm, and the black dashed line corresponds to $r=3$ fm in the evolving picture. $S_B$ follows the similar trend in both the plots. The numerical value of $S_B$ increases as a function of $\tau$. The effect of $\tau_B$ is identical as observed in the Bjorken flow case.

\begin{figure*}
     \centering
     \begin{subfigure}[b]{0.47\textwidth}
         \centering
         \includegraphics[width=\textwidth]{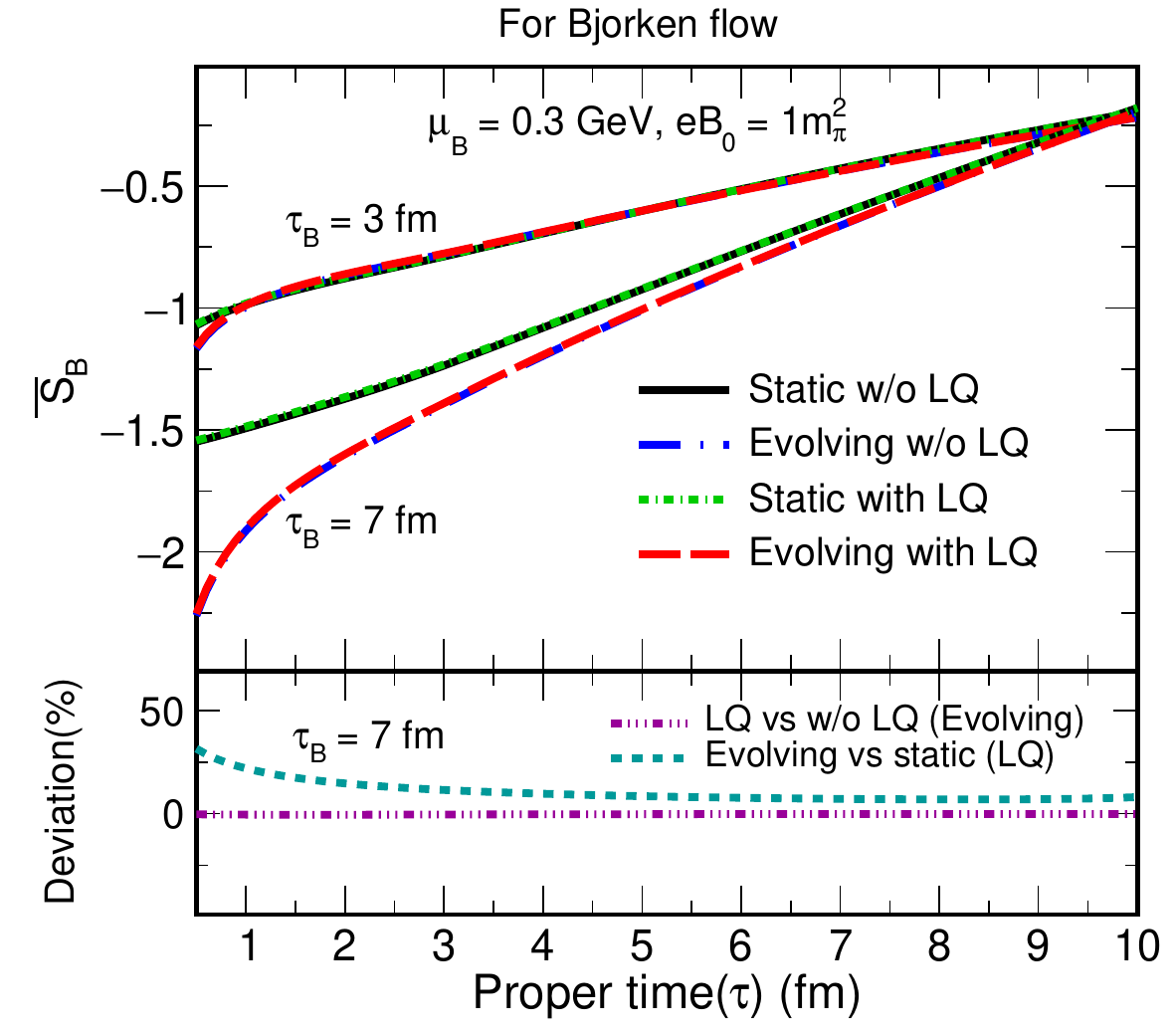}
          \caption{}
     \end{subfigure}
     \hfill
     \begin{subfigure}[b]{0.47\textwidth}
         \centering
         \includegraphics[width=\textwidth]{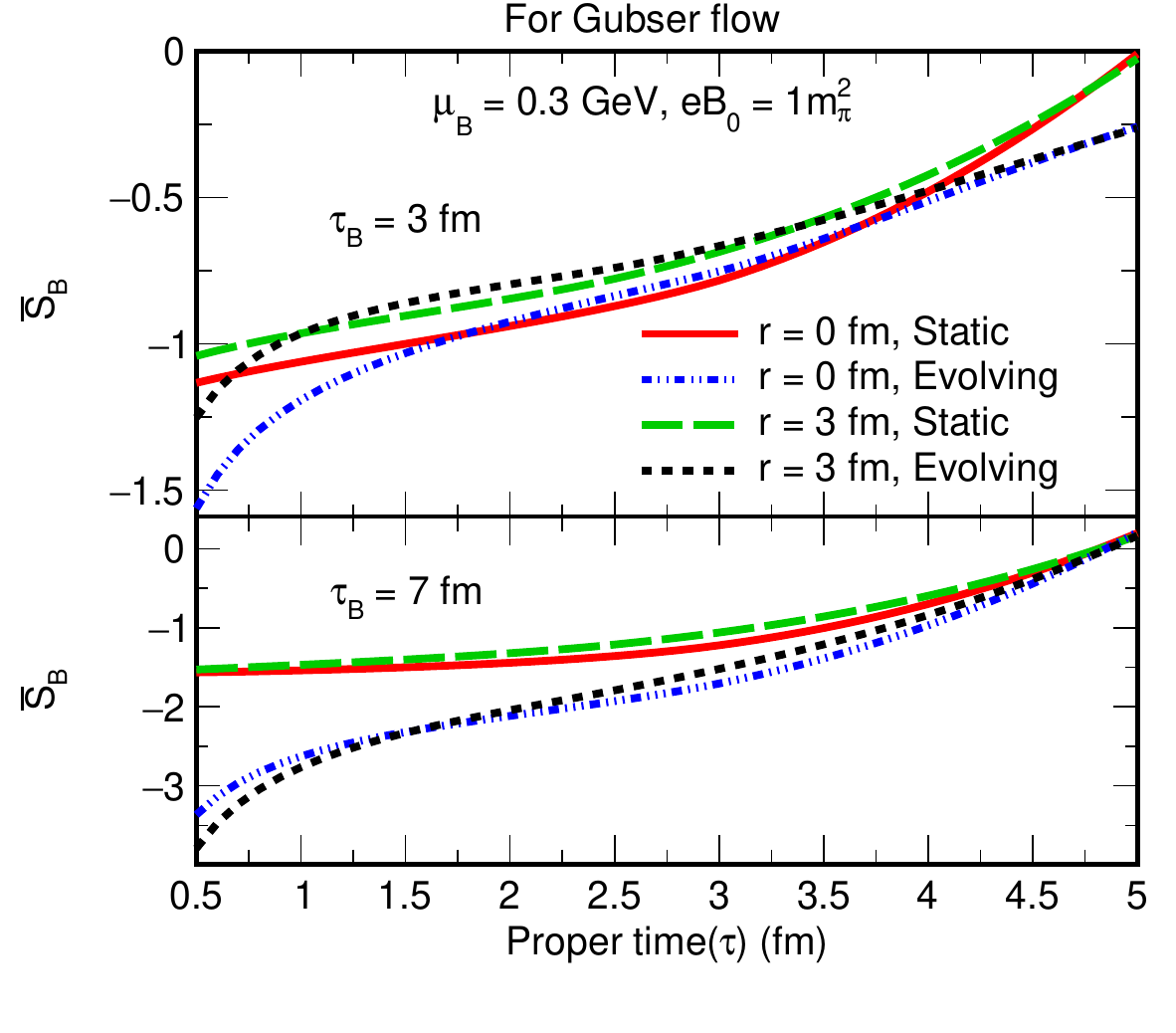}
          \caption{}
     \end{subfigure}
     \hfill
     \begin{subfigure}[b]{0.47\textwidth}
         \centering
         \includegraphics[width=\textwidth]{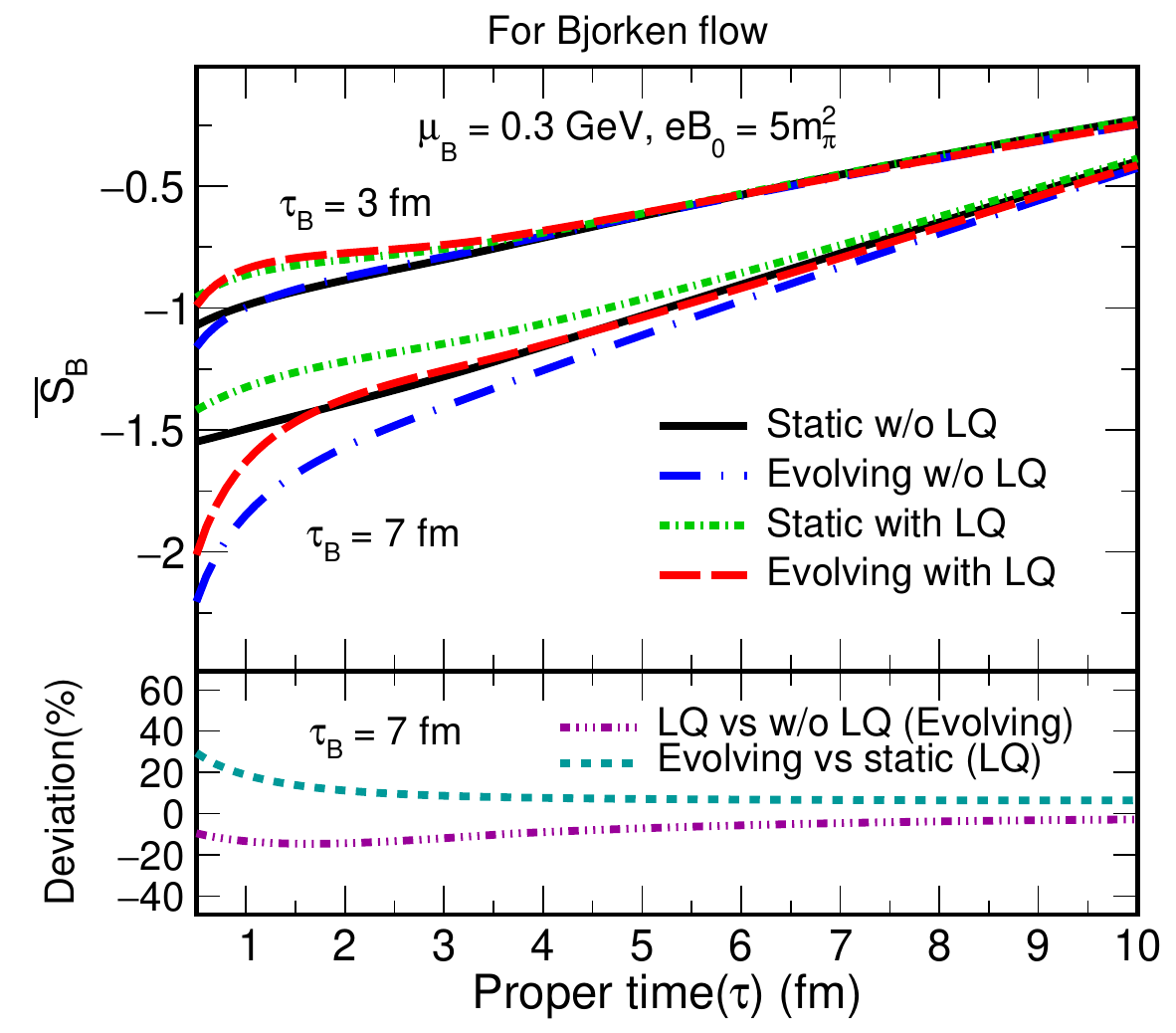}
          \caption{}
     \end{subfigure}
      \hfill
     \begin{subfigure}[b]{0.47\textwidth}
         \centering
         \includegraphics[width=\textwidth]{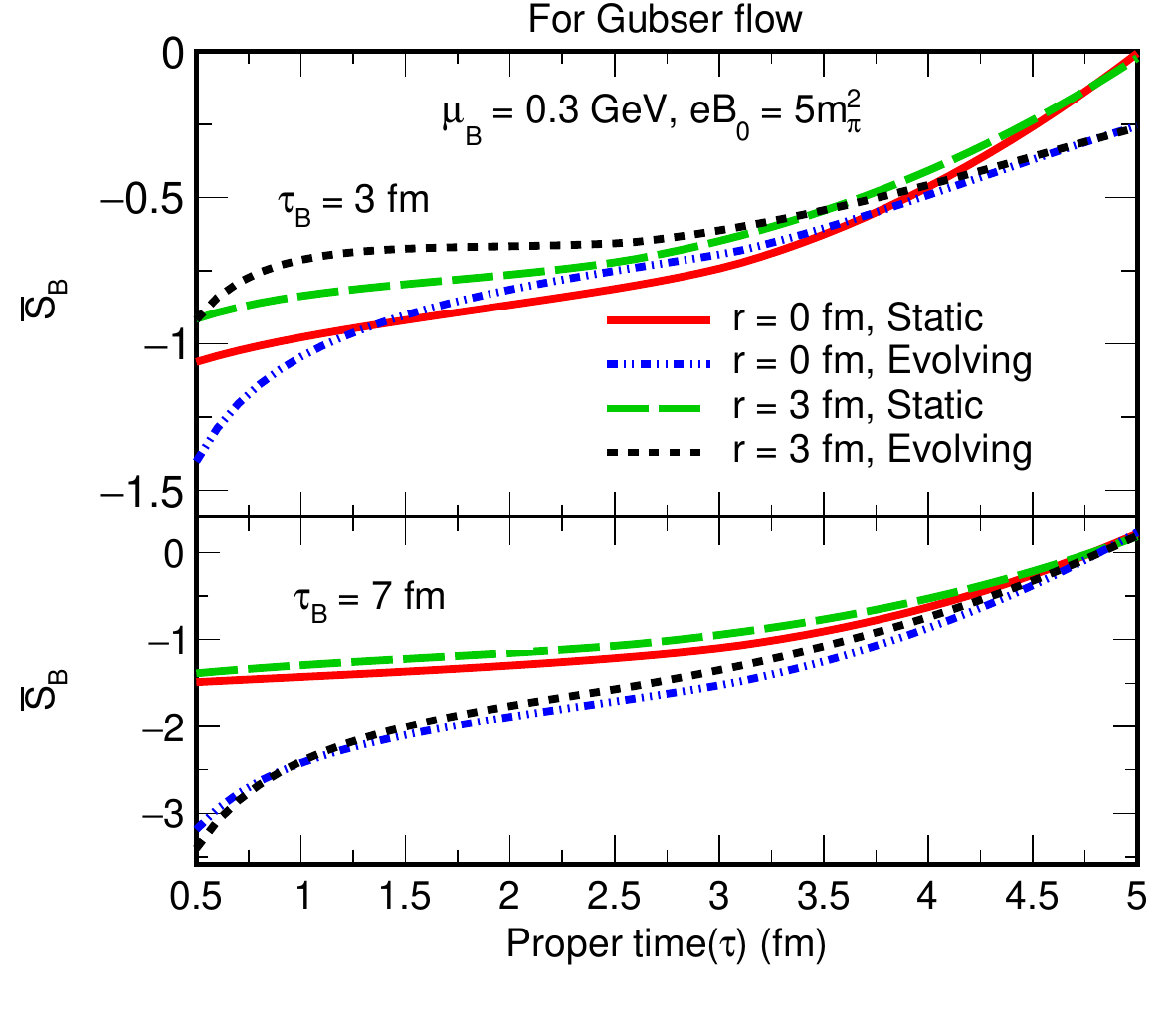}
          \caption{}
     \end{subfigure}
        \caption{Magneto-Seebeck coefficient ($\overline{S}_B$) as a function of proper time ($\tau$) at $\mu_{B} = 0.3$ GeV.
            (a) $\overline{S}_B$ for static, evolving, with and without (w/o) Landau quantization (LQ), four cases are plotted here in the upper panel for $\tau_B = 3,~7$~fm for Bjorken flow at $1~m_\pi^2$. The lower panel represents their percentage deviation at $\tau_B = 7~$fm. 
            (b) $\overline{S}_B$ in Gubser flow for static and evolving picture at $r=0$ and $3$~fm with LQ is plotted at $\tau_B = 3~$fm (upper panel) and $\tau_B = 7~$fm (lower panel) at $eB_0 = 1~m_\pi^2$. 
            (c) and (d) are the same as (a) and (b), respectively, but at $eB_0 = 5~m_\pi^2$.}
	\label{Fig-seebar}
\end{figure*}
Figure~(\ref{Fig-seebar}) is the same as Fig.~(\ref{Fig-temp2}), but for  $\overline{S}_B$. From Eq.~(\ref{sb}), at vanishing Hall like components $\sigma_H \rightarrow 0$ and $I_{56} \rightarrow 0$, $\overline{S}_B = S_B$. Expression of $\overline{S}_B$ is similar to that of Seebeck $S$ as in $eB_0 = 0$ case, except the fact that in magnetic field effective relaxation time modifies which is a function of $\tau_R$ and $\tau_B$.
The Hall-like components ($\sigma_H, I_{56}$) are approximately one order smaller than their leading order counterparts ($\sigma_e, I_{34}$). Therefore, at the early evolution, $\overline{S}_B$ is close to $S_B$. In the later time, as the magnetic field significantly reduces, $\overline{S}_B$ approaches the Seebeck coefficient ($S$) as for the $eB_0 = 0$ case.      


\begin{figure*}
     \centering
     \begin{subfigure}[b]{0.47\textwidth}
         \centering
         \includegraphics[width=\textwidth]{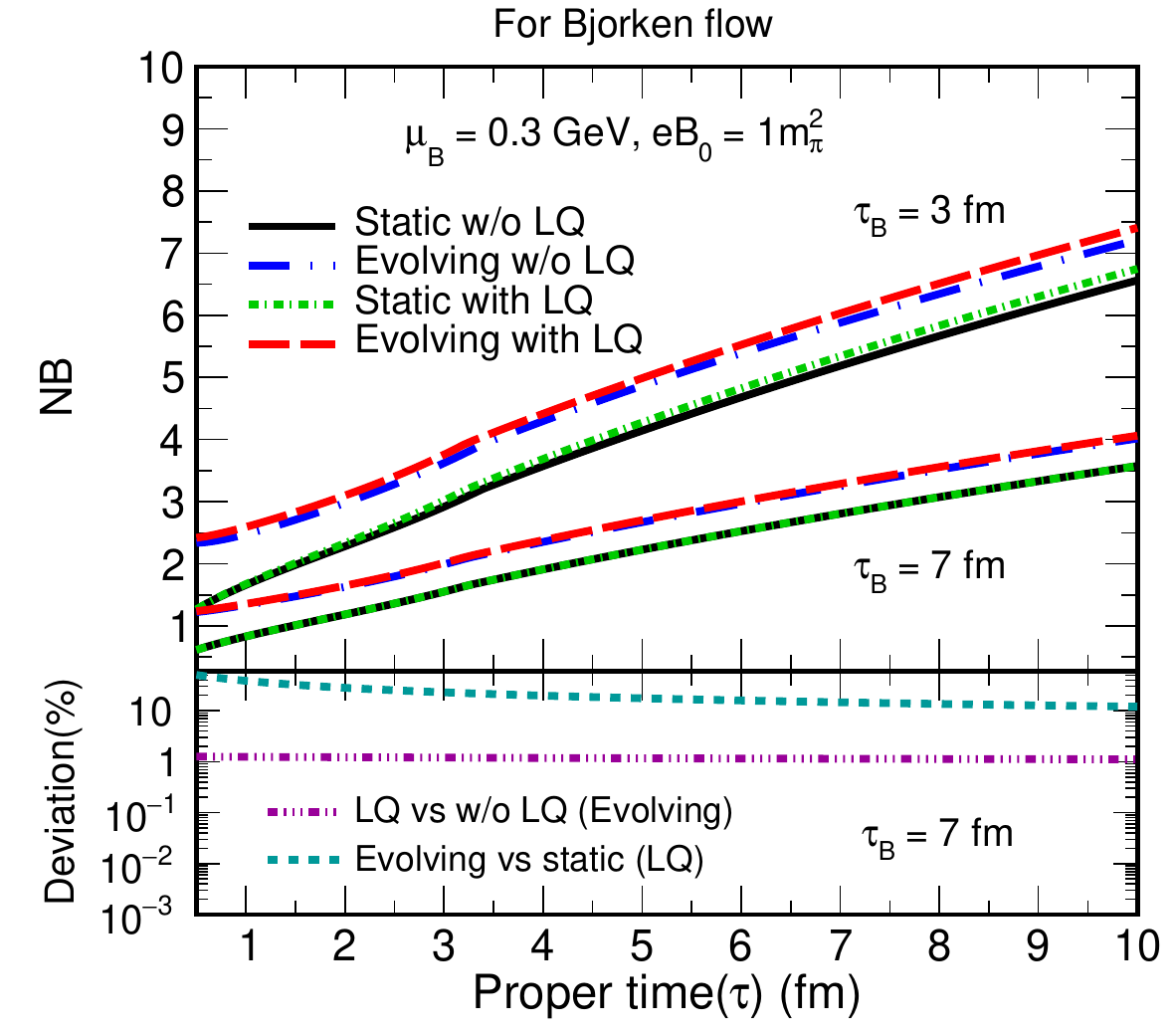}
          \caption{}
     \end{subfigure}
     \hfill
     \begin{subfigure}[b]{0.47\textwidth}
         \centering
         \includegraphics[width=\textwidth]{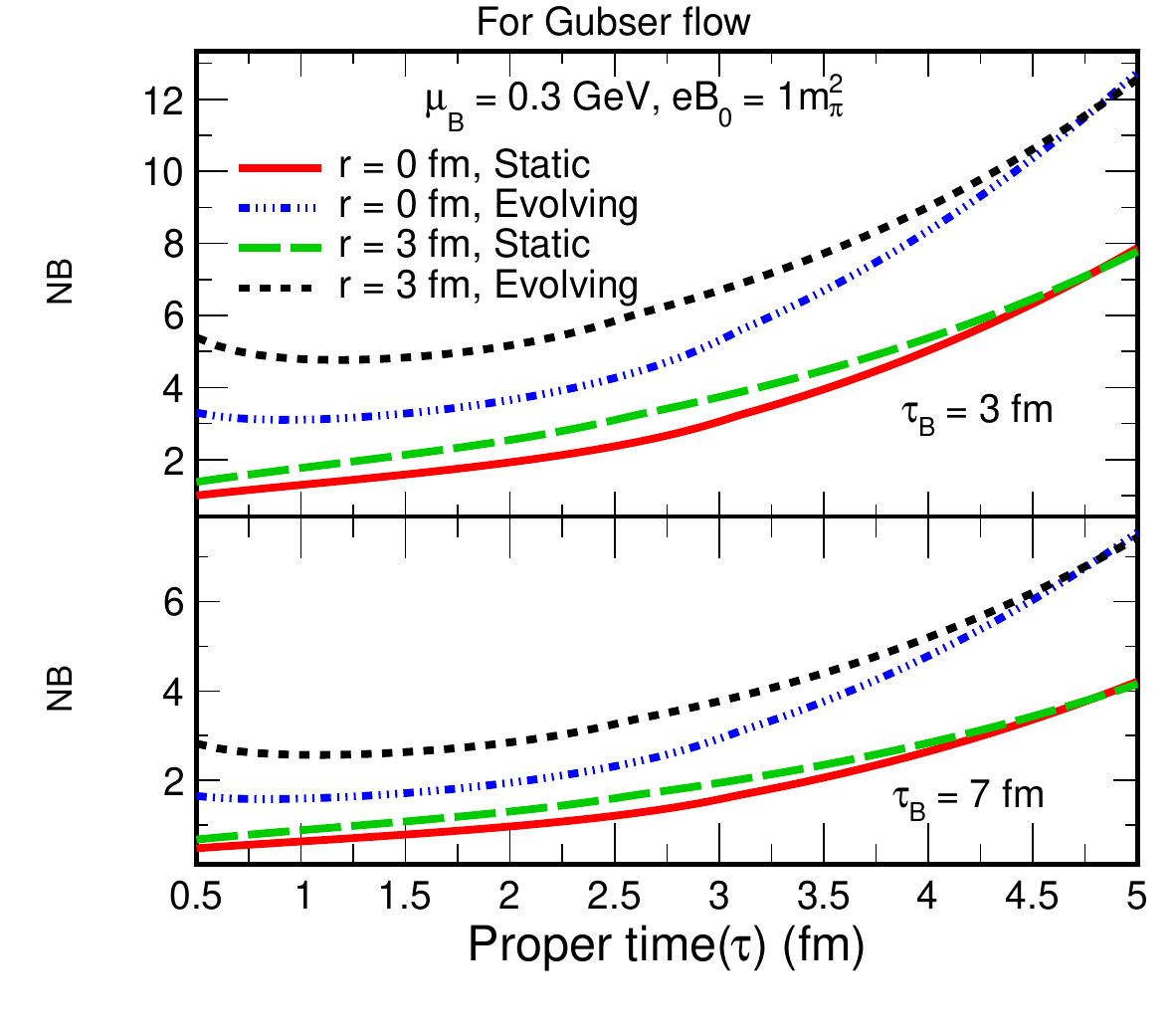}
          \caption{}
     \end{subfigure}
     \hfill
     \begin{subfigure}[b]{0.47\textwidth}
         \centering
         \includegraphics[width=\textwidth]{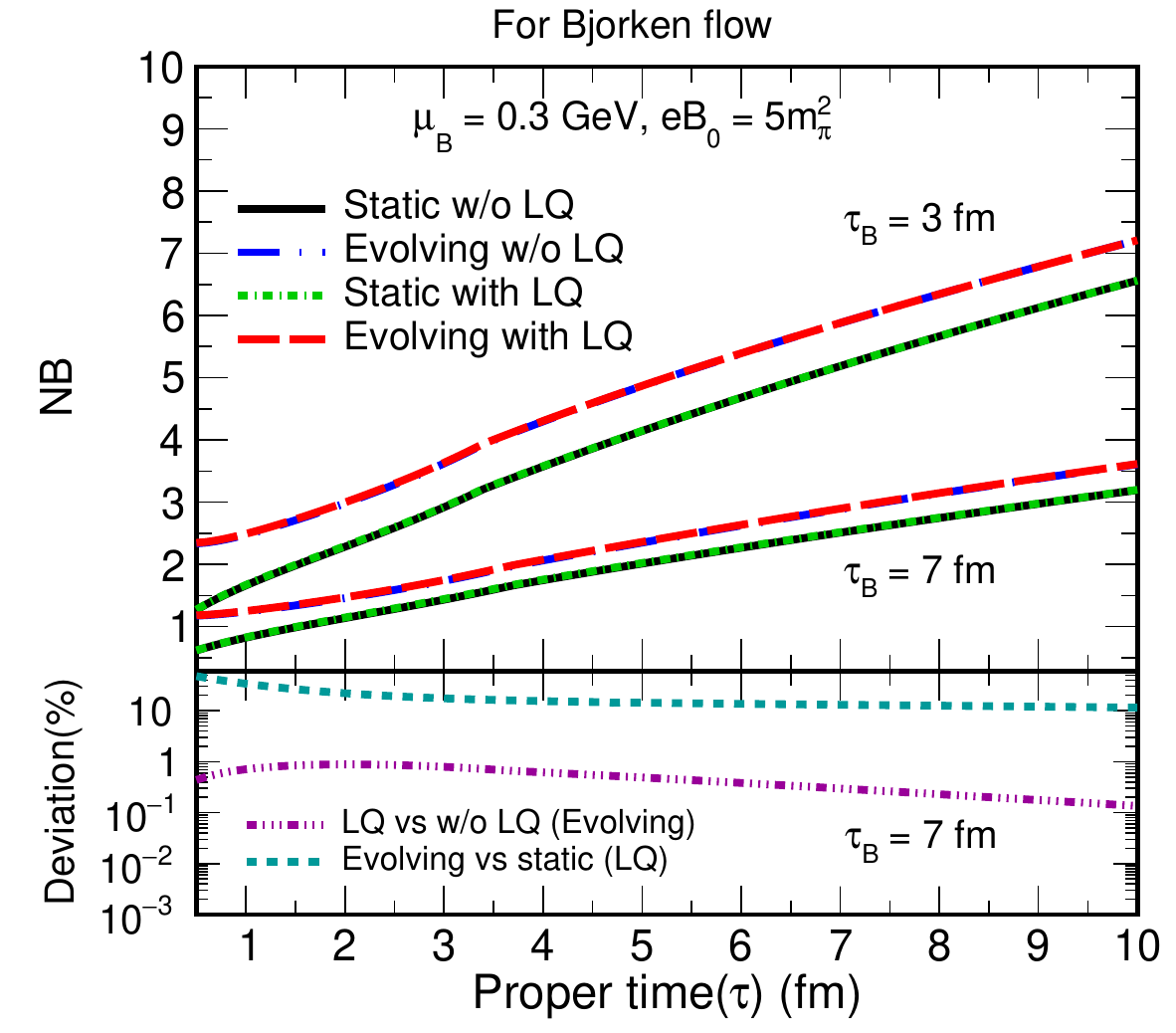}
          \caption{}
     \end{subfigure}
      \hfill
     \begin{subfigure}[b]{0.47\textwidth}
         \centering
         \includegraphics[width=\textwidth]{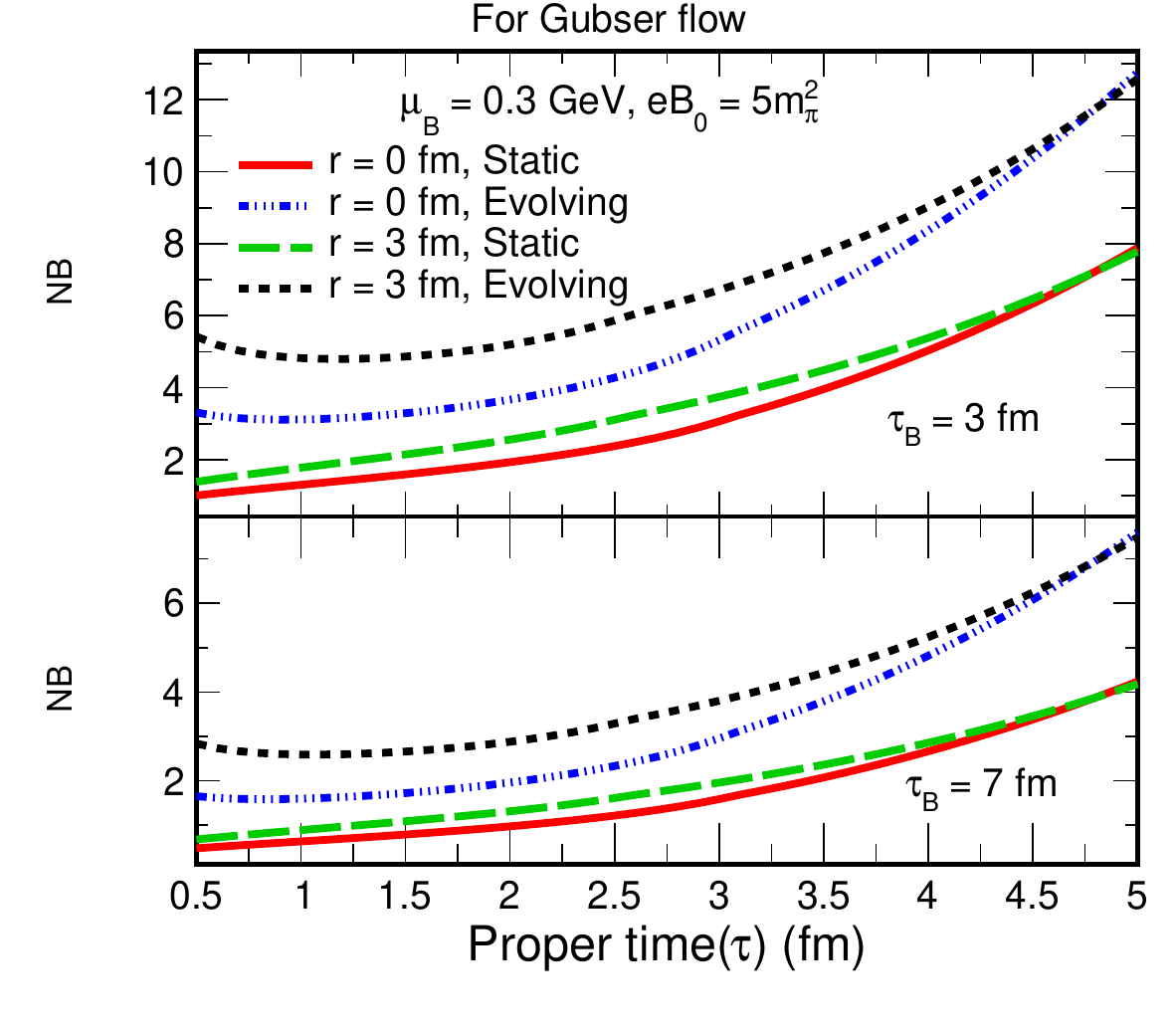}
          \caption{}
     \end{subfigure}
        \caption{Dimensionless Nernst coefficient ($NB$) as a function of proper time ($\tau$) at $\mu_{B} = 0.3$ GeV.
            (a) $NB$ for static, evolving, with and without (w/o) Landau quantization (LQ), four cases are plotted here in the upper panel for $\tau_B = 3,~7$~fm for Bjorken flow at $1~m_\pi^2$. The lower panel represents their percentage deviation at $\tau_B = 7~$fm. 
            (b) $NB$ in Gubser flow for static and evolving picture at $r=0$ and $3$~fm with LQ is plotted at $\tau_B = 3~$fm (upper panel) and $\tau_B = 7~$fm (lower panel) at $eB_0 = 1~m_\pi^2$. 
            (c) and (d) are the same as (a) and (b), respectively, but at $eB_0 = 5~m_\pi^2$.}
	\label{Fig-temp4}
\end{figure*}
In Fig.~(\ref{Fig-temp4}), dimensionless Nernst coefficient ($NB$) as a function of proper time ($\tau$) is plotted. Representation is same as in Fig.~(\ref{Fig-temp2}) and (\ref{Fig-seebar}). 
Unlike magneto-Seebeck coefficients, $NB$ is positive and increases with evolution. The effect of the magnetic field is similar as in $S_B$ and $\overline{S}_B$. In the Gubser flow case, $NB$ increases as we go away from the center, as we see for $r = 0, ~3$~fm. $NB$ enhances in the evolution picture. Therefore, all thermoelectric coefficients improved or enhanced due to the temperature evolution, and the effect is prominent in the early evolution.

\subsection{Induced electric field in anisotropic system}
\begin{figure*}
	\centering
	\includegraphics[scale=0.75]{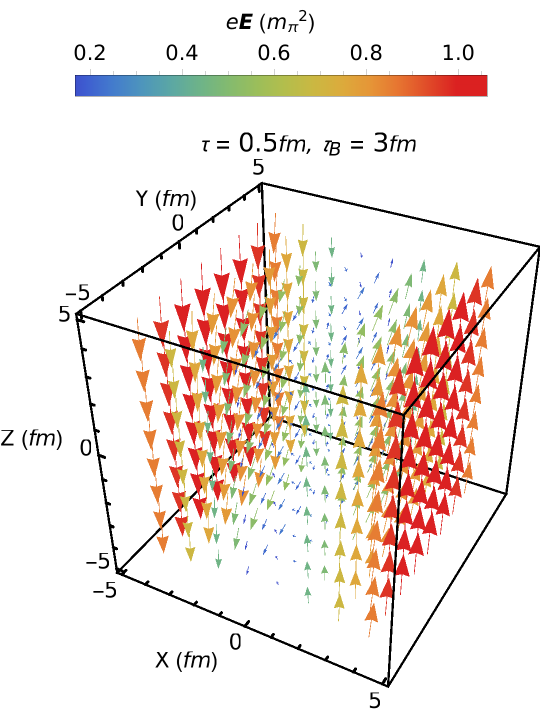}
        \includegraphics[scale=0.75]{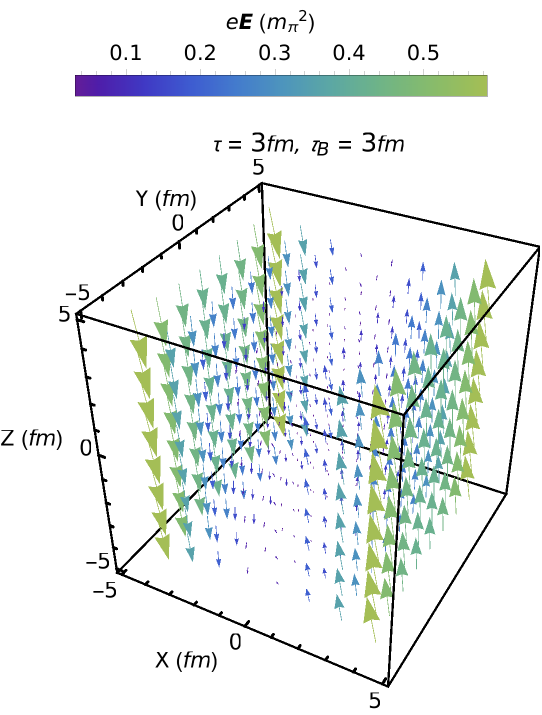}\\
        \includegraphics[scale=0.75]{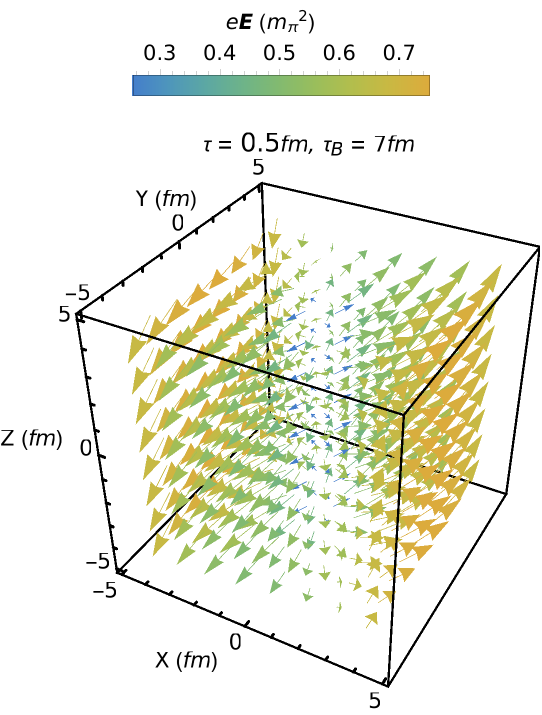}
        \includegraphics[scale=0.75]{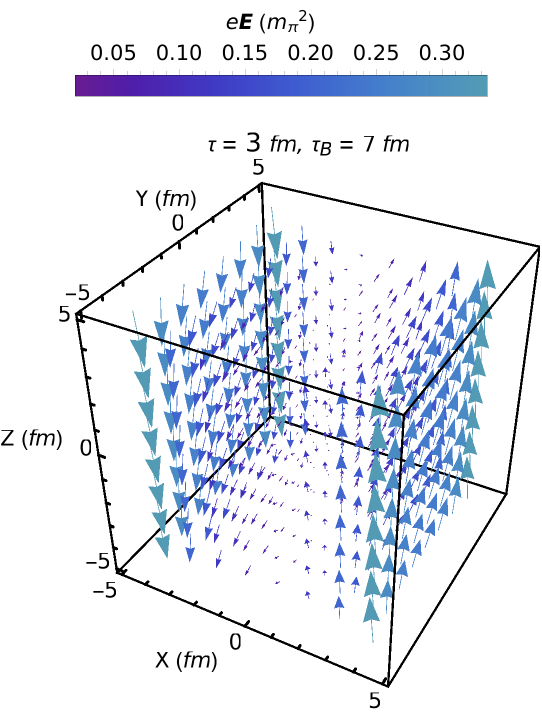}
  \caption{Time evolution of the induced electric field in the QGP in the peripheral collisions with $eB_0 = 5~m_\pi^2$. Upper left: at = $\tau = 0.5~$fm, $\tau_B = 3~$fm, upper right: at $\tau = 3~$fm, $\tau_B = 3~$fm, Bottom left: at $\tau = 0.5~$fm, $\tau_B = 7~$fm, and bottom right: $\tau = 3~$fm, $\tau_B = 7~$fm.}\label{E3D}
\end{figure*}
\begin{figure}
	\centering
	\includegraphics[scale=0.75]{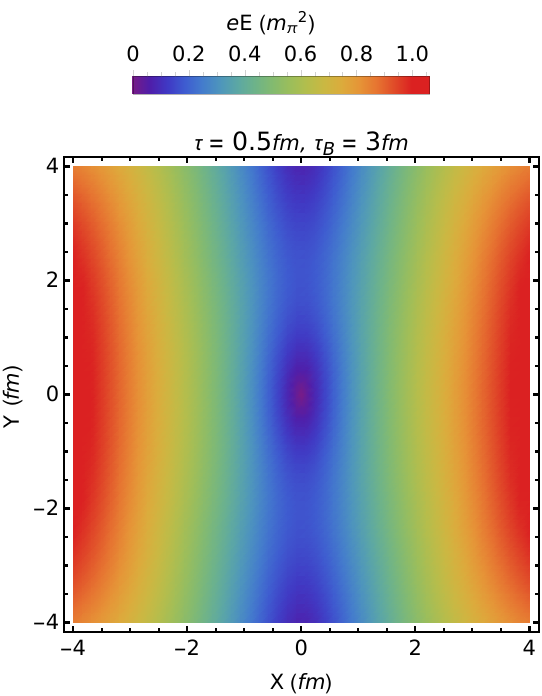}
        \includegraphics[scale=0.75]{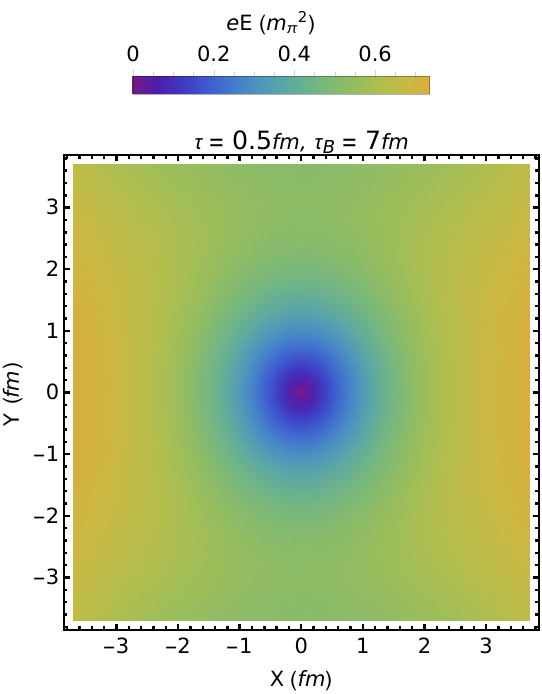}       
  \caption{Magnitude of the electric field in the transverse plane with $eB_0 = 5~m_\pi^2$.}
  \label{Eabs}
\end{figure}
\begin{figure}
    \centering
    \includegraphics[scale=0.75]{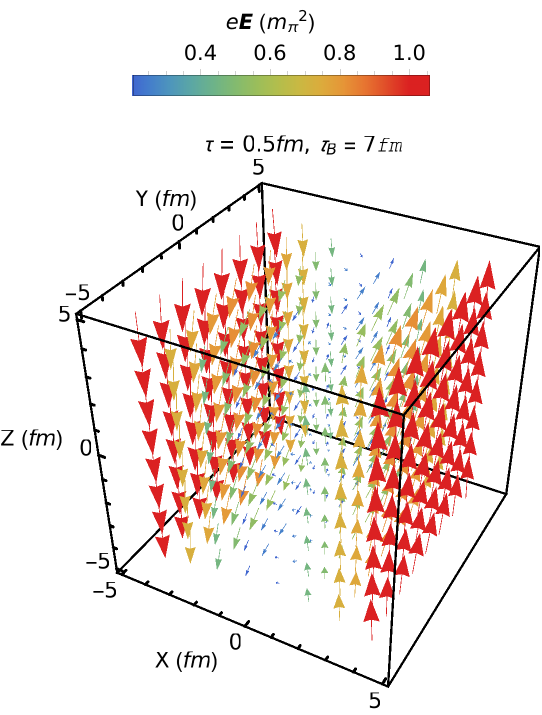}
    \includegraphics[scale=0.75]{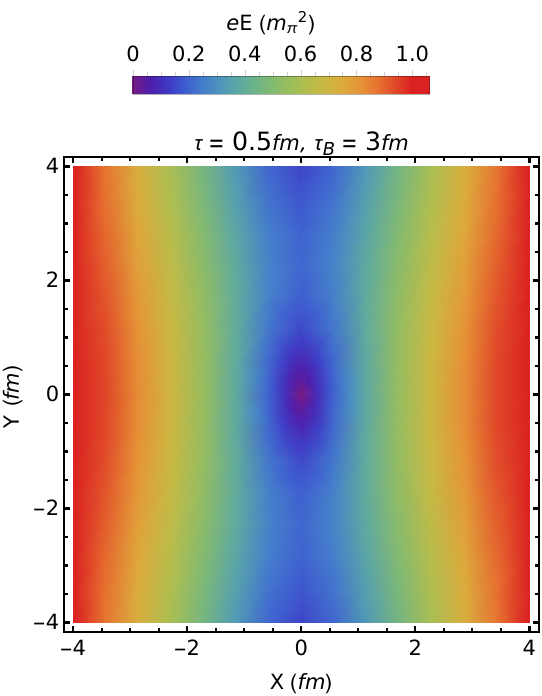}
    \caption{Induced electric field vector (upper figure) and magnitude (bottom figure) for $eB_0 = 1$~m$_\pi^2$.}
    \label{fig:EleceB1}
\end{figure}
Here, we estimate the induced electric field in the peripheral collisions for the different magnetic fields at $\mu_B = 0.3$~GeV. To have a more realistic picture of the induced field,  we used evolving and Landau quantized values of thermoelectric coefficients.
In peripheral collisions, the initial magnetic field created by moving spectators can break the momentum isotropy of the medium. As a result, in Eq.~(\ref{matrix2}), we obtained an anisotropic coefficient matrix, which further produces an anisotropic electric field due to the thermoelectric effect. 

In Fig.~(\ref{E3D}), we plotted the space-time profile of electric field $\vec E$, induced in QGP created in peripheral collision with the initial magnetic field of $eB_0 = 5~m_\pi^2$. Upper panels are for magnetic field decay parameter $\tau_B = 3$~fm at two different times $\tau = 0.5,~3$~fm, and lower panels are for $\tau_B = 7$~fm. 
 In the bottom left panel at $\tau = 0.5$~fm and $\tau_B = 7$~fm, the produced electric field direction differs from other results. Which needs a careful observation to understand. From the Eq.~\ref{matrix2}, components of electric field $E_x,~E_y,~E_z$ are directly proportional to $S_B,~\Bar{S}_B,~NB$, respectively. Therefore, the relative strength of the thermoelectric coefficients is sensitive to the direction of the induced electric field. For $\tau_B = 7$~fm at $\tau = 0.5$~fm, $S_B$ and $\Bar{S}_B$ dominates over $-NB$. As a result, $x$ and $y$ components of the electric field dominate. However, in other electric field results, $NB$ dominates, leading to electric field directing along the $z$ axis.   
Unlike the head-on collision case, $E_z \neq 0$ in the peripheral collision. Due to boost invariant symmetry, $\vec E$ is symmetric along the $z$ axis. However, symmetry in the transverse plane is now broken due to the magnetic field along the $y$ axis. This can be easily seen from the magnitude plot in Fig.~(\ref{Eabs}). In the early time during the QGP evolution, the electric field could go as high as $eE \approx 1~m_\pi^2$, close to the QCD scale. We see that increasing $\tau_B$ from $3$ to $7$~fm, the strength of $\vec E$ reduces at an early time; however, at a later time, it is quite the same for $\tau_B = 3$ and $7$~fm. This signifies that in the current formalism, the electric field induced by the thermoelectric effect increases for faster decay of the initially produced magnetic field. This can be understood as follows. For QGP with lower electrical conductivity, the decay of the magnetic field is comparatively fast. Now, from Eq.~(\ref{sb}) thermoelectric coefficients varies inversely with electrical conductivity ($S_B/~\Bar{S}_B/~NB \sim \frac{1}{\sigma_{e}}$). Therefore, smaller $\sigma_e$ means fast decay of magnetic field or smaller value of $\tau_B$, which enhances Seebeck and Nernst coefficients. As a result, $\vec E$ increases. In Fig.~(\ref{fig:EleceB1}), we plotted $\vec E$ for $eB_0 = 1~m_\pi^2$ to see the effect of different magnetic field strengths. There is a minor change in the induced field upon changing the initial magnetic field strength from $eB = 5$ to $1~m_\pi^2$. A similar thing was already observed in the thermoelectric coefficients in Figs.~(\ref{Fig-temp2},~\ref{Fig-seebar}), and \ref{Fig-temp4}. 

\subsubsection{Significance of the findings}
Studies~\cite{PhysRevC.89.054905,
DAS2017260, 
PhysRevC.98.055201} show that charge-dependent directed flow ($v_1$) can be used as a probe to the electromagnetic field in the QGP. Recent experimental study at RHIC~\cite{PhysRevX.14.011028} has explored this with $v_1$ of $p^\pm$ (proton), $k^\pm$ (kaon), and $\pi^\pm$ (pion). In peripheral collisions, $v_1$ for positive (negative) charged particles get enhanced (reduced) in rapidity $y>0$, due to {\it Hall effect} from Lorentz force. This is reversed in rapidity $y<0$. Experimental data~\cite{PhysRevX.14.011028} show that $v_1$ for positively charged particles ($p^+,~k^+,~\pi^+$) dominates over their negative counterparts (or antiparticles). These findings support our results of the induced electric field. From Fig.~\ref{E3D}, electric field in $x>0$ is directed along $+z$. Whereas, for $x<0$ it is directed along $-z$. This induced electric field enhances (reduces) the {\it Hall effect} in positively (negatively) charged particles by increasing (decreasing) their velocity from Coulombic force (not from spectators). As a result, $v_1$ for positively charged particles is higher than their antiparticles (see Fig.~5 of Ref.~\cite{PhysRevX.14.011028}). In future studies, we will extend this work to include a detailed quantitative analysis of the thermoelectric effect's impact on $v_1$.      

It is important to note that we did not account for the effect of the magnetic field on the Gubser cooling rate. A more realistic picture of the cooling rate should be calculated with a finite and time-varying magnetic field. Moreover, according to Maxwell's electromagnetic relation, the induced space-time-dependent electric field will give rise to a magnetic field. However, this would require the calculation of permittivity and permeability of the evolving QGP, which is subjected to further research and beyond the scope of the current work.
In the present study, for the first time, we have estimated the induced electric field in QGP due to the thermoelectric effect. We used the hydrodynamic theory, where analytical solutions are achievable. This study can be quickly extended to incorporate in (3+1)D hydrodynamics or magnetohydrodynamics such as ECHO QGP by dividing temperature into space-time grids.     
%

	%
 
\section{Summary and Conclusion} \label{sec-summary}
In summary, for the first time, we estimated the induced electric field in QGP solely using the thermoelectric effect. We employed the kinetic theory-based RTA approach to calculate thermoelectric coefficients and induced field. For numerical estimation, we used a quasiparticle model that reproduces the lattice QCD EoS of QGP. An analytical expression of the cooling rate is necessary to calculate the induced electric field, for which we use Gubser hydrodynamics. Moreover, considering the realistic scenario of the temperature evolution of QGP, we introduced a term containing the time derivative of temperature gradient in the calculation of the total single-particle distribution function in the Boltzmann equation. We named it the evolving picture, and the scenario where it is not considered is termed a static picture. We have also studied their relative significance. Heat conduction demands a conserved charge (or particle number). Here, we have baryon number conservation. In this instance, a finite baryon chemical potential is necessary. Otherwise, thermoelectric coefficients diverge up at vanishing baryon density. All results here are presented for the baryon chemical potential $\mu_B = 0.3$~GeV.

In a fluid of charged particles with finite thermal conductivity, thermoelectric phenomena can create or induce an electromagnetic field. QGP consists of quarks, which are electrically charged and have finite baryon quantum numbers. Therefore, they are capable of electrical (charge) and thermal (baryon) conduction. As a result, thermoelectric transport in QGP at finite baryon chemical potential leads to the induction of an electric field. QGP medium is governed by the QCD interaction. Studies show that the EM field of the order of $\Lambda_{\rm QCD}$ scale can interfere with the interactions, contributing to many interesting phenomena~\cite{PhysRevD.78.074033,PhysRevLett.73.3499,PhysRevD.56.5251}. It is widely accepted that such a huge EM field was created in peripheral collisions~\cite{Tuchin2013ie}. The present work shows that an EM field can also be created during the QGP evolution due to the thermoelectric effect, even in head-on collisions. The maximum electric field produced in head-on and peripheral collisions is $\sim 0.7~m_\pi^2$ and $1.0~m_\pi^2$, respectively, which are close to the QCD scale ($\sim 1 m_\pi^2$). However, this can contribute to many interesting phenomena, as mentioned before. Nevertheless, the impact might be small. Moreover, we have considered the temperature evolution in calculating thermoelectric coefficients and estimating induced fields. Evolution contributes significantly at the early stage, enhancing the induced electric field compared to the static picture. 
In peripheral collisions, we consider an external time-varying magnetic field. The induced electric field is further enhanced by the external magnetic field. In other words, the magneto-thermoelectric effect induces a stronger electric field than the thermoelectric effect. The impact of the magnitude of the initial magnetic field and its decay parameter are also studied. At the early time of evolution, the magnetic decay parameter has a substantial impact; however, the strength of the initial magnetic field has a modest effect. 
Furthermore, we have accounted quantum modification of particles' energy state due to Landau quantization in the presence of the magnetic field.
Considering RHICs, the external magnetic field is taken along the $y$ direction for the numerical estimation, for which the induced field becomes anisotropic in the transverse plane. However, we have also generalized the calculations, where the direction of the magnetic field was not preferred, and we obtained the thermoelectric coefficient matrix for the same.   

Despite the interesting findings, our approach has a few limitations. We need to have an analytical solution for the cooling rate to calculate the thermoelectric coefficients for an evolving system. However, the hydrodynamic equations in ($3+1$) D result in coupled differential equations, which can only be solved numerically. Therefore, we considered simplistic hydrodynamic pictures (Bjorken and Gubser flow), where an analytical solution for the cooling rate is achievable. Nevertheless, the results of thermoelectric coefficients obtained here can be used in ($3+1$) D hydrodynamic simulation for a better understanding of the space-time profile of the induced EM field.  



\label{sec-summary}

\acknowledgments
K.S. acknowledges the doctoral fellowship from UGC, Government of India. J.D. and R.S. gratefully acknowledge the DAE-DST, Govt. of India funding under the mega-science project – “Indian participation in the ALICE experiment at CERN” bearing Project No. SR/MF/PS-02/2021-
IITI (E-37123).

\appendix
\setcounter{equation}{0}
\renewcommand{\thesubsection}{\Alph{subsection}}
\renewcommand{\theequation}{\thesubsection.\arabic{equation}}
\section{Appendix}

\subsection{THERMOELECTRIC COEFFICIENTS}
\label{appendix2}
\subsubsection{Case-I (B = 0)}

To find $\delta f_i$ in Eq.~(\ref{Cur-den}), we solve the Boltzmann transport equation (BTE) with the help of relaxation time approximation (RTA). In the presence of temperature gradients, the BTE under RTA can be expressed as~\cite{PhysRevD.102.016016}
\begin{align}\label{ABTE-RTA0}
	\frac{\del f_i}{\del t} + \frac{\vk_i}{\om_i} \cdot \frac{\del f_i}{\del \vec{x_i}} + q_i \vec{E}\cdot \frac{\del f_i}{\del \vec{k_i}}
	= -\frac{\delta f_i}{\tau^i_R}~,
\end{align}
where $\tau^i_R$ is the relaxation time of the particle. $\Vec{E}$ is an external electric field. 
We can solve Eq.~(\ref{ABTE-RTA0})  using Eqs.~(\ref{delta-f0}) and (\ref{Omega00}), and get the expression of $\alpha_j$'s. 
Here, the first term on left-hand side (lhs) of the Eq.~(\ref{ABTE-RTA0}) becomes,
\begin{align}\label{5.10}
&  -(\om_i - {\rm b}_{i}\mu)\frac{\dot T}{T}\frac{\partial f^0_i}{\partial \om_i} +\om_i {\vec{v}_i}.\Big\{\dot{\alpha_1}{\vec{E}}+\alpha_1 {\dot {\vec E}}+\dot{\alpha_2} {\vec {\nabla} T} \nn\\
&+ \alpha_2 {\vec {\nabla} \dot T}\Big\}\frac{\partial f^0_i}{\partial \om_i}.
\end{align}
The second term in the lhs of the Boltzmann equation leads to, 
\begin{align}\label{5.11}
\frac{\partial f_i}{\partial x_i} &= \frac{\partial}{\partial x_i}(f^{0}_i +\delta f_i)\nonumber = \frac{\partial}{\partial x_i}(f^0_i) + \frac{\partial f_i}{\partial x_i}(\delta f_i) \nonumber\\
&=\frac{\partial}{\partial x_i}\Big(\frac{1}{1\pm\exp{\big(\beta( \om_i - {\rm b}_i \mu_i)\big)}}\Big) + 0\nn\\
&= -(\om_i - {\rm b}_ih)\frac{\vec{\nabla}T}{T}\frac{\partial f^0_i}{\partial \om_i}.
\end{align}
The third term in the lhs leads to,
\begin{align}\label{5.12}
\frac{\partial f_i}{\partial k_i} &= \frac{\partial f^0_i}{\partial k_i} + \frac{\partial \delta f_i}{\partial k_i} = \vec{v_i}\frac{\partial f^0_i}{\partial \om_i} + 
\vec {\Omega}\frac{\partial f^0_i}{\partial \om_i}.
\end{align}
Hence,
\begin{align}\label{5.13}
&q_i \vec{E} \cdot \frac{\del f_i}{\del \vec{k_i}} = q_i\vec{v_i} \cdot \vec{E}\frac{\partial f^0_i}{\partial \om_i} + q_iE \big(\alpha_1 \vec{E} +\alpha_2 \vec{\nabla}T + \alpha_3 \vec{\nabla}\dot{T} \big)\frac{\partial f^0_i}{\partial \om_i}. 
 \end{align}
Finally, after the substitution of the above results in both sides of Eq.~(\ref{ABTE-RTA0}), we get
\begin{align}\label{5.10A}
&  -(\om_i - {\rm b}_i\mu)\frac{\dot T}{T}\frac{\partial f^0_i}{\partial \om_i} + \om_i {\vec{v}_i}.\Big(\dot{\alpha_1}{\vec{E}}+\alpha_1 {\dot {\vec E}}+\dot{\alpha_2} {\vec {\nabla} T} + \alpha_2 {\vec {\nabla} \dot T}\Big)\nn\\
&\frac{\partial f^0_i}{\partial \om_i} -(\om_i - {\rm b}_ih)\frac{\vec{\nabla}T}{T}\frac{\partial f^0_i}{\partial \om_i}
 + q_i\vec{v_i} \cdot \vec{E}\frac{\partial f^0_i}{\partial \om_i}\nn\\ 
&= -\frac{\om_i}{\tau_R^i}\Big(\alpha_1 \vec{E} +\alpha_2 \vec{\nabla}T + \alpha_3 \vec{\nabla}\dot{T}\Big)\frac{\partial f^0_i}{\partial \om_i}.
\end{align}
In the current analysis, we consider only the terms with first-order derivatives of the fields and neglect higher-order derivative terms. The comparison of the coefficients of ${\vec {v}}\cdot{\vec {E} }$, ${\vec {v}}\cdot{\vec {\nabla} T}$ and ${\vec {v}}\cdot{\vec {\nabla} \dot T}$ on both sides of the above equation leads to,
\begin{align}\label{alphas0}
 \dot{\alpha_1} &= -\Bigg(\frac{1}{\tau_R^i}\alpha_1+ \frac{q_i}{\om_i}\Bigg),~~~ 
  \dot{\alpha_2} =  -\bigg(\frac{1}{\tau_R^i}\alpha_2  -\frac{\om_i - {\rm b}_ih}{\om_iT}\bigg),\nn\\
  {\alpha_3} &=-{\tau_R^i}\alpha_2, 
\end{align}
Here, $\dot{\alpha_1}$ and $\dot{\alpha_3}$  from Eq.~(\ref{alphas0}) can be expressed in terms of matrix equation as
\begin{equation}\label{5.160}
 \frac{d X}{d t}= AX +G,   
\end{equation}
where the matrices take the following forms
\begin{align}
 X=\begin{pmatrix}
\alpha_1\\\alpha_2\\
\end{pmatrix},~~~
A=\begin{pmatrix}
-\frac{1}{\tau_R^i} &0 \\
 0 &-\frac{1}{\tau_R^i} \\
\end{pmatrix},~~~
G =\begin{pmatrix}
-\frac{q_i}{\om_i}\\ \frac{\om_i - {\rm b}_ih}{\om_iT}
\end{pmatrix}.
\end{align}
Eq.~(\ref{5.160}) can be solved by using the method of the variation of constants. The eigenvalues corresponding to matrix $A$ are,\\
$\lambda = \lambda_1 = \lambda_2 $ = $-\frac{1}{\tau_R^i}.$
Hence, one can write the linearly independent solutions corresponding to the homogeneous part of differential  Eq.~ (\ref{5.160}) in terms of its eigenvectors as
\begin{align}
y_1=\begin{pmatrix}
e^{\eta}\\ 0 
\end{pmatrix},~~~
y_2=\begin{pmatrix}
0\\ e^{\eta}
\end{pmatrix};~~{\rm where} ~~\eta &= -\frac{\tau}{\tau_R^i} .
\end{align}
Therefore, the fundamental matrix for Eq.~(\ref{5.160}) is 
\begin{align}\label{5.200}
&Y=\begin{pmatrix}
e^{\eta}& 0 \\
0 &  e^{\eta} \\
\end{pmatrix},
\end{align}
We seek a particular solution of the equation with a given form of Eq.~(\ref{5.160})
is
\begin{align}\label{5.220}
Y_p &= YU,
\end{align}
where $U$ is a column matrix of unknowns. 
\begin{align}\label{5.230}
U=\begin{pmatrix}
u1\\ u2\\ .\\ .\\ .\\ u_n\\
\end{pmatrix}.
\end{align}
From Eq.~(\ref{5.220}) and Eq.~(\ref{5.160}), we can see that $Y_p$ is a column matrix with the same coefficients as that of matrix X. Further, the differentiation of Eq.~(\ref{5.220}) with respect to time gives us
$Y_p^{'}$ = $Y^{'}U + YU^{'}$, where $Y^{'} = AY$. 
Hence, 
$Y_p^{'}$ = $AY_p + YU^{'}$.
Comparison of above equation with Eq.~(\ref{5.160}) tells us 
\begin{align}\label{5.24}
 G = YU^{'}.   
\end{align}
The determinant of matrix $Y$ is $e^{2\eta}$. Then, 
\begin{align}\label{}
u^{'}_1 &= \frac{1}{e^{2\eta}}
det
\begin{pmatrix}
-\frac{q_i}{\om_i}& 0 \\
\frac{\om_i - {\rm b}_ih}{\om_iT} &  e^{\eta} \\
\end{pmatrix},
~~~u^{'}_2 = \frac{1}{e^{2\eta}} 
 det
\begin{pmatrix}
{e^\eta}& -\frac{q_i}{\om_i} \\
0 & \frac{\om_i - {\rm b}_ih}{\om_iT} \\
\end{pmatrix}.\nonumber
\end{align}
After integrating $u^{'}_1$ and $u^{'}_2$ with respect to time, we get the matrix $U$ as\\
\bea
U =
\begin{pmatrix}
\frac{-q_i}{\om_i} \xi_1\vspace{2mm}\\
 \frac{\om_i - {\rm b}_ih}{\om_iT} \xi_2\\
 \end{pmatrix}.
 \eea
Where $\xi = \int e^{-\eta} d\tau$. \\
\\After substituting the above value of $U$ in Eq.~(\ref{5.22}) one obtain,

\begin{align}
Y_p &= 
\begin{pmatrix}
{e^\eta}& 0 &  \\
0 & e^{\eta} \\
\end{pmatrix}
\begin{pmatrix}
\frac{-q_i}{\om_i} \xi_1\\
 \frac{\om_i - {\rm b}_ih}{\om_iT} \xi_2\\
\end{pmatrix}, \nonumber\\
\begin{pmatrix}
\alpha_1\\\alpha_2\\
\end{pmatrix}
 &= 
\begin{pmatrix}
e^{\eta}& 0 &  \\
0 & e^{\eta} \\
\end{pmatrix} 
\begin{pmatrix}
u_1\\
u_2\\
\end{pmatrix}.\nonumber
\end{align}
Hence,
\begin{align}\label{talpha3}
\alpha_1 = u_1 e^{\eta},~~~
\alpha_2 = u_2e^{\eta}.
\end{align}
The functions $u_1(\tau)$ and $u_2(\tau)$ can be defined as $u_1 =-\frac{q_i}{\om_i} \xi_1$,  $u_2 =i \frac{\om_i - {\rm b}_ih}{\om_iT} \xi_2$.
Hence, the alphas in Eq.~(\ref{alphas0}) get the forms,
\begin{align}
\alpha_1 = -\frac{q_i\tau_R^i}{\om_i},~~~
\alpha_2 = \frac{\tau_R^i(\om_i - {\rm b}_i h)}{T\om_i} ,~~~
\alpha_3 = -\tau_R^i\alpha_2. \nn\\ 
\end{align}
After getting the expressions for $\alpha_j$'s, the simplified form of $\delta f_i$ is,
\begin{align}
 \delta f_i &= -q_i \tau_R^i
 (\vec{v_i}.\vec{E})\frac{\partial f^0_i}{\partial \om_i}
 + \frac{(\om_i - {\rm b}_ih) \tau_R^i}{T}\Big[\Big(\vec{v_i}\cdot\vec{\nabla}{T}\Big)\nn\\
&-\tau_R^i\Big(\vec{v_i}\cdot\vec{\nabla}{\dot T}\Big)\Big]\frac{\partial f^0_i}{\partial \om_i}.
\end{align}
Here, $\pm$ indicates a positive sign for positively charged particles(antiparticles) and a negative sign for negatively charged particles (antiparticles).
\subsubsection{Case-II (B $\not=$ 0)}
To find the expression of $\delta f_i$ in Eq.~(\ref{Cur-den}), we solve the Boltzmann transport equation (BTE) with the help of relaxation time approximation (RTA). In the presence of an external electromagnetic field,  BTE under RTA can be expressed as~\cite{PhysRevD.102.016016}
\begin{align}\label{ABTE-RTA}
	\frac{\del f_i}{\del t} + \frac{\vk_i}{\om_i} \cdot \frac{\del f_i}{\del \vec{x_i}} + q_i \left(\vec{E} + \frac{\vk_i}{\om_i} \times \vec{B}\right) \cdot \frac{\del f_i}{\del \vec{k_i}}
	= -\frac{\delta f_i}{\tau^i_R}~,
\end{align}
where $\tau^i_R$ is the relaxation time of the particle. We can solve Eq.~(\ref{ABTE-RTA})  using Eqs.~(\ref{delta-f}) and (\ref{Omega}), and get the expression of $\alpha_j$'s. 
Here, the first term on the lhs of the Eq.~(\ref{ABTE-RTA}) becomes,
\begin{align}\label{5.10}
&\om_i {\vec{v}_i}.\Big\{\dot{\alpha_1 }{\vec {E}}+\alpha_1 \dot{{\vec {E}}}+\dot{\alpha_3} {\vec {\nabla}T} + {\alpha_3} {\vec {\nabla}\dot T}+\alpha_5 ({\vec {\nabla}T}\times \dot{{\vec {B}}})\nn\\
&+\dot{\alpha_5 }({\vec {\nabla}T} \times {\vec {B}})
+ {\alpha_5 }({\vec {\nabla} \dot T} \times {\vec {B}})
+\alpha_5 (\vec{\nabla}\dot T \times{\vec{B}}) + \alpha_8 ({\vec E}\times \dot{{\vec {B}}})\nn\\ &+\dot{\alpha_8 }({\vec E} \times  {\vec {B}})+ \alpha_8 (\dot{\vec{E}}\times  {\vec {B}})-(\om_i - {\rm b}_i\mu)\frac{\dot T}{T}\Big\}\frac{\partial f^0_i}{\partial \om_i}.
\end{align}
The second term in the lhs of the Boltzmann equation leads to, 
\begin{align}\label{5.11}
\frac{\partial f_i}{\partial x_i} &= \frac{\partial}{\partial x_i}(f^{0}_i +\delta f_i) =\frac{\partial}{\partial x_i}(f^0_i) + \frac{\partial f_i}{\partial x_i}(\delta f_i) \nn\\
&=\frac{\partial}{\partial x_i}\Big\{\frac{1}{1\pm\exp{\big(\beta( \om_i - {\rm b}_i \mu_i)\big)}}\Big\} + 0\nonumber\\
&= -(\om_i - {\rm b}_ih)\frac{\vec{\nabla}T}{T}\frac{\partial f^0_i}{\partial \om_i}.
\end{align}
The third term in the lhs leads to,
\begin{align}\label{5.12}
\frac{\partial f_i}{\partial k_i} &= \frac{\partial f^0_i}{\partial k_i} + \frac{\partial \delta f_i}{\partial k_i} = \vec{v_i}\frac{\partial f^0_i}{\partial \om_i} + 
\vec {\Omega}\frac{\partial f^0_i}{\partial \om_i}.
\end{align}
Hence,
\begin{align}\label{5.13}
&q_i \left(\vec{E} + \frac{\vk_i}{\om_i} \times \vec{B}\right) \cdot \frac{\del f_i}{\del \vec{k_i}} = q_i\vec{v_i} \cdot \vec{E}\frac{\partial f^0_i}{\partial \om_i} + \big(\vec{v} \times \vec{B} \big) \nn\\
&\big\{\alpha_1 \vec{E} + \alpha_2 \dot{\vec E} +\alpha_3 \vec{\nabla}T + \alpha_4 \vec{\nabla}\dot{T} + \alpha_5 (\vec{\nabla}T \times \vec{B})\nn\\
 &+ \alpha_6 (\vec{\nabla}T\times \dot{\vec{B}}) 
 + \alpha_7 (\vec{\nabla}\dot{T} \times \vec{B})  + \alpha_8 (\vec{E} \times {\vec{B}})+ \alpha_9 (\vec{E} \times {\dot{\vec{B}}})\nn\\ 
 &+ \alpha_{10}(\dot{\vec{E}} \times {\vec{B}})\big\}\frac{\partial f^0_i}{\partial \om_i}. 
 \end{align}
Finally, after the substitution of the above results on both the sides of Eq.~(\ref{ABTE-RTA}), the left side becomes
\begin{widetext}
\begin{align}\label{5.14}
&\om_i {\vec{v}_i}.\Big\{\dot{\alpha_1 }{\vec {E}}+\alpha_1 \dot{{\vec {E}}}+\dot{\alpha_3} {\vec {\nabla}T} + {\alpha_3} {\vec {\nabla}\dot T}+\alpha_5 ({\vec {\nabla}T}\times \dot{{\vec {B}}})+\dot{\alpha_5 }({\vec {\nabla}T} \times {\vec {B}})+ {\alpha_5 }({\vec {\nabla} \dot T} \times {\vec {B}})
+\alpha_5 (\vec{\nabla}\dot T \times{\vec{B}}) + \alpha_8 ({\vec E}\times \dot{{\vec {B}}})\nn\\ &+\dot{\alpha_8 }({\vec E} \times  {\vec {B}})
+ \alpha_8 (\dot{\vec{E}}\times  {\vec {B}})
-(\om_i - {\rm b}_i\mu)\frac{\dot T}{T}\Big\}\frac{\partial f^0_i}{\partial \om_i}.
-(\om_i -{\rm b}_i h)\vec{v_i}\cdot\frac{\vec{\nabla}T}{T}\frac{\partial f^0_i}{\partial \om_i} + q_i\vec{v_i} \cdot \vec{E}\frac{\partial f^0_i}{\partial \om_i}
 +\big\{-\alpha_1 q_{i}\vec{v_i}\cdot(\vec{E} \times \vec{B})\nn\\
 &-\alpha_2 q_{i}\vec{v_i}\cdot(\dot{\vec E} \times \vec{B})
 -\alpha_3 q_{i}\vec{v_i}\cdot(\vec{\nabla}T \times \vec{B}) -\alpha_4 q_{i}\vec{v_i}\cdot(\vec{\nabla}\dot T \times \vec{B}) +\alpha_5 q_{i}\{-(\vec{v_i}\cdot\vec{B})(\vec{\nabla T}\cdot\vec{B}) + (\vec{v_i}\cdot\vec{\nabla} T)({\vec{B}}\cdot\vec{B})\}\nn\\
 &+\alpha_6 q_{i}\{-(\vec{v_i}\cdot\dot{\vec B})(\vec{\nabla T}\cdot\vec{B}) 
 + (\vec{v_i}\cdot\vec{\nabla} T)(\dot{\vec{B}}\cdot\vec{B})\}
 +\alpha_7 q_{i}\{-(\vec{v_i}\cdot\vec{B})(\vec{\nabla} \dot{T}\cdot\vec{B}) + (\vec{v_i}\cdot  \vec{\nabla} \dot T)({\vec{B}}\cdot\vec{B})\}\nn\\
 &+\alpha_8 q_{i}\{-(\vec{v_i}\cdot\vec{B})(\vec{E} \cdot\vec{B}) + (\vec{v_i}\cdot  \vec{E})({\vec{B}}\cdot\vec{B})\} 
 +\alpha_9 q_{i}\{-(\vec{v_i}\cdot \dot{\vec{B}})(\vec{E} \cdot\vec{B}) + (\vec{v_i}\cdot  \vec{\dot E})({\vec{B}}\cdot\dot{\vec B})\}
  +\alpha_9 q_{i}\{-(\vec{v_i}\cdot \dot{\vec{B}})(\vec{E} \cdot\vec{B}) \nn\\
  &+ (\vec{v_i}\cdot  \dot{\vec E})({\vec{B}}\cdot\dot{\vec B})\} +\alpha_{10} q_{i}\{-(\vec{v_i}\cdot\vec{B})(\dot{\vec{E}} \cdot\vec{B})
  + (\vec{v_i}\cdot  \dot{\vec{E}})({\vec{B}}\cdot\vec{B})\} \Big\}\frac{\partial f^0_i}{\partial \om_i},
\end{align}   
and the right side becomes
\begin{align}
&= -\frac{\om_i}{\tau_R^i}\alpha_1 \vec{E} + \alpha_2 \dot{\vec E} +\alpha_3 \vec{\nabla}T + \alpha_4 \vec{\nabla}\dot{T} + \alpha_5 (\vec{\nabla}T \times \vec{B})
 + \alpha_6 (\vec{\nabla}T\times \dot{\vec{B}}) 
 + \alpha_7 (\vec{\nabla}\dot{T} \times \vec{B})  + \alpha_8 (\vec{E} \times {\vec{B}})+ \alpha_9 (\vec{E} \times {\dot{\vec{B}}})\nn\\
 &+ \alpha_{10}(\dot{\vec{E}} \times {\vec{B}})\frac{\partial f^0_i}{\partial \om_i}.    
\end{align}
\end{widetext}
In the current analysis, we consider only the terms with first-order derivatives of the fields and neglect higher-order derivative terms. The comparison of the coefficients of ${\vec {v}}\cdot{\vec {E}}$, ${\vec {v}}\cdot{\dot{\vec E}}$, ${\vec {v_i}}\cdot{\vec {\nabla}T}$, ${\vec {v_i}}\cdot{\vec {\nabla} \dot T}$, ${\vec {v_i}}\cdot({\vec {\nabla}T \times {\vec{B}}})$, ${\vec {v_i}}\cdot({\vec {\nabla}T\times \dot {\vec{B}}})$,  ${\vec {v_i}}\cdot({\vec {\nabla}\dot T\times \vec{B}})$, ${\vec {v_i}}\cdot({\vec {E} \times {\vec{B}}})$, ${\vec {v_i}}\cdot( {\vec E} \times \dot{\vec B})$ and ${\vec {v_i}}\cdot(\dot {\vec E} \times {\vec B})$   on both sides, gives us $\dot{\alpha_1}$, ${\alpha_2}$, $\dot{\alpha_3}$, $\alpha_4$, $\dot {\alpha_5}$, $\dot{\alpha_3}$, $\alpha_6$, $\alpha_7$, $\dot \alpha_8$, $\alpha_9$ and $\alpha_{10}$ respectively as 
\begin{align}\label{alphaA}
  \dot{\alpha_1} &= -\Bigg[\frac{1}{\tau_R^i}\alpha_1+\frac{q_{i}\big\{(\vec{B}\cdot\vec{B} - \tau_R^i\vec{B}\cdot \dot{\vec{B}})\alpha_7\big\}}{\om_i} + \frac{q_i}{\om_i}\Bigg], \nn\\
  \alpha_2 &= -\tau_R^i \alpha_1 - \frac{\tau_R^i q_{i} \vec{B}^2}{\om_i} \alpha_{10} ,\nn\\  
  \dot{\alpha_3} &=  -\bigg\{\frac{1}{\tau_R^i}\alpha_3 +(\frac{q_{i}(\vec{B}\cdot\vec{B}-\tau_R^i\vec{B}\cdot\dot{\vec{B}})}{\om_i})\alpha_5 -\frac{\om_i - {\rm b}_ih}{T\om_i}\bigg\},\nn \\
  {\alpha_4} &=-{\tau_R^i}\Big\{\alpha_3+\frac{q_{i} (\vec{B}\cdot\vec{B})}{\om_i} \alpha_7 \Big\},\nn\\ 
  \dot{\alpha_5} &= -\frac{1}{\tau_R^i}\alpha_5 +\frac{q_{i}}{\om_i}\alpha_3,\nn \\
  \alpha_6 &= -\tau_R^i \alpha_5,\nn\\
  \alpha_7 &= -\tau_R^i \alpha_5 + \frac{\tau_R^i q_{i}}{\om_i} \alpha_4,\nn\\ 
   \dot{\alpha_8} &= -\frac{1}{\tau_R^i}\alpha_8 +\frac{q_{i}}{\om_i}\alpha_1,\nn \\
   \alpha_9 &= -\tau_R^i \alpha_8,\nn\\
\alpha_{10} &= -\tau_R^i \alpha_8 + \frac{\tau_R^i q_{i}}{\om_i} \alpha_2.  
\end{align}
Here, $\dot{\alpha_1}$, $\dot{\alpha_3}$, $\dot{\alpha_5}$ and $\dot{\alpha_8}$  from Eq.~(\ref{alphaA}) can be expressed in terms of matrix equation as
\begin{equation}\label{5.16}
 \frac{d X}{d t}= AX +G,   
\end{equation}
where the matrices take the following forms
\begin{align}
 X&=\begin{pmatrix}
\alpha_1\\\alpha_3\\\alpha_5\\\alpha_8
\end{pmatrix},~~~
A=\begin{pmatrix}
-\frac{1}{\tau_R^i} &0 &0 &-\frac{q_{i} F^2}{\om_i}\\
 0 &-\frac{1}{\tau_R^i} & -\frac{q_{i} F^2}{\om_i} &0\\
0 &\frac{q_{i}}{\om_i} &  -\frac{1}{\tau_R^i} &0\\
\frac{q_{i}}{\om_i} &0 &0 & -\frac{1}{\tau_R^i},
\end{pmatrix},\nn\\
G &=\begin{pmatrix}
-\frac{q_i}{\om_i}\\ \frac{\om_i - {\rm b}_ih}{T\om_i}\\ 0\\0
\end{pmatrix},\nonumber
\end{align}
with $F = \sqrt{B(B-\tau_R^i \dot{B})}$. Eq.~(\ref{5.16}) can be solved by diagonalizing the matrix $A$ and using the method of the variation of constants. The eigenvalues corresponding to matrix $A$ are,\\
$\lambda _ j $ = $-\frac{1}{\tau_R^i} + a_j i\frac{q_{i} F}{\om_i}$,~~
with $a_1=$ $a_2 = -1$, $a_3 =$ $a_4 = 1$.\\
Hence, one can write the linearly independent solutions corresponding to the homogeneous part of differential  Eq.~ (\ref{5.16}) in terms of its eigenvectors as
\begin{align}
y_1&=\begin{pmatrix}
-iF e^{\eta_1}\\ 0 \\ 0 \\ e^{\eta_1}
\end{pmatrix},~~
y_2=\begin{pmatrix}
0\\ -iF e^{\eta_2} \\e^{\eta_2}\\ 0
\end{pmatrix},\nn\\
y_3&=\begin{pmatrix}
iF e^{\eta_3} \\0 \\0\\ e^{\eta_3}
\end{pmatrix},~~~
y_4=\begin{pmatrix}
0\\ iF e^{\eta_4} \\e^{\eta_4}\\ 0
\end{pmatrix}.\nonumber
\end{align}
Where,
\begin{align}\label{5.19}
\eta_j &= -\frac{\tau}{\tau_R^i} +a_j\frac{q_{i} i}{\om}\int F d\tau.
\end{align}
Therefore, the fundamental matrix for Eq.~(\ref{5.16}) is 
\begin{align}\label{5.20}
&Y=\begin{pmatrix}
-iFe^{\eta_1}& 0 &  iFe^{\eta_3} & 0\\
0 & -iF e^{\eta_2} & 0 & iF e^{\eta_4}\\
0 & e^{\eta_2} & 0 & e^{\eta_4}\\
e^{\eta_1}& 0 &  e^{\eta_3} & 0\\
\end{pmatrix},
\end{align}
We seek a particular solution of the equation with a given form of Eq.~(\ref{5.16})
is
\begin{align}\label{5.22}
Y_p &= YU,
\end{align}
where $U$ is a column matrix of unknowns. 
\begin{align}\label{5.23}
U=\begin{pmatrix}
u1\\ u2\\ .\\ .\\ .\\ u_n\\
\end{pmatrix}.
\end{align}
From Eq.~(\ref{5.22}) and Eq.~(\ref{5.16}), we can see that $Y_p$ is a column matrix with the same coefficients as that of matrix X. Further, the differentiation of Eq.~(\ref{5.22}) with respect to time gives us
$Y_p^{'}$ = $Y^{'}U + YU^{'}$, where $Y^{'} = AY$. 
Hence, 
$Y_p^{'}$ = $AY_p + YU^{'}$.
Comparison of above equation with Eq.~(\ref{5.16}) tells us 
\begin{align}\label{5.24}
 G = YU^{'}.   
\end{align}
The determinant of matrix $Y$ is $4F^{2}e^{\eta}$. Then, 
\begin{align}
u^{'}_1 &= \frac{1}{4F^{2}e^{\eta}}
det
\begin{pmatrix}
-\frac{q_i}{\om_i}& 0 &  iFe^{\eta_3} & 0\\
\frac{\om_i - {\rm b}_ih}{T\om_i} & -iF e^{\eta_2} & 0 & iF e^{\eta_4}\\
0 & e^{\eta_2} & 0 & e^{\eta_4}\\
0& 0 &  e^{\eta_3} & 0\\
\end{pmatrix}, \nonumber
\\u^{'}_2 &= \frac{1}{4F^{2}e^{\eta}} 
 det
\begin{pmatrix}
-iFe^{\eta_1}& -\frac{q_i}{\om_i} &  iFe^{\eta_3} & 0\\
0 & \frac{\om_i - {\rm b}_ih}{T\om_i} & 0 & iF e^{\eta_4}\\
0 & 0 & 0 & e^{\eta_4}\\
e^{\eta_1}& 0 &  e^{\eta_3} & 0\\
\end{pmatrix},\nonumber
\\u^{'}_3 &= \frac{1}{4F^{2}e^{\eta}}
det
\begin{pmatrix}
-iFe^{\eta_1}& 0 &  -\frac{q_i}{\om_i} & 0\\
0 & -iF e^{\eta_2} & \frac{\om_i - {\rm b}_ih}{T\om_i} & iF e^{\eta_4}\\
0 & e^{\eta_2} & 0 & e^{\eta_4}\\
e^{\eta_1}& 0 &  0& 0\\
\end{pmatrix},\nonumber
\\u^{'}_4 &= \frac{1}{4F^{2}e^{\eta}}
det
\begin{pmatrix}
-iFe^{\eta_1}& 0 &  iFe^{\eta_3} & -\frac{q_i}{\om_i}\\
0 & -iF e^{\eta_2} & 0 & \frac{\om_i - {\rm b}_ih}{T\om_i}\\
0 & e^{\eta_2} & 0 & 0\\
e^{\eta_1}& 0 &  e^{\eta_3} & 0\\
\end{pmatrix}.
\end{align}
After integrating $u^{'}_1$, $u^{'}_2$, $u^{'}_3$ and $u^{'}_4$ with respect to time , we get the matrix $U$ as\\
\begin{align}\label{5.26}
U &=
\begin{pmatrix}
\frac{q_i}{2i\om_i} \xi_1\\
i \frac{\om_i - {\rm b}_ih}{2\om_iT} \xi_2\\
-\frac{q_i}{2i\om_i} \xi_3\\
-i \frac{\om_i - {\rm b}_ih}{2\om_i} \xi_4\\
 \end{pmatrix},~~~
 \xi_j = \int \frac{e^{-\eta_{j}}}{F} d\tau. 
 \end{align}
After substituting the above value of $U$ in Eq.~(\ref{5.22}) one obtain,

\begin{align}
Y_p &= 
\begin{pmatrix}
-iFe^{\eta_1}& 0 &  iFe^{\eta_3} & 0\\
0 & -iF e^{\eta_2} & 0 & iF e^{\eta_4}\\
0 & e^{\eta_2} & 0 & e^{\eta_4}\\
e^{\eta_1}& 0 &  e^{\eta_3} & 0\\
\end{pmatrix}
\begin{pmatrix}
\frac{q_i}{2i\om_i} \xi_1\\
i \frac{\om_i - {\rm b}_ih}{2\om_iT} \xi_2\\
-\frac{q_i}{2i\om_i} \xi_3\\
-i \frac{\om_i - {\rm b}_ih}{2\om_i} \xi_4\\
\end{pmatrix}, \nonumber\\
\begin{pmatrix}
\alpha_1\\\alpha_3\\\alpha_5\\\alpha_8
\end{pmatrix}
 &= 
\begin{pmatrix}
-iFe^{\eta_1}& 0 &  iFe^{\eta_3} & 0\\
0 & -iF e^{\eta_2} & 0 & iF e^{\eta_4}\\
0 & e^{\eta_2} & 0 & e^{\eta_4}\\
e^{\eta_1}& 0 &  e^{\eta_3} & 0\\
\end{pmatrix} 
\begin{pmatrix}
u_1\\
u_2\\
u_3\\
u_4\\
\end{pmatrix}.\nonumber
\end{align}
Hence,\\
\begin{align}\label{talpha3}
\alpha_1 = -u_1 iFe^{\eta_1} + u_3 iF e^{\eta_3},~~
\alpha_3 &= -u_2iFe^{\eta_2} + u_4iFe^{\eta_4},\nn\\
\alpha_5 = u_2e^{\eta_2} + u_4e^{\eta_4},~~  
\alpha_8 &= u_1e^{\eta_1} + u_3e^{\eta_3}.
\end{align}
The functions $u_1(\tau)$, $u_2(\tau)$, $u_3(\tau)$ and $u_4(\tau)$ can be defined as $u_1 =\frac{q_i}{2i\om_i} \xi_1$,  $u_2 =i \frac{\om_i - {\rm b}_ih}{2\om_iT} \xi_2$, $u_3 =\frac{q_i}{2i\om_i} \xi_3$, and $u_4 =-i\frac{(\om_i - {\rm b}_ih)}{2\om_i} \xi_4$.
 For time-varying field, we get\\
$ F = B\sqrt{1+\frac{\tau_R^i}{\tau_B}}$ , ~~$\int F d\tau$ = $B\tau\sqrt{1+\frac{\tau_R^i}{\tau_B}}$.\\
Hence, it leads to the given form, 
\begin{align}\label{5.29}
    \eta_j &= -\frac{\tau}{\tau_R^i} +a_ji\frac{\sqrt{1+\frac{\tau_R^i}{\tau_B}}}{\tau_B}~\tau,\nn \\ 
\xi_j &= \frac{1}{\sqrt{1+\frac{\tau_R^i}{\tau_B}} B_0} \frac{e^{\Big(\frac{1}{\tau_R^i} +\frac{1}{\tau_B}-a_ji\frac{\sqrt{1+\frac{\tau_R^i}{\tau_B}}}{\tau_B} \Big)\tau}}{\Big( \frac{1}{\tau_R^i} +\frac{1}{\tau_B}-a_ji\frac{\sqrt{1+\frac{\tau_R^i}{\tau_B}}}{\tau_B} \Big)}.
\end{align}
Finally, the alphas in Eq.~(\ref{alphaA}) are,
\begin{align}
\alpha_1 &= -\frac{q_i}{\om_i}  ~\frac{\tau_R^i}{(1+\chi_i + \chi_i^2)},  \nn \\
\alpha_2 &= \frac{-\tau_R^i (1+\chi_i - \chi_i^2)}{(1+\chi_i)  (1 + \chi_i^2)}\alpha_1,  \nn\\
\alpha_3 &= \frac{(\om_i - {\rm b}_i h)}{\om_i}  ~\frac{\tau_R^i}{(1+\chi_i + \chi_i^2)},  \nn \\ 
\alpha_4 &= \frac{-\tau_R^i (1+\chi_i - \chi_i^2)}{(1+\chi_i)  (1 + \chi_i^2)}\alpha_3,  \nn\\ 
\alpha_5 &= \frac{\chi_i}{B (1+\chi_i)} \alpha_3,  \nn\\
\alpha_6 &= -\tau_R^i\frac{\chi_i}{B (1+\chi_i)} \alpha_3,  \nn\\
\alpha_7 &= -\tau_R^i\frac{\chi_i(2+\chi_i)}{B (1+\chi_i)(1 + \chi_i^2)} \alpha_3,  \nn\\
\alpha_8 &= \frac{\chi_i}{B (1+\chi_i)} \alpha_1,  \nn\\
\alpha_9 &= -\tau_R^i\frac{\chi_i}{B (1+\chi_i)} \alpha_1,  \nn\\
\alpha_{10}&= -\tau_R^i\frac{\chi_i(2+\chi_i)}{B (1+\chi_i)(1 + \chi_i^2)} \alpha_1.
\end{align}
 After getting the expressions for $\alpha_j$'s, the simplified form of $\delta f_i$ in Eq.~(\ref{delta-f}) is,
\begin{align}
 \delta f_i &= \frac{-q_i \tau_R^i}{(1+\chi_i)(1+ \chi_i + \chi_i^2)}\Big[\Big\{(1+\chi_i)+\frac{\chi_i(1+ \chi_i - \chi_i^2)}{(1+ \chi_i^2)}\Big\}\nn\\
 &(\vec{v_i}.\vec{E}) \pm \Big\{\chi_i(1+\chi_i)
 + \frac{\chi_i^2(2+ \chi_i)}{(1+ \chi_i^2)}\Big\}\Big(\vec{v_i}.(\vec{E}\times \hat{b})\Big)\Big]\nn\\ 
 &+ \frac{(\om_i - {\rm b}_ih) \tau_R^i}{T(1+\chi_i)(1+ \chi_i + \chi_i^2)}\Big[(1+ \chi_i)\Big(\vec{v_i}\cdot\vec{\nabla}{T}\Big)\nn\\
 &-\tau_R^i\frac{(1+ \chi_i - \chi_i^2)}{(1+ \chi_i^2)}\Big(\vec{v_i}\cdot\vec{\nabla}{\dot T}\Big)\pm\chi_i(1+\chi_i)
 \Big(\vec{v_i}\cdot(\vec{\nabla}{T} \times \hat{b})\Big)\nn\\
 &\mp\tau_R^i \frac{\chi_i(2+\chi_i)}{(1+ \chi_i^2)}
 \Big(\vec{v_i}\cdot(\vec{\nabla}{\dot T} \times \hat{b})\Big)\Big]\frac{\partial f^0_i}{\partial \om_i}.
\end{align}
For simplicity we have considered $\tau_E = \tau_B$ and $\chi_i = \frac{\tau_R^i}{\tau_B} = \frac{\tau_R^i}{\tau_E}$. Here $\pm$ indicates a positive sign for positively charged particles(antiparticles) and a negative sign for negatively charged particles (antiparticles).

\bibliographystyle{apsrev4-2}
\bibliography{reference_1}

\begin{thebibliography}{73}%
\makeatletter
\providecommand \@ifxundefined [1]{%
 \@ifx{#1\undefined}
}%
\providecommand \@ifnum [1]{%
 \ifnum #1\expandafter \@firstoftwo
 \else \expandafter \@secondoftwo
 \fi
}%
\providecommand \@ifx [1]{%
 \ifx #1\expandafter \@firstoftwo
 \else \expandafter \@secondoftwo
 \fi
}%
\providecommand \natexlab [1]{#1}%
\providecommand \enquote  [1]{``#1''}%
\providecommand \bibnamefont  [1]{#1}%
\providecommand \bibfnamefont [1]{#1}%
\providecommand \citenamefont [1]{#1}%
\providecommand \href@noop [0]{\@secondoftwo}%
\providecommand \href [0]{\begingroup \@sanitize@url \@href}%
\providecommand \@href[1]{\@@startlink{#1}\@@href}%
\providecommand \@@href[1]{\endgroup#1\@@endlink}%
\providecommand \@sanitize@url [0]{\catcode `\\12\catcode `\$12\catcode
  `\&12\catcode `\#12\catcode `\^12\catcode `\_12\catcode `\%12\relax}%
\providecommand \@@startlink[1]{}%
\providecommand \@@endlink[0]{}%
\providecommand \url  [0]{\begingroup\@sanitize@url \@url }%
\providecommand \@url [1]{\endgroup\@href {#1}{\urlprefix }}%
\providecommand \urlprefix  [0]{URL }%
\providecommand \Eprint [0]{\href }%
\providecommand \doibase [0]{https://doi.org/}%
\providecommand \selectlanguage [0]{\@gobble}%
\providecommand \bibinfo  [0]{\@secondoftwo}%
\providecommand \bibfield  [0]{\@secondoftwo}%
\providecommand \translation [1]{[#1]}%
\providecommand \BibitemOpen [0]{}%
\providecommand \bibitemStop [0]{}%
\providecommand \bibitemNoStop [0]{.\EOS\space}%
\providecommand \EOS [0]{\spacefactor3000\relax}%
\providecommand \BibitemShut  [1]{\csname bibitem#1\endcsname}%
\let\auto@bib@innerbib\@empty
\bibitem [{\citenamefont {Brewer}\ \emph {et~al.}(2021)\citenamefont {Brewer},
  \citenamefont {Yan},\ and\ \citenamefont {Yin}}]{BREWER2021136189}%
  \BibitemOpen
  \bibfield  {author} {\bibinfo {author} {\bibfnamefont {J.}~\bibnamefont
  {Brewer}}, \bibinfo {author} {\bibfnamefont {L.}~\bibnamefont {Yan}},\ and\
  \bibinfo {author} {\bibfnamefont {Y.}~\bibnamefont {Yin}},\ }\href
  {https://doi.org/https://doi.org/10.1016/j.physletb.2021.136189} {\bibfield
  {journal} {\bibinfo  {journal} {Physics Letters B}\ }\textbf {\bibinfo
  {volume} {816}},\ \bibinfo {pages} {136189} (\bibinfo {year}
  {2021})}\BibitemShut {NoStop}%
\bibitem [{\citenamefont {Busza}\ \emph {et~al.}(2018)\citenamefont {Busza},
  \citenamefont {Rajagopal},\ and\ \citenamefont {van~der
  Schee}}]{Busza:2018rrf}%
  \BibitemOpen
  \bibfield  {author} {\bibinfo {author} {\bibfnamefont {W.}~\bibnamefont
  {Busza}}, \bibinfo {author} {\bibfnamefont {K.}~\bibnamefont {Rajagopal}},\
  and\ \bibinfo {author} {\bibfnamefont {W.}~\bibnamefont {van~der Schee}},\
  }\href {https://doi.org/10.1146/annurev-nucl-101917-020852} {\bibfield
  {journal} {\bibinfo  {journal} {Ann. Rev. Nucl. Part. Sci.}\ }\textbf
  {\bibinfo {volume} {68}},\ \bibinfo {pages} {339} (\bibinfo {year}
  {2018})}\BibitemShut {NoStop}%
\bibitem [{\citenamefont {Elfner}\ and\ \citenamefont
  {M\"uller}(2023)}]{Elfner:2022iae}%
  \BibitemOpen
  \bibfield  {author} {\bibinfo {author} {\bibfnamefont {H.}~\bibnamefont
  {Elfner}}\ and\ \bibinfo {author} {\bibfnamefont {B.}~\bibnamefont
  {M\"uller}},\ }\href {https://doi.org/10.1088/1361-6471/ace824} {\bibfield
  {journal} {\bibinfo  {journal} {J. Phys. G}\ }\textbf {\bibinfo {volume}
  {50}},\ \bibinfo {pages} {103001} (\bibinfo {year} {2023})}\BibitemShut
  {NoStop}%
\bibitem [{\citenamefont {Alzhrani}\ \emph {et~al.}(2022)\citenamefont
  {Alzhrani}, \citenamefont {Ryu},\ and\ \citenamefont
  {Shen}}]{PhysRevC.106.014905}%
  \BibitemOpen
  \bibfield  {author} {\bibinfo {author} {\bibfnamefont {S.}~\bibnamefont
  {Alzhrani}}, \bibinfo {author} {\bibfnamefont {S.}~\bibnamefont {Ryu}},\ and\
  \bibinfo {author} {\bibfnamefont {C.}~\bibnamefont {Shen}},\ }\href
  {https://doi.org/10.1103/PhysRevC.106.014905} {\bibfield  {journal} {\bibinfo
   {journal} {Phys. Rev. C}\ }\textbf {\bibinfo {volume} {106}},\ \bibinfo
  {pages} {014905} (\bibinfo {year} {2022})}\BibitemShut {NoStop}%
\bibitem [{\citenamefont {Liao}(2016)}]{LIAO201699}%
  \BibitemOpen
  \bibfield  {author} {\bibinfo {author} {\bibfnamefont {J.}~\bibnamefont
  {Liao}},\ }\href
  {https://doi.org/https://doi.org/10.1016/j.nuclphysa.2016.02.027} {\bibfield
  {journal} {\bibinfo  {journal} {Nuclear Physics A}\ }\textbf {\bibinfo
  {volume} {956}},\ \bibinfo {pages} {99} (\bibinfo {year} {2016})},\ \bibinfo
  {note} {the XXV International Conference on Ultrarelativistic Nucleus-Nucleus
  Collisions: Quark Matter 2015}\BibitemShut {NoStop}%
\bibitem [{\citenamefont {Adler}\ \emph {et~al.}(2001)\citenamefont {Adler}
  \emph {et~al.}}]{PhysRevLett.87.182301}%
  \BibitemOpen
  \bibfield  {author} {\bibinfo {author} {\bibfnamefont {C.}~\bibnamefont
  {Adler}} \emph {et~al.} (\bibinfo {collaboration} {STAR Collaboration}),\
  }\href {https://doi.org/10.1103/PhysRevLett.87.182301} {\bibfield  {journal}
  {\bibinfo  {journal} {Phys. Rev. Lett.}\ }\textbf {\bibinfo {volume} {87}},\
  \bibinfo {pages} {182301} (\bibinfo {year} {2001})}\BibitemShut {NoStop}%
\bibitem [{\citenamefont {Abelev}\ \emph {et~al.}(2008)\citenamefont {Abelev}
  \emph {et~al.}}]{PhysRevC.77.054901}%
  \BibitemOpen
  \bibfield  {author} {\bibinfo {author} {\bibfnamefont {B.~I.}\ \bibnamefont
  {Abelev}} \emph {et~al.} (\bibinfo {collaboration} {STAR Collaboration}),\
  }\href {https://doi.org/10.1103/PhysRevC.77.054901} {\bibfield  {journal}
  {\bibinfo  {journal} {Phys. Rev. C}\ }\textbf {\bibinfo {volume} {77}},\
  \bibinfo {pages} {054901} (\bibinfo {year} {2008})}\BibitemShut {NoStop}%
\bibitem [{\citenamefont {Del~Zanna}\ \emph {et~al.}(2013)\citenamefont
  {Del~Zanna}, \citenamefont {Joshi}, \citenamefont {Inghirami}, \citenamefont
  {Rolando}, \citenamefont {Beraudo}, \citenamefont {Pace}, \citenamefont
  {Pagliara}, \citenamefont {Drago},\ and\ \citenamefont
  {Becattini}}]{article}%
  \BibitemOpen
  \bibfield  {author} {\bibinfo {author} {\bibfnamefont {L.}~\bibnamefont
  {Del~Zanna}}, \bibinfo {author} {\bibfnamefont {V.~C.}\ \bibnamefont
  {Joshi}}, \bibinfo {author} {\bibfnamefont {G.}~\bibnamefont {Inghirami}},
  \bibinfo {author} {\bibfnamefont {V.}~\bibnamefont {Rolando}}, \bibinfo
  {author} {\bibfnamefont {A.}~\bibnamefont {Beraudo}}, \bibinfo {author}
  {\bibfnamefont {A.}~\bibnamefont {Pace}}, \bibinfo {author} {\bibfnamefont
  {G.}~\bibnamefont {Pagliara}}, \bibinfo {author} {\bibfnamefont
  {A.}~\bibnamefont {Drago}},\ and\ \bibinfo {author} {\bibfnamefont
  {F.}~\bibnamefont {Becattini}},\ }\href
  {https://doi.org/10.1140/epjc/s10052-013-2524-5} {\bibfield  {journal}
  {\bibinfo  {journal} {European Physical Journal C}\ }\textbf {\bibinfo
  {volume} {73}} (\bibinfo {year} {2013})}\BibitemShut {NoStop}%
\bibitem [{\citenamefont {Panda}\ \emph {et~al.}(2023)\citenamefont {Panda},
  \citenamefont {Gangadharan},\ and\ \citenamefont {Roy}}]{Panda_2023}%
  \BibitemOpen
  \bibfield  {author} {\bibinfo {author} {\bibfnamefont {A.~K.}\ \bibnamefont
  {Panda}}, \bibinfo {author} {\bibfnamefont {R.}~\bibnamefont {Gangadharan}},\
  and\ \bibinfo {author} {\bibfnamefont {V.}~\bibnamefont {Roy}},\ }\href
  {https://doi.org/10.1088/1361-6471/acd39b} {\bibfield  {journal} {\bibinfo
  {journal} {Journal of Physics G: Nuclear and Particle Physics}\ }\textbf
  {\bibinfo {volume} {50}},\ \bibinfo {pages} {075102} (\bibinfo {year}
  {2023})}\BibitemShut {NoStop}%
\bibitem [{\citenamefont {Khaidukov}\ and\ \citenamefont
  {Simonov}(2019)}]{PhysRevD.100.076009}%
  \BibitemOpen
  \bibfield  {author} {\bibinfo {author} {\bibfnamefont {Z.~V.}\ \bibnamefont
  {Khaidukov}}\ and\ \bibinfo {author} {\bibfnamefont {Y.~A.}\ \bibnamefont
  {Simonov}},\ }\href {https://doi.org/10.1103/PhysRevD.100.076009} {\bibfield
  {journal} {\bibinfo  {journal} {Phys. Rev. D}\ }\textbf {\bibinfo {volume}
  {100}},\ \bibinfo {pages} {076009} (\bibinfo {year} {2019})}\BibitemShut
  {NoStop}%
\bibitem [{\citenamefont {Koothottil}\ and\ \citenamefont
  {Bannur}(2019)}]{PhysRevC.99.035210}%
  \BibitemOpen
  \bibfield  {author} {\bibinfo {author} {\bibfnamefont {S.}~\bibnamefont
  {Koothottil}}\ and\ \bibinfo {author} {\bibfnamefont {V.~M.}\ \bibnamefont
  {Bannur}},\ }\href {https://doi.org/10.1103/PhysRevC.99.035210} {\bibfield
  {journal} {\bibinfo  {journal} {Phys. Rev. C}\ }\textbf {\bibinfo {volume}
  {99}},\ \bibinfo {pages} {035210} (\bibinfo {year} {2019})}\BibitemShut
  {NoStop}%
\bibitem [{\citenamefont {Goswami}\ \emph {et~al.}(2024)\citenamefont
  {Goswami}, \citenamefont {Sahu}, \citenamefont {Dey}, \citenamefont {Sahoo},\
  and\ \citenamefont {Stock}}]{Goswami:2023eol}%
  \BibitemOpen
  \bibfield  {author} {\bibinfo {author} {\bibfnamefont {K.}~\bibnamefont
  {Goswami}}, \bibinfo {author} {\bibfnamefont {D.}~\bibnamefont {Sahu}},
  \bibinfo {author} {\bibfnamefont {J.}~\bibnamefont {Dey}}, \bibinfo {author}
  {\bibfnamefont {R.}~\bibnamefont {Sahoo}},\ and\ \bibinfo {author}
  {\bibfnamefont {R.}~\bibnamefont {Stock}},\ }\href
  {https://doi.org/10.1103/PhysRevD.109.074012} {\bibfield  {journal} {\bibinfo
   {journal} {Phys. Rev. D}\ }\textbf {\bibinfo {volume} {109}},\ \bibinfo
  {pages} {074012} (\bibinfo {year} {2024})}\BibitemShut {NoStop}%
\bibitem [{\citenamefont {Gavin}(1985)}]{Gavin:1985ph}%
  \BibitemOpen
  \bibfield  {author} {\bibinfo {author} {\bibfnamefont {S.}~\bibnamefont
  {Gavin}},\ }\href {https://doi.org/10.1016/0375-9474(85)90190-3} {\bibfield
  {journal} {\bibinfo  {journal} {Nucl. Phys. A}\ }\textbf {\bibinfo {volume}
  {435}},\ \bibinfo {pages} {826} (\bibinfo {year} {1985})}\BibitemShut
  {NoStop}%
\bibitem [{\citenamefont {Das}\ \emph {et~al.}(2019)\citenamefont {Das},
  \citenamefont {Mishra},\ and\ \citenamefont
  {Mohapatra}}]{PhysRevD.100.114004}%
  \BibitemOpen
  \bibfield  {author} {\bibinfo {author} {\bibfnamefont {A.}~\bibnamefont
  {Das}}, \bibinfo {author} {\bibfnamefont {H.}~\bibnamefont {Mishra}},\ and\
  \bibinfo {author} {\bibfnamefont {R.~K.}\ \bibnamefont {Mohapatra}},\ }\href
  {https://doi.org/10.1103/PhysRevD.100.114004} {\bibfield  {journal} {\bibinfo
   {journal} {Phys. Rev. D}\ }\textbf {\bibinfo {volume} {100}},\ \bibinfo
  {pages} {114004} (\bibinfo {year} {2019})}\BibitemShut {NoStop}%
\bibitem [{\citenamefont {Dash}\ \emph {et~al.}(2020)\citenamefont {Dash},
  \citenamefont {Samanta}, \citenamefont {Dey}, \citenamefont {Gangopadhyaya},
  \citenamefont {Ghosh},\ and\ \citenamefont {Roy}}]{PhysRevD.102.016016}%
  \BibitemOpen
  \bibfield  {author} {\bibinfo {author} {\bibfnamefont {A.}~\bibnamefont
  {Dash}}, \bibinfo {author} {\bibfnamefont {S.}~\bibnamefont {Samanta}},
  \bibinfo {author} {\bibfnamefont {J.}~\bibnamefont {Dey}}, \bibinfo {author}
  {\bibfnamefont {U.}~\bibnamefont {Gangopadhyaya}}, \bibinfo {author}
  {\bibfnamefont {S.}~\bibnamefont {Ghosh}},\ and\ \bibinfo {author}
  {\bibfnamefont {V.}~\bibnamefont {Roy}},\ }\href
  {https://doi.org/10.1103/PhysRevD.102.016016} {\bibfield  {journal} {\bibinfo
   {journal} {Phys. Rev. D}\ }\textbf {\bibinfo {volume} {102}},\ \bibinfo
  {pages} {016016} (\bibinfo {year} {2020})}\BibitemShut {NoStop}%
\bibitem [{\citenamefont {Gowthama}\ \emph {et~al.}(2022)\citenamefont
  {Gowthama}, \citenamefont {Kurian},\ and\ \citenamefont
  {Chandra}}]{PhysRevD.106.034008}%
  \BibitemOpen
  \bibfield  {author} {\bibinfo {author} {\bibfnamefont {K.~K.}\ \bibnamefont
  {Gowthama}}, \bibinfo {author} {\bibfnamefont {M.}~\bibnamefont {Kurian}},\
  and\ \bibinfo {author} {\bibfnamefont {V.}~\bibnamefont {Chandra}},\ }\href
  {https://doi.org/10.1103/PhysRevD.106.034008} {\bibfield  {journal} {\bibinfo
   {journal} {Phys. Rev. D}\ }\textbf {\bibinfo {volume} {106}},\ \bibinfo
  {pages} {034008} (\bibinfo {year} {2022})}\BibitemShut {NoStop}%
\bibitem [{\citenamefont {Singh}\ \emph {et~al.}(2023)\citenamefont {Singh},
  \citenamefont {Dey}, \citenamefont {Sahoo},\ and\ \citenamefont
  {Ghosh}}]{PhysRevD.108.094007}%
  \BibitemOpen
  \bibfield  {author} {\bibinfo {author} {\bibfnamefont {K.}~\bibnamefont
  {Singh}}, \bibinfo {author} {\bibfnamefont {J.}~\bibnamefont {Dey}}, \bibinfo
  {author} {\bibfnamefont {R.}~\bibnamefont {Sahoo}},\ and\ \bibinfo {author}
  {\bibfnamefont {S.}~\bibnamefont {Ghosh}},\ }\href
  {https://doi.org/10.1103/PhysRevD.108.094007} {\bibfield  {journal} {\bibinfo
   {journal} {Phys. Rev. D}\ }\textbf {\bibinfo {volume} {108}},\ \bibinfo
  {pages} {094007} (\bibinfo {year} {2023})}\BibitemShut {NoStop}%
\bibitem [{\citenamefont {Tuchin}(2016)}]{Tuchin:2015oka}%
  \BibitemOpen
  \bibfield  {author} {\bibinfo {author} {\bibfnamefont {K.}~\bibnamefont
  {Tuchin}},\ }\href {https://doi.org/10.1103/PhysRevC.93.014905} {\bibfield
  {journal} {\bibinfo  {journal} {Phys. Rev. C}\ }\textbf {\bibinfo {volume}
  {93}},\ \bibinfo {pages} {014905} (\bibinfo {year} {2016})}\BibitemShut
  {NoStop}%
\bibitem [{\citenamefont {Satow}(2014{\natexlab{a}})}]{PhysRevD.90.034018}%
  \BibitemOpen
  \bibfield  {author} {\bibinfo {author} {\bibfnamefont {D.}~\bibnamefont
  {Satow}},\ }\href {https://doi.org/10.1103/PhysRevD.90.034018} {\bibfield
  {journal} {\bibinfo  {journal} {Phys. Rev. D}\ }\textbf {\bibinfo {volume}
  {90}},\ \bibinfo {pages} {034018} (\bibinfo {year}
  {2014}{\natexlab{a}})}\BibitemShut {NoStop}%
\bibitem [{\citenamefont {Acharya}\ \emph {et~al.}(2020)\citenamefont {Acharya}
  \emph {et~al.}}]{PhysRevLett.125.022301}%
  \BibitemOpen
  \bibfield  {author} {\bibinfo {author} {\bibfnamefont {S.}~\bibnamefont
  {Acharya}} \emph {et~al.} (\bibinfo {collaboration} {ALICE Collaboration}),\
  }\href {https://doi.org/10.1103/PhysRevLett.125.022301} {\bibfield  {journal}
  {\bibinfo  {journal} {Phys. Rev. Lett.}\ }\textbf {\bibinfo {volume} {125}},\
  \bibinfo {pages} {022301} (\bibinfo {year} {2020})}\BibitemShut {NoStop}%
\bibitem [{\citenamefont {Kharzeev}\ \emph {et~al.}(2008)\citenamefont
  {Kharzeev}, \citenamefont {McLerran},\ and\ \citenamefont
  {Warringa}}]{KHARZEEV2008227}%
  \BibitemOpen
  \bibfield  {author} {\bibinfo {author} {\bibfnamefont {D.~E.}\ \bibnamefont
  {Kharzeev}}, \bibinfo {author} {\bibfnamefont {L.~D.}\ \bibnamefont
  {McLerran}},\ and\ \bibinfo {author} {\bibfnamefont {H.~J.}\ \bibnamefont
  {Warringa}},\ }\href
  {https://doi.org/https://doi.org/10.1016/j.nuclphysa.2008.02.298} {\bibfield
  {journal} {\bibinfo  {journal} {Nuclear Physics A}\ }\textbf {\bibinfo
  {volume} {803}},\ \bibinfo {pages} {227} (\bibinfo {year}
  {2008})}\BibitemShut {NoStop}%
\bibitem [{\citenamefont {Abdulhamid}\ \emph {et~al.}(2024)\citenamefont
  {Abdulhamid} \emph {et~al.}}]{PhysRevX.14.011028}%
  \BibitemOpen
  \bibfield  {author} {\bibinfo {author} {\bibfnamefont {M.~I.}\ \bibnamefont
  {Abdulhamid}} \emph {et~al.} (\bibinfo {collaboration} {STAR
  Collaboration}),\ }\href {https://doi.org/10.1103/PhysRevX.14.011028}
  {\bibfield  {journal} {\bibinfo  {journal} {Phys. Rev. X}\ }\textbf {\bibinfo
  {volume} {14}},\ \bibinfo {pages} {011028} (\bibinfo {year}
  {2024})}\BibitemShut {NoStop}%
\bibitem [{\citenamefont {Kumar}\ \emph {et~al.}(2024)\citenamefont {Kumar},
  \citenamefont {Jain}, \citenamefont {Bangotra}, \citenamefont {Kumar},
  \citenamefont {Singh},\ and\ \citenamefont {Rajouria}}]{Kumar2024}%
  \BibitemOpen
  \bibfield  {author} {\bibinfo {author} {\bibfnamefont {Y.}~\bibnamefont
  {Kumar}}, \bibinfo {author} {\bibfnamefont {P.}~\bibnamefont {Jain}},
  \bibinfo {author} {\bibfnamefont {P.}~\bibnamefont {Bangotra}}, \bibinfo
  {author} {\bibfnamefont {V.}~\bibnamefont {Kumar}}, \bibinfo {author}
  {\bibfnamefont {D.~V.}\ \bibnamefont {Singh}},\ and\ \bibinfo {author}
  {\bibfnamefont {S.~K.}\ \bibnamefont {Rajouria}},\ }\href
  {https://doi.org/10.1155/2024/1870528} {\bibfield  {journal} {\bibinfo
  {journal} {Advances in High Energy Physics}\ }\textbf {\bibinfo {volume}
  {2024}},\ \bibinfo {pages} {1870528} (\bibinfo {year} {2024})}\BibitemShut
  {NoStop}%
\bibitem [{\citenamefont {Brandt}\ \emph {et~al.}(2023)\citenamefont {Brandt},
  \citenamefont {Cuteri},\ and\ \citenamefont {Endr{\H{o}}di}}]{Brandt2023}%
  \BibitemOpen
  \bibfield  {author} {\bibinfo {author} {\bibfnamefont {B.~B.}\ \bibnamefont
  {Brandt}}, \bibinfo {author} {\bibfnamefont {F.}~\bibnamefont {Cuteri}},\
  and\ \bibinfo {author} {\bibfnamefont {G.}~\bibnamefont {Endr{\H{o}}di}},\
  }\href {https://doi.org/10.1007/JHEP07(2023)055} {\bibfield  {journal}
  {\bibinfo  {journal} {Journal of High Energy Physics}\ }\textbf {\bibinfo
  {volume} {2023}},\ \bibinfo {pages} {55} (\bibinfo {year}
  {2023})}\BibitemShut {NoStop}%
\bibitem [{\citenamefont {Bors{\'a}nyi}\ \emph {et~al.}(2012)\citenamefont
  {Bors{\'a}nyi}, \citenamefont {Endr{\H{o}}di}, \citenamefont {Fodor},
  \citenamefont {Katz}, \citenamefont {Krieg}, \citenamefont {Ratti},\ and\
  \citenamefont {Szab{\'o}}}]{Borsanyi2012}%
  \BibitemOpen
  \bibfield  {author} {\bibinfo {author} {\bibfnamefont {S.}~\bibnamefont
  {Bors{\'a}nyi}}, \bibinfo {author} {\bibfnamefont {G.}~\bibnamefont
  {Endr{\H{o}}di}}, \bibinfo {author} {\bibfnamefont {Z.}~\bibnamefont
  {Fodor}}, \bibinfo {author} {\bibfnamefont {S.~D.}\ \bibnamefont {Katz}},
  \bibinfo {author} {\bibfnamefont {S.}~\bibnamefont {Krieg}}, \bibinfo
  {author} {\bibfnamefont {C.}~\bibnamefont {Ratti}},\ and\ \bibinfo {author}
  {\bibfnamefont {K.~K.}\ \bibnamefont {Szab{\'o}}},\ }\href
  {https://doi.org/10.1007/JHEP08(2012)053} {\bibfield  {journal} {\bibinfo
  {journal} {Journal of High Energy Physics}\ }\textbf {\bibinfo {volume}
  {2012}},\ \bibinfo {pages} {53} (\bibinfo {year} {2012})}\BibitemShut
  {NoStop}%
\bibitem [{\citenamefont {Bali}\ \emph {et~al.}(2014)\citenamefont {Bali},
  \citenamefont {Bruckmann}, \citenamefont {Endr{\H{o}}di}, \citenamefont
  {Katz},\ and\ \citenamefont {Sch{\"a}fer}}]{Bali2014}%
  \BibitemOpen
  \bibfield  {author} {\bibinfo {author} {\bibfnamefont {G.~S.}\ \bibnamefont
  {Bali}}, \bibinfo {author} {\bibfnamefont {F.}~\bibnamefont {Bruckmann}},
  \bibinfo {author} {\bibfnamefont {G.}~\bibnamefont {Endr{\H{o}}di}}, \bibinfo
  {author} {\bibfnamefont {S.~D.}\ \bibnamefont {Katz}},\ and\ \bibinfo
  {author} {\bibfnamefont {A.}~\bibnamefont {Sch{\"a}fer}},\ }\href
  {https://doi.org/10.1007/JHEP08(2014)177} {\bibfield  {journal} {\bibinfo
  {journal} {Journal of High Energy Physics}\ }\textbf {\bibinfo {volume}
  {2014}},\ \bibinfo {pages} {177} (\bibinfo {year} {2014})}\BibitemShut
  {NoStop}%
\bibitem [{\citenamefont {Li}\ and\ \citenamefont
  {Yee}(2018)}]{PhysRevD.97.056024}%
  \BibitemOpen
  \bibfield  {author} {\bibinfo {author} {\bibfnamefont {S.}~\bibnamefont
  {Li}}\ and\ \bibinfo {author} {\bibfnamefont {H.-U.}\ \bibnamefont {Yee}},\
  }\href {https://doi.org/10.1103/PhysRevD.97.056024} {\bibfield  {journal}
  {\bibinfo  {journal} {Phys. Rev. D}\ }\textbf {\bibinfo {volume} {97}},\
  \bibinfo {pages} {056024} (\bibinfo {year} {2018})}\BibitemShut {NoStop}%
\bibitem [{\citenamefont {Gusynin}\ \emph {et~al.}(1994)\citenamefont
  {Gusynin}, \citenamefont {Miransky},\ and\ \citenamefont
  {Shovkovy}}]{PhysRevLett.73.3499}%
  \BibitemOpen
  \bibfield  {author} {\bibinfo {author} {\bibfnamefont {V.~P.}\ \bibnamefont
  {Gusynin}}, \bibinfo {author} {\bibfnamefont {V.~A.}\ \bibnamefont
  {Miransky}},\ and\ \bibinfo {author} {\bibfnamefont {I.~A.}\ \bibnamefont
  {Shovkovy}},\ }\href {https://doi.org/10.1103/PhysRevLett.73.3499} {\bibfield
   {journal} {\bibinfo  {journal} {Phys. Rev. Lett.}\ }\textbf {\bibinfo
  {volume} {73}},\ \bibinfo {pages} {3499} (\bibinfo {year}
  {1994})}\BibitemShut {NoStop}%
\bibitem [{\citenamefont {Lee}\ \emph {et~al.}(1997)\citenamefont {Lee},
  \citenamefont {Leung},\ and\ \citenamefont {Ng}}]{PhysRevD.55.6504}%
  \BibitemOpen
  \bibfield  {author} {\bibinfo {author} {\bibfnamefont {D.-S.}\ \bibnamefont
  {Lee}}, \bibinfo {author} {\bibfnamefont {C.~N.}\ \bibnamefont {Leung}},\
  and\ \bibinfo {author} {\bibfnamefont {Y.~J.}\ \bibnamefont {Ng}},\ }\href
  {https://doi.org/10.1103/PhysRevD.55.6504} {\bibfield  {journal} {\bibinfo
  {journal} {Phys. Rev. D}\ }\textbf {\bibinfo {volume} {55}},\ \bibinfo
  {pages} {6504} (\bibinfo {year} {1997})}\BibitemShut {NoStop}%
\bibitem [{\citenamefont {Avancini}\ \emph {et~al.}(2012)\citenamefont
  {Avancini}, \citenamefont {Menezes}, \citenamefont {Pinto},\ and\
  \citenamefont {Provid\^encia}}]{PhysRevD.85.091901}%
  \BibitemOpen
  \bibfield  {author} {\bibinfo {author} {\bibfnamefont {S.~S.}\ \bibnamefont
  {Avancini}}, \bibinfo {author} {\bibfnamefont {D.~P.}\ \bibnamefont
  {Menezes}}, \bibinfo {author} {\bibfnamefont {M.~B.}\ \bibnamefont {Pinto}},\
  and\ \bibinfo {author} {\bibfnamefont {C.~m.~c.}\ \bibnamefont
  {Provid\^encia}},\ }\href {https://doi.org/10.1103/PhysRevD.85.091901}
  {\bibfield  {journal} {\bibinfo  {journal} {Phys. Rev. D}\ }\textbf {\bibinfo
  {volume} {85}},\ \bibinfo {pages} {091901} (\bibinfo {year}
  {2012})}\BibitemShut {NoStop}%
\bibitem [{\citenamefont {Endr{\H{o}}di}\ \emph {et~al.}(2011)\citenamefont
  {Endr{\H{o}}di}, \citenamefont {Fodor}, \citenamefont {Katz}, \citenamefont
  {{Szab{\'o}}},\ and\ \citenamefont {K.}}]{Endrodi2011}%
  \BibitemOpen
  \bibfield  {author} {\bibinfo {author} {\bibfnamefont {G.}~\bibnamefont
  {Endr{\H{o}}di}}, \bibinfo {author} {\bibfnamefont {Z.}~\bibnamefont
  {Fodor}}, \bibinfo {author} {\bibfnamefont {S.~D.}\ \bibnamefont {Katz}},
  \bibinfo {author} {\bibnamefont {{Szab{\'o}}}},\ and\ \bibinfo {author}
  {\bibfnamefont {K.}~\bibnamefont {K.}},\ }\href
  {https://doi.org/10.1007/JHEP04(2011)001} {\bibfield  {journal} {\bibinfo
  {journal} {Journal of High Energy Physics}\ }\textbf {\bibinfo {volume}
  {2011}},\ \bibinfo {pages} {1} (\bibinfo {year} {2011})}\BibitemShut
  {NoStop}%
\bibitem [{\citenamefont {Wang}\ and\ \citenamefont
  {Shovkovy}(2021)}]{Wang:2021eud}%
  \BibitemOpen
  \bibfield  {author} {\bibinfo {author} {\bibfnamefont {X.}~\bibnamefont
  {Wang}}\ and\ \bibinfo {author} {\bibfnamefont {I.}~\bibnamefont
  {Shovkovy}},\ }\href {https://doi.org/10.1140/epjc/s10052-021-09650-3}
  {\bibfield  {journal} {\bibinfo  {journal} {Eur. Phys. J. C}\ }\textbf
  {\bibinfo {volume} {81}},\ \bibinfo {pages} {901} (\bibinfo {year}
  {2021})}\BibitemShut {NoStop}%
\bibitem [{\citenamefont {McInnes}(2016)}]{MCINNES2016173}%
  \BibitemOpen
  \bibfield  {author} {\bibinfo {author} {\bibfnamefont {B.}~\bibnamefont
  {McInnes}},\ }\href
  {https://doi.org/https://doi.org/10.1016/j.nuclphysb.2016.08.001} {\bibfield
  {journal} {\bibinfo  {journal} {Nuclear Physics B}\ }\textbf {\bibinfo
  {volume} {911}},\ \bibinfo {pages} {173} (\bibinfo {year}
  {2016})}\BibitemShut {NoStop}%
\bibitem [{\citenamefont {Simonov}(2021)}]{Simonov:2021eyt}%
  \BibitemOpen
  \bibfield  {author} {\bibinfo {author} {\bibfnamefont {Y.~A.}\ \bibnamefont
  {Simonov}},\ }\href {https://doi.org/10.1134/S1063778821130342} {\bibfield
  {journal} {\bibinfo  {journal} {Phys. Atom. Nucl.}\ }\textbf {\bibinfo
  {volume} {84}},\ \bibinfo {pages} {1195} (\bibinfo {year}
  {2021})}\BibitemShut {NoStop}%
\bibitem [{\citenamefont {Dey}\ \emph {et~al.}(2023)\citenamefont {Dey},
  \citenamefont {Bandyopadhyay}, \citenamefont {Gupta}, \citenamefont
  {Pujari},\ and\ \citenamefont {Ghosh}}]{Dey:2021fbo}%
  \BibitemOpen
  \bibfield  {author} {\bibinfo {author} {\bibfnamefont {J.}~\bibnamefont
  {Dey}}, \bibinfo {author} {\bibfnamefont {A.}~\bibnamefont {Bandyopadhyay}},
  \bibinfo {author} {\bibfnamefont {A.}~\bibnamefont {Gupta}}, \bibinfo
  {author} {\bibfnamefont {N.}~\bibnamefont {Pujari}},\ and\ \bibinfo {author}
  {\bibfnamefont {S.}~\bibnamefont {Ghosh}},\ }\href
  {https://doi.org/10.1016/j.nuclphysa.2023.122654} {\bibfield  {journal}
  {\bibinfo  {journal} {Nucl. Phys. A}\ }\textbf {\bibinfo {volume} {1034}},\
  \bibinfo {pages} {122654} (\bibinfo {year} {2023})}\BibitemShut {NoStop}%
\bibitem [{\citenamefont {Dash}\ and\ \citenamefont
  {Panda}(2024)}]{DASH2024138342}%
  \BibitemOpen
  \bibfield  {author} {\bibinfo {author} {\bibfnamefont {A.}~\bibnamefont
  {Dash}}\ and\ \bibinfo {author} {\bibfnamefont {A.~K.}\ \bibnamefont
  {Panda}},\ }\href
  {https://doi.org/https://doi.org/10.1016/j.physletb.2023.138342} {\bibfield
  {journal} {\bibinfo  {journal} {Physics Letters B}\ }\textbf {\bibinfo
  {volume} {848}},\ \bibinfo {pages} {138342} (\bibinfo {year}
  {2024})}\BibitemShut {NoStop}%
\bibitem [{\citenamefont {Goldsmid}\ \emph {et~al.}()\citenamefont {Goldsmid}
  \emph {et~al.}}]{goldsmid2010introduction}%
  \BibitemOpen
  \bibfield  {author} {\bibinfo {author} {\bibfnamefont {H.~J.}\ \bibnamefont
  {Goldsmid}} \emph {et~al.},\ }\href
  {https://doi.org/https://doi.org/10.1007/978-3-662-49256-7} {\emph {\bibinfo
  {title} {Introduction to thermoelectricity}}},\ Vol.\ \bibinfo {volume}
  {121}\ (\bibinfo  {publisher} {Springer})\BibitemShut {NoStop}%
\bibitem [{\citenamefont {Das}\ and\ \citenamefont {Mishra}(2021)}]{Das2021}%
  \BibitemOpen
  \bibfield  {author} {\bibinfo {author} {\bibfnamefont {A.}~\bibnamefont
  {Das}}\ and\ \bibinfo {author} {\bibfnamefont {H.}~\bibnamefont {Mishra}},\
  }\href {https://doi.org/10.1140/epjs/s11734-021-00022-2} {\bibfield
  {journal} {\bibinfo  {journal} {The European Physical Journal Special
  Topics}\ }\textbf {\bibinfo {volume} {230}},\ \bibinfo {pages} {607}
  (\bibinfo {year} {2021})}\BibitemShut {NoStop}%
\bibitem [{\citenamefont {Dey}\ and\ \citenamefont
  {Patra}(2020)}]{PhysRevD.102.096011}%
  \BibitemOpen
  \bibfield  {author} {\bibinfo {author} {\bibfnamefont {D.}~\bibnamefont
  {Dey}}\ and\ \bibinfo {author} {\bibfnamefont {B.~K.}\ \bibnamefont
  {Patra}},\ }\href {https://doi.org/10.1103/PhysRevD.102.096011} {\bibfield
  {journal} {\bibinfo  {journal} {Phys. Rev. D}\ }\textbf {\bibinfo {volume}
  {102}},\ \bibinfo {pages} {096011} (\bibinfo {year} {2020})}\BibitemShut
  {NoStop}%
\bibitem [{\citenamefont {Das}\ \emph {et~al.}(2020)\citenamefont {Das},
  \citenamefont {Mishra},\ and\ \citenamefont
  {Mohapatra}}]{PhysRevD.102.014030}%
  \BibitemOpen
  \bibfield  {author} {\bibinfo {author} {\bibfnamefont {A.}~\bibnamefont
  {Das}}, \bibinfo {author} {\bibfnamefont {H.}~\bibnamefont {Mishra}},\ and\
  \bibinfo {author} {\bibfnamefont {R.~K.}\ \bibnamefont {Mohapatra}},\ }\href
  {https://doi.org/10.1103/PhysRevD.102.014030} {\bibfield  {journal} {\bibinfo
   {journal} {Phys. Rev. D}\ }\textbf {\bibinfo {volume} {102}},\ \bibinfo
  {pages} {014030} (\bibinfo {year} {2020})}\BibitemShut {NoStop}%
\bibitem [{\citenamefont {Bhatt}\ \emph {et~al.}(2019)\citenamefont {Bhatt},
  \citenamefont {Das},\ and\ \citenamefont {Mishra}}]{PhysRevD.99.014015}%
  \BibitemOpen
  \bibfield  {author} {\bibinfo {author} {\bibfnamefont {J.~R.}\ \bibnamefont
  {Bhatt}}, \bibinfo {author} {\bibfnamefont {A.}~\bibnamefont {Das}},\ and\
  \bibinfo {author} {\bibfnamefont {H.}~\bibnamefont {Mishra}},\ }\href
  {https://doi.org/10.1103/PhysRevD.99.014015} {\bibfield  {journal} {\bibinfo
  {journal} {Phys. Rev. D}\ }\textbf {\bibinfo {volume} {99}},\ \bibinfo
  {pages} {014015} (\bibinfo {year} {2019})}\BibitemShut {NoStop}%
\bibitem [{\citenamefont {Singh}\ \emph {et~al.}(2024)\citenamefont {Singh},
  \citenamefont {Dey},\ and\ \citenamefont {Sahoo}}]{PhysRevD.109.014018}%
  \BibitemOpen
  \bibfield  {author} {\bibinfo {author} {\bibfnamefont {K.}~\bibnamefont
  {Singh}}, \bibinfo {author} {\bibfnamefont {J.}~\bibnamefont {Dey}},\ and\
  \bibinfo {author} {\bibfnamefont {R.}~\bibnamefont {Sahoo}},\ }\href
  {https://doi.org/10.1103/PhysRevD.109.014018} {\bibfield  {journal} {\bibinfo
   {journal} {Phys. Rev. D}\ }\textbf {\bibinfo {volume} {109}},\ \bibinfo
  {pages} {014018} (\bibinfo {year} {2024})}\BibitemShut {NoStop}%
\bibitem [{\citenamefont {Rischke}\ \emph {et~al.}(1995)\citenamefont
  {Rischke}, \citenamefont {Pursun},\ and\ \citenamefont
  {Maruhn}}]{Rischke:1995mt}%
  \BibitemOpen
  \bibfield  {author} {\bibinfo {author} {\bibfnamefont {D.~H.}\ \bibnamefont
  {Rischke}}, \bibinfo {author} {\bibfnamefont {Y.}~\bibnamefont {Pursun}},\
  and\ \bibinfo {author} {\bibfnamefont {J.~A.}\ \bibnamefont {Maruhn}},\
  }\href {https://doi.org/10.1016/0375-9474(95)00356-3} {\bibfield  {journal}
  {\bibinfo  {journal} {Nucl. Phys. A}\ }\textbf {\bibinfo {volume} {595}},\
  \bibinfo {pages} {383} (\bibinfo {year} {1995})},\ \bibinfo {note} {[Erratum:
  Nucl.Phys.A 596, 717--717 (1996)]}\BibitemShut {NoStop}%
\bibitem [{\citenamefont {Okamoto}\ \emph {et~al.}(2016)\citenamefont
  {Okamoto}, \citenamefont {Akamatsu},\ and\ \citenamefont
  {Nonaka}}]{Okamoto2016}%
  \BibitemOpen
  \bibfield  {author} {\bibinfo {author} {\bibfnamefont {K.}~\bibnamefont
  {Okamoto}}, \bibinfo {author} {\bibfnamefont {Y.}~\bibnamefont {Akamatsu}},\
  and\ \bibinfo {author} {\bibfnamefont {C.}~\bibnamefont {Nonaka}},\ }\href
  {https://doi.org/10.1140/epjc/s10052-016-4433-x} {\bibfield  {journal}
  {\bibinfo  {journal} {The European Physical Journal C}\ }\textbf {\bibinfo
  {volume} {76}},\ \bibinfo {pages} {579} (\bibinfo {year} {2016})}\BibitemShut
  {NoStop}%
\bibitem [{\citenamefont {Bjorken}(1983)}]{PhysRevD.27.140}%
  \BibitemOpen
  \bibfield  {author} {\bibinfo {author} {\bibfnamefont {J.~D.}\ \bibnamefont
  {Bjorken}},\ }\href {https://doi.org/10.1103/PhysRevD.27.140} {\bibfield
  {journal} {\bibinfo  {journal} {Phys. Rev. D}\ }\textbf {\bibinfo {volume}
  {27}},\ \bibinfo {pages} {140} (\bibinfo {year} {1983})}\BibitemShut
  {NoStop}%
\bibitem [{\citenamefont {Gubser}(2010)}]{PhysRevD.82.085027}%
  \BibitemOpen
  \bibfield  {author} {\bibinfo {author} {\bibfnamefont {S.~S.}\ \bibnamefont
  {Gubser}},\ }\href {https://doi.org/10.1103/PhysRevD.82.085027} {\bibfield
  {journal} {\bibinfo  {journal} {Phys. Rev. D}\ }\textbf {\bibinfo {volume}
  {82}},\ \bibinfo {pages} {085027} (\bibinfo {year} {2010})}\BibitemShut
  {NoStop}%
\bibitem [{\citenamefont {Mitra}(2021)}]{PhysRevC.103.014905}%
  \BibitemOpen
  \bibfield  {author} {\bibinfo {author} {\bibfnamefont {S.}~\bibnamefont
  {Mitra}},\ }\href {https://doi.org/10.1103/PhysRevC.103.014905} {\bibfield
  {journal} {\bibinfo  {journal} {Phys. Rev. C}\ }\textbf {\bibinfo {volume}
  {103}},\ \bibinfo {pages} {014905} (\bibinfo {year} {2021})}\BibitemShut
  {NoStop}%
\bibitem [{\citenamefont {Romatschke}\ and\ \citenamefont
  {Romatschke}(2007)}]{PhysRevLett.99.172301}%
  \BibitemOpen
  \bibfield  {author} {\bibinfo {author} {\bibfnamefont {P.}~\bibnamefont
  {Romatschke}}\ and\ \bibinfo {author} {\bibfnamefont {U.}~\bibnamefont
  {Romatschke}},\ }\href {https://doi.org/10.1103/PhysRevLett.99.172301}
  {\bibfield  {journal} {\bibinfo  {journal} {Phys. Rev. Lett.}\ }\textbf
  {\bibinfo {volume} {99}},\ \bibinfo {pages} {172301} (\bibinfo {year}
  {2007})}\BibitemShut {NoStop}%
\bibitem [{\citenamefont {Roy}\ \emph {et~al.}(2015)\citenamefont {Roy},
  \citenamefont {Pu}, \citenamefont {Rezzolla},\ and\ \citenamefont
  {Rischke}}]{ROY201545}%
  \BibitemOpen
  \bibfield  {author} {\bibinfo {author} {\bibfnamefont {V.}~\bibnamefont
  {Roy}}, \bibinfo {author} {\bibfnamefont {S.}~\bibnamefont {Pu}}, \bibinfo
  {author} {\bibfnamefont {L.}~\bibnamefont {Rezzolla}},\ and\ \bibinfo
  {author} {\bibfnamefont {D.}~\bibnamefont {Rischke}},\ }\href
  {https://doi.org/https://doi.org/10.1016/j.physletb.2015.08.046} {\bibfield
  {journal} {\bibinfo  {journal} {Physics Letters B}\ }\textbf {\bibinfo
  {volume} {750}},\ \bibinfo {pages} {45} (\bibinfo {year} {2015})}\BibitemShut
  {NoStop}%
\bibitem [{\citenamefont {Panda}\ \emph {et~al.}(2021)\citenamefont {Panda},
  \citenamefont {Dash}, \citenamefont {Biswas},\ and\ \citenamefont
  {Roy}}]{Panda2021}%
  \BibitemOpen
  \bibfield  {author} {\bibinfo {author} {\bibfnamefont {A.~K.}\ \bibnamefont
  {Panda}}, \bibinfo {author} {\bibfnamefont {A.}~\bibnamefont {Dash}},
  \bibinfo {author} {\bibfnamefont {R.}~\bibnamefont {Biswas}},\ and\ \bibinfo
  {author} {\bibfnamefont {V.}~\bibnamefont {Roy}},\ }\href
  {https://doi.org/10.1007/JHEP03(2021)216} {\bibfield  {journal} {\bibinfo
  {journal} {Journal of High Energy Physics}\ }\textbf {\bibinfo {volume}
  {2021}},\ \bibinfo {pages} {216} (\bibinfo {year} {2021})}\BibitemShut
  {NoStop}%
\bibitem [{\citenamefont {Satow}(2014{\natexlab{b}})}]{Satow:2014lia}%
  \BibitemOpen
  \bibfield  {author} {\bibinfo {author} {\bibfnamefont {D.}~\bibnamefont
  {Satow}},\ }\href {https://doi.org/10.1103/PhysRevD.90.034018} {\bibfield
  {journal} {\bibinfo  {journal} {Phys. Rev. D}\ }\textbf {\bibinfo {volume}
  {90}},\ \bibinfo {pages} {034018} (\bibinfo {year}
  {2014}{\natexlab{b}})}\BibitemShut {NoStop}%
\bibitem [{\citenamefont {Stewart}\ and\ \citenamefont
  {Tuchin}(2018)}]{PhysRevC.97.044906}%
  \BibitemOpen
  \bibfield  {author} {\bibinfo {author} {\bibfnamefont {E.}~\bibnamefont
  {Stewart}}\ and\ \bibinfo {author} {\bibfnamefont {K.}~\bibnamefont
  {Tuchin}},\ }\href {https://doi.org/10.1103/PhysRevC.97.044906} {\bibfield
  {journal} {\bibinfo  {journal} {Phys. Rev. C}\ }\textbf {\bibinfo {volume}
  {97}},\ \bibinfo {pages} {044906} (\bibinfo {year} {2018})}\BibitemShut
  {NoStop}%
\bibitem [{\citenamefont {Tuchin}(2013)}]{Tuchin2013ie}%
  \BibitemOpen
  \bibfield  {author} {\bibinfo {author} {\bibfnamefont {K.}~\bibnamefont
  {Tuchin}},\ }\href {https://doi.org/10.1155/2013/490495} {\bibfield
  {journal} {\bibinfo  {journal} {Advances in High Energy Physics}\ }\textbf
  {\bibinfo {volume} {2013}},\ \bibinfo {pages} {490495} (\bibinfo {year}
  {2013})}\BibitemShut {NoStop}%
\bibitem [{\citenamefont {Huang}(2016)}]{Huang_2016}%
  \BibitemOpen
  \bibfield  {author} {\bibinfo {author} {\bibfnamefont {X.-G.}\ \bibnamefont
  {Huang}},\ }\href {https://doi.org/10.1088/0034-4885/79/7/076302} {\bibfield
  {journal} {\bibinfo  {journal} {Reports on Progress in Physics}\ }\textbf
  {\bibinfo {volume} {79}},\ \bibinfo {pages} {076302} (\bibinfo {year}
  {2016})}\BibitemShut {NoStop}%
\bibitem [{\citenamefont {Hongo}\ \emph {et~al.}(2017)\citenamefont {Hongo},
  \citenamefont {Hirono},\ and\ \citenamefont {Hirano}}]{Hongo:2013cqa}%
  \BibitemOpen
  \bibfield  {author} {\bibinfo {author} {\bibfnamefont {M.}~\bibnamefont
  {Hongo}}, \bibinfo {author} {\bibfnamefont {Y.}~\bibnamefont {Hirono}},\ and\
  \bibinfo {author} {\bibfnamefont {T.}~\bibnamefont {Hirano}},\ }\href
  {https://doi.org/10.1016/j.physletb.2017.10.028} {\bibfield  {journal}
  {\bibinfo  {journal} {Phys. Lett. B}\ }\textbf {\bibinfo {volume} {775}},\
  \bibinfo {pages} {266} (\bibinfo {year} {2017})}\BibitemShut {NoStop}%
\bibitem [{\citenamefont {Abhishek}\ \emph {et~al.}(2022)\citenamefont
  {Abhishek}, \citenamefont {Das}, \citenamefont {Kumar},\ and\ \citenamefont
  {Mishra}}]{Abhishek:2020wjm}%
  \BibitemOpen
  \bibfield  {author} {\bibinfo {author} {\bibfnamefont {A.}~\bibnamefont
  {Abhishek}}, \bibinfo {author} {\bibfnamefont {A.}~\bibnamefont {Das}},
  \bibinfo {author} {\bibfnamefont {D.}~\bibnamefont {Kumar}},\ and\ \bibinfo
  {author} {\bibfnamefont {H.}~\bibnamefont {Mishra}},\ }\href
  {https://doi.org/10.1140/epjc/s10052-022-09999-z} {\bibfield  {journal}
  {\bibinfo  {journal} {Eur. Phys. J. C}\ }\textbf {\bibinfo {volume} {82}},\
  \bibinfo {pages} {71} (\bibinfo {year} {2022})}\BibitemShut {NoStop}%
\bibitem [{\citenamefont {Kurian}(2021)}]{Kurian:2021zyb}%
  \BibitemOpen
  \bibfield  {author} {\bibinfo {author} {\bibfnamefont {M.}~\bibnamefont
  {Kurian}},\ }\href {https://doi.org/10.1103/PhysRevD.103.054024} {\bibfield
  {journal} {\bibinfo  {journal} {Phys. Rev. D}\ }\textbf {\bibinfo {volume}
  {103}},\ \bibinfo {pages} {054024} (\bibinfo {year} {2021})}\BibitemShut
  {NoStop}%
\bibitem [{\citenamefont {Gowthama}\ \emph {et~al.}(2021)\citenamefont
  {Gowthama}, \citenamefont {Kurian},\ and\ \citenamefont
  {Chandra}}]{PhysRevD.104.094037}%
  \BibitemOpen
  \bibfield  {author} {\bibinfo {author} {\bibfnamefont {K.~K.}\ \bibnamefont
  {Gowthama}}, \bibinfo {author} {\bibfnamefont {M.}~\bibnamefont {Kurian}},\
  and\ \bibinfo {author} {\bibfnamefont {V.}~\bibnamefont {Chandra}},\ }\href
  {https://doi.org/10.1103/PhysRevD.104.094037} {\bibfield  {journal} {\bibinfo
   {journal} {Phys. Rev. D}\ }\textbf {\bibinfo {volume} {104}},\ \bibinfo
  {pages} {094037} (\bibinfo {year} {2021})}\BibitemShut {NoStop}%
\bibitem [{\citenamefont {Gorenstein}\ and\ \citenamefont
  {Yang}(1995)}]{Gorenstein:1995vm}%
  \BibitemOpen
  \bibfield  {author} {\bibinfo {author} {\bibfnamefont {M.~I.}\ \bibnamefont
  {Gorenstein}}\ and\ \bibinfo {author} {\bibfnamefont {S.-N.}\ \bibnamefont
  {Yang}},\ }\href {https://doi.org/10.1103/PhysRevD.52.5206} {\bibfield
  {journal} {\bibinfo  {journal} {Phys. Rev. D}\ }\textbf {\bibinfo {volume}
  {52}},\ \bibinfo {pages} {5206} (\bibinfo {year} {1995})}\BibitemShut
  {NoStop}%
\bibitem [{\citenamefont {Srivastava}\ \emph {et~al.}(2010)\citenamefont
  {Srivastava}, \citenamefont {Tiwari},\ and\ \citenamefont
  {Singh}}]{Srivastava:2010xa}%
  \BibitemOpen
  \bibfield  {author} {\bibinfo {author} {\bibfnamefont {P.~K.}\ \bibnamefont
  {Srivastava}}, \bibinfo {author} {\bibfnamefont {S.~K.}\ \bibnamefont
  {Tiwari}},\ and\ \bibinfo {author} {\bibfnamefont {C.~P.}\ \bibnamefont
  {Singh}},\ }\href {https://doi.org/10.1103/PhysRevD.82.014023} {\bibfield
  {journal} {\bibinfo  {journal} {Phys. Rev. D}\ }\textbf {\bibinfo {volume}
  {82}},\ \bibinfo {pages} {014023} (\bibinfo {year} {2010})}\BibitemShut
  {NoStop}%
\bibitem [{\citenamefont {Berrehrah}\ \emph {et~al.}(2014)\citenamefont
  {Berrehrah}, \citenamefont {Bratkovskaya}, \citenamefont {Cassing},
  \citenamefont {Gossiaux}, \citenamefont {Aichelin},\ and\ \citenamefont
  {Bleicher}}]{Berrehrah:2013mua}%
  \BibitemOpen
  \bibfield  {author} {\bibinfo {author} {\bibfnamefont {H.}~\bibnamefont
  {Berrehrah}}, \bibinfo {author} {\bibfnamefont {E.}~\bibnamefont
  {Bratkovskaya}}, \bibinfo {author} {\bibfnamefont {W.}~\bibnamefont
  {Cassing}}, \bibinfo {author} {\bibfnamefont {P.~B.}\ \bibnamefont
  {Gossiaux}}, \bibinfo {author} {\bibfnamefont {J.}~\bibnamefont {Aichelin}},\
  and\ \bibinfo {author} {\bibfnamefont {M.}~\bibnamefont {Bleicher}},\ }\href
  {https://doi.org/10.1103/PhysRevC.89.054901} {\bibfield  {journal} {\bibinfo
  {journal} {Phys. Rev. C}\ }\textbf {\bibinfo {volume} {89}},\ \bibinfo
  {pages} {054901} (\bibinfo {year} {2014})}\BibitemShut {NoStop}%
\bibitem [{\citenamefont {Plumari}\ \emph {et~al.}(2011)\citenamefont
  {Plumari}, \citenamefont {Alberico}, \citenamefont {Greco},\ and\
  \citenamefont {Ratti}}]{PhysRevD.84.094004}%
  \BibitemOpen
  \bibfield  {author} {\bibinfo {author} {\bibfnamefont {S.}~\bibnamefont
  {Plumari}}, \bibinfo {author} {\bibfnamefont {W.~M.}\ \bibnamefont
  {Alberico}}, \bibinfo {author} {\bibfnamefont {V.}~\bibnamefont {Greco}},\
  and\ \bibinfo {author} {\bibfnamefont {C.}~\bibnamefont {Ratti}},\ }\href
  {https://doi.org/10.1103/PhysRevD.84.094004} {\bibfield  {journal} {\bibinfo
  {journal} {Phys. Rev. D}\ }\textbf {\bibinfo {volume} {84}},\ \bibinfo
  {pages} {094004} (\bibinfo {year} {2011})}\BibitemShut {NoStop}%
\bibitem [{\citenamefont {Bannur}(2007)}]{Bannur:2006hp}%
  \BibitemOpen
  \bibfield  {author} {\bibinfo {author} {\bibfnamefont {V.~M.}\ \bibnamefont
  {Bannur}},\ }\href {https://doi.org/10.1016/j.physletb.2007.02.030}
  {\bibfield  {journal} {\bibinfo  {journal} {Phys. Lett. B}\ }\textbf
  {\bibinfo {volume} {647}},\ \bibinfo {pages} {271} (\bibinfo {year}
  {2007})}\BibitemShut {NoStop}%
\bibitem [{\citenamefont {Peshier}\ \emph {et~al.}(1996)\citenamefont
  {Peshier}, \citenamefont {Kampfer}, \citenamefont {Pavlenko},\ and\
  \citenamefont {Soff}}]{Peshier:1995ty}%
  \BibitemOpen
  \bibfield  {author} {\bibinfo {author} {\bibfnamefont {A.}~\bibnamefont
  {Peshier}}, \bibinfo {author} {\bibfnamefont {B.}~\bibnamefont {Kampfer}},
  \bibinfo {author} {\bibfnamefont {O.~P.}\ \bibnamefont {Pavlenko}},\ and\
  \bibinfo {author} {\bibfnamefont {G.}~\bibnamefont {Soff}},\ }\href
  {https://doi.org/10.1103/PhysRevD.54.2399} {\bibfield  {journal} {\bibinfo
  {journal} {Phys. Rev. D}\ }\textbf {\bibinfo {volume} {54}},\ \bibinfo
  {pages} {2399} (\bibinfo {year} {1996})}\BibitemShut {NoStop}%
\bibitem [{\citenamefont {Peshier}\ \emph {et~al.}(2000)\citenamefont
  {Peshier}, \citenamefont {Kampfer},\ and\ \citenamefont
  {Soff}}]{Peshier:1999ww}%
  \BibitemOpen
  \bibfield  {author} {\bibinfo {author} {\bibfnamefont {A.}~\bibnamefont
  {Peshier}}, \bibinfo {author} {\bibfnamefont {B.}~\bibnamefont {Kampfer}},\
  and\ \bibinfo {author} {\bibfnamefont {G.}~\bibnamefont {Soff}},\ }\href
  {https://doi.org/10.1103/PhysRevC.61.045203} {\bibfield  {journal} {\bibinfo
  {journal} {Phys. Rev. C}\ }\textbf {\bibinfo {volume} {61}},\ \bibinfo
  {pages} {045203} (\bibinfo {year} {2000})}\BibitemShut {NoStop}%
\bibitem [{\citenamefont {Sambataro}\ \emph {et~al.}(2024)\citenamefont
  {Sambataro}, \citenamefont {Greco}, \citenamefont {Parisi},\ and\
  \citenamefont {Plumari}}]{Sambataro:2024mkr}%
  \BibitemOpen
  \bibfield  {author} {\bibinfo {author} {\bibfnamefont {M.~L.}\ \bibnamefont
  {Sambataro}}, \bibinfo {author} {\bibfnamefont {V.}~\bibnamefont {Greco}},
  \bibinfo {author} {\bibfnamefont {G.}~\bibnamefont {Parisi}},\ and\ \bibinfo
  {author} {\bibfnamefont {S.}~\bibnamefont {Plumari}},\ }\href
  {https://doi.org/10.1140/epjc/s10052-024-13276-6} {\bibfield  {journal}
  {\bibinfo  {journal} {Eur. Phys. J. C}\ }\textbf {\bibinfo {volume} {84}},\
  \bibinfo {pages} {881} (\bibinfo {year} {2024})}\BibitemShut {NoStop}%
\bibitem [{\citenamefont {Hosoya}\ and\ \citenamefont
  {Kajantie}(1985)}]{Hosoya:1983xm}%
  \BibitemOpen
  \bibfield  {author} {\bibinfo {author} {\bibfnamefont {A.}~\bibnamefont
  {Hosoya}}\ and\ \bibinfo {author} {\bibfnamefont {K.}~\bibnamefont
  {Kajantie}},\ }\href {https://doi.org/10.1016/0550-3213(85)90499-7}
  {\bibfield  {journal} {\bibinfo  {journal} {Nucl. Phys. B}\ }\textbf
  {\bibinfo {volume} {250}},\ \bibinfo {pages} {666} (\bibinfo {year}
  {1985})}\BibitemShut {NoStop}%
\bibitem [{\citenamefont {Dey}\ \emph {et~al.}(2022)\citenamefont {Dey},
  \citenamefont {Samanta}, \citenamefont {Ghosh},\ and\ \citenamefont
  {Satapathy}}]{PhysRevC.106.044914}%
  \BibitemOpen
  \bibfield  {author} {\bibinfo {author} {\bibfnamefont {J.}~\bibnamefont
  {Dey}}, \bibinfo {author} {\bibfnamefont {S.}~\bibnamefont {Samanta}},
  \bibinfo {author} {\bibfnamefont {S.}~\bibnamefont {Ghosh}},\ and\ \bibinfo
  {author} {\bibfnamefont {S.}~\bibnamefont {Satapathy}},\ }\href
  {https://doi.org/10.1103/PhysRevC.106.044914} {\bibfield  {journal} {\bibinfo
   {journal} {Phys. Rev. C}\ }\textbf {\bibinfo {volume} {106}},\ \bibinfo
  {pages} {044914} (\bibinfo {year} {2022})}\BibitemShut {NoStop}%
\bibitem [{\citenamefont {G\"ursoy}\ \emph {et~al.}(2014)\citenamefont
  {G\"ursoy}, \citenamefont {Kharzeev},\ and\ \citenamefont
  {Rajagopal}}]{PhysRevC.89.054905}%
  \BibitemOpen
  \bibfield  {author} {\bibinfo {author} {\bibfnamefont {U.}~\bibnamefont
  {G\"ursoy}}, \bibinfo {author} {\bibfnamefont {D.}~\bibnamefont {Kharzeev}},\
  and\ \bibinfo {author} {\bibfnamefont {K.}~\bibnamefont {Rajagopal}},\ }\href
  {https://doi.org/10.1103/PhysRevC.89.054905} {\bibfield  {journal} {\bibinfo
  {journal} {Phys. Rev. C}\ }\textbf {\bibinfo {volume} {89}},\ \bibinfo
  {pages} {054905} (\bibinfo {year} {2014})}\BibitemShut {NoStop}%
\bibitem [{\citenamefont {Das}\ \emph {et~al.}(2017)\citenamefont {Das},
  \citenamefont {Plumari}, \citenamefont {Chatterjee}, \citenamefont {Alam},
  \citenamefont {Scardina},\ and\ \citenamefont {Greco}}]{DAS2017260}%
  \BibitemOpen
  \bibfield  {author} {\bibinfo {author} {\bibfnamefont {S.~K.}\ \bibnamefont
  {Das}}, \bibinfo {author} {\bibfnamefont {S.}~\bibnamefont {Plumari}},
  \bibinfo {author} {\bibfnamefont {S.}~\bibnamefont {Chatterjee}}, \bibinfo
  {author} {\bibfnamefont {J.}~\bibnamefont {Alam}}, \bibinfo {author}
  {\bibfnamefont {F.}~\bibnamefont {Scardina}},\ and\ \bibinfo {author}
  {\bibfnamefont {V.}~\bibnamefont {Greco}},\ }\href
  {https://doi.org/https://doi.org/10.1016/j.physletb.2017.02.046} {\bibfield
  {journal} {\bibinfo  {journal} {Physics Letters B}\ }\textbf {\bibinfo
  {volume} {768}},\ \bibinfo {pages} {260} (\bibinfo {year}
  {2017})}\BibitemShut {NoStop}%
\bibitem [{\citenamefont {G\"ursoy}\ \emph {et~al.}(2018)\citenamefont
  {G\"ursoy}, \citenamefont {Kharzeev}, \citenamefont {Marcus}, \citenamefont
  {Rajagopal},\ and\ \citenamefont {Shen}}]{PhysRevC.98.055201}%
  \BibitemOpen
  \bibfield  {author} {\bibinfo {author} {\bibfnamefont {U.}~\bibnamefont
  {G\"ursoy}}, \bibinfo {author} {\bibfnamefont {D.}~\bibnamefont {Kharzeev}},
  \bibinfo {author} {\bibfnamefont {E.}~\bibnamefont {Marcus}}, \bibinfo
  {author} {\bibfnamefont {K.}~\bibnamefont {Rajagopal}},\ and\ \bibinfo
  {author} {\bibfnamefont {C.}~\bibnamefont {Shen}},\ }\href
  {https://doi.org/10.1103/PhysRevC.98.055201} {\bibfield  {journal} {\bibinfo
  {journal} {Phys. Rev. C}\ }\textbf {\bibinfo {volume} {98}},\ \bibinfo
  {pages} {055201} (\bibinfo {year} {2018})}\BibitemShut {NoStop}%
\bibitem [{\citenamefont {Fukushima}\ \emph {et~al.}(2008)\citenamefont
  {Fukushima}, \citenamefont {Kharzeev},\ and\ \citenamefont
  {Warringa}}]{PhysRevD.78.074033}%
  \BibitemOpen
  \bibfield  {author} {\bibinfo {author} {\bibfnamefont {K.}~\bibnamefont
  {Fukushima}}, \bibinfo {author} {\bibfnamefont {D.~E.}\ \bibnamefont
  {Kharzeev}},\ and\ \bibinfo {author} {\bibfnamefont {H.~J.}\ \bibnamefont
  {Warringa}},\ }\href {https://doi.org/10.1103/PhysRevD.78.074033} {\bibfield
  {journal} {\bibinfo  {journal} {Phys. Rev. D}\ }\textbf {\bibinfo {volume}
  {78}},\ \bibinfo {pages} {074033} (\bibinfo {year} {2008})}\BibitemShut
  {NoStop}%
\bibitem [{\citenamefont {Gusynin}\ and\ \citenamefont
  {Shovkovy}(1997)}]{PhysRevD.56.5251}%
  \BibitemOpen
  \bibfield  {author} {\bibinfo {author} {\bibfnamefont {V.~P.}\ \bibnamefont
  {Gusynin}}\ and\ \bibinfo {author} {\bibfnamefont {I.~A.}\ \bibnamefont
  {Shovkovy}},\ }\href {https://doi.org/10.1103/PhysRevD.56.5251} {\bibfield
  {journal} {\bibinfo  {journal} {Phys. Rev. D}\ }\textbf {\bibinfo {volume}
  {56}},\ \bibinfo {pages} {5251} (\bibinfo {year} {1997})}\BibitemShut
  {NoStop}%
\end{thebibliography}%
\end{document}